\documentclass[12pt]{article}
\usepackage{lscape}
\usepackage{pdflscape}
\usepackage{natbib}
\usepackage{graphicx}
\usepackage{amsmath,amssymb}
\usepackage{tabularx}
\usepackage{booktabs}
\usepackage{rotating}
\usepackage{dcolumn}
\usepackage{multirow}
\usepackage{nicefrac}
\usepackage{float}
\usepackage{pdflscape}
\usepackage{subcaption}
\usepackage{comment}
\usepackage{xcolor}
\usepackage[singlespacing,nodisplayskipstretch]{setspace}
\usepackage{color}
\usepackage{url}
\usepackage[margin=1in]{geometry}
\usepackage{varioref}
\usepackage{eurosym}
\usepackage{mdframed}
\usepackage[pdfpagemode=UseThumbs,%
     colorlinks=true,%
     urlcolor=navy,%
     citecolor=black,%
     linkcolor=black]{hyperref}
\usepackage{cleveref}
\usepackage{rotating}
\usepackage{enumitem}
\usepackage{nameref}
\usepackage{array,makecell,booktabs}
\newcolumntype{R}[1]{>{\raggedleft\arraybackslash}p{#1}}   
\newcolumntype{C}[1]{>{\centering\arraybackslash}p{#1}}    

\usepackage[font=bf]{caption}

\newtheorem{theorem}{Theorem}

\newtheorem{assumption}[theorem]{Assumption}

\newtheorem{definition}[theorem]{Definition}

\newcolumntype{d}[1]{D{.}{.}{#1}}
\newcolumntype{e}{D{.}{.}{-1}}
\newcolumntype{.}{D{.}{.}{3}}
\definecolor{navy}{rgb}{0,.3,.7}

\graphicspath{{figs/}}

\begin{document}

\title{\large Do Test Scores Help Teachers Give Better Track Advice to Students? \\
 A Principal Stratification Analysis\thanks{
 Ichino: U. Bologna, European University Institute and CEPR (andrea.ichino@unibo.it); Mealli: European University Institute and U. Firenze (fabrizia.mealli@eui.eu); Viviens: European University Institute (javier.viviens@eui.eu). We are grateful to Gustav Ax{\'e}n, Kosuke Imai, Matthijs Osterveen, Lina Segers and Dinand Webbink for insightful discussions.
}
\vspace{.8 cm}
}
\vspace{.8 cm}
\author{
Andrea Ichino \and Fabrizia Mealli \and  Javier Viviens}

\date{\vspace{.8 cm} February 13,  2026 \\
}
\maketitle

\begin{abstract}
Every year, over one million EU students choose a secondary school track based on teacher recommendations, yet little evidence shows this yields optimal assignments. Using Dutch data, we examine whether access to standardized test scores improves recommendation quality. We develop a Principal-Stratification metric in a quasi-randomized setting, conduct a welfare analysis that flexibly weights short- and long-term losses, and assess principal fairness by examining whether test-score access affects equity across protected attributes. Results are robust to replacing the Exclusion Restriction assumption underlying our main identification strategy with alternative assumptions. Allowing recommendation upgrades when test scores exceed expectations increases successful placement in more demanding tracks by at least 6\%, while misplacing 7\% of weaker students. Only unrealistically high weights on short-term losses would justify banning such upgrades. Test-score access also yields fairer recommendations for immigrant and low-SES students. Our methodology and findings contribute to the literature on algorithm-assisted human decisions.

\vspace{0.5 cm}
\noindent JEL-Code: I2

\vspace{0.2cm}
\noindent Keywords: principal stratification; secondary school track recommendations.
\end{abstract}

\thispagestyle{empty}

\newpage

\begin{spacing}{1.5}

\setcounter{page}{1}
\section{Introduction}

Every year, more than one million EU students, upon completing primary education, choose a secondary school track based on recommendations received from their teachers. In some institutional settings—such as the Dutch school system examined in this paper—only under exceptional circumstances students may begin secondary education in a more demanding track than the one recommended to them. Despite the potentially dramatic consequences that bad track recommendations may have, there is no clear evidence that assigning to teachers alone the role of providing these recommendations is the best solution and generates an optimal track assignment.\footnote{There is a wide literature on teachers' biases that may affect the quality of secondary school track recommendations. Recent examples are: 
\cite{Driessen2008},
\cite{Burgess2013},
\cite{Gershenson2016}, 
\cite{Osikominu2021}, 
\cite{vanLeest2021},
\cite{Geven2021},
\cite{Carlana2022b}, 
\cite{Carlana2022c}, 
\cite{Carlana2022a}, 
\cite{Ferman2022},
\cite{Batruch2023}
\cite{Bach2023},  
\cite{Alesina2024},
\cite{Huizen2021}, and 
\cite{Falk2020}.
\label{f:bias_teacher}}   
 Our goal in this paper is to establish, using a well-defined metric and a quasi-randomized experiment, whether teachers can ``improve" the quality of their recommendations after seeing the scores of standardized tests taken by their students.

To the best of our knowledge, there is no consensus on what the goal of secondary school track advice should be or what defines a ``better" recommendation.
The metric we adopt is the one based on Principal Stratification \citep{frangakis2002principal}\footnote{For a discussion on Principal Stratification and how it is rooted in the Instrumental Variable literature,  see \cite{MealliMattei2012Refreshing}.}  that \cite{Imai2023} use to study whether algorithms can help judges in deciding which arrested individuals should be released while waiting for their trial: in this case, a decision is better than another if it avoids releasing subjects at risk of repeating a crime.
In the Dutch setting, a reform was introduced that gives teachers the right (but not the obligation) to revise their initial track assignment upwards if the student performs better on a standardized test than the initial recommendation would imply. This reform offers an ideal setting to apply the metric that we adopt\footnote{The setting offered by the reform was also used in \cite{Oosterveen2022}. }.

The nature of this metric in the context of track advising can be intuitively described with reference to an education system featuring two tracks: low (e.g., vocational) and high (e.g., academically oriented).  We make
two assumptions that we maintain throughout the analysis: (1) if a student is able to complete a more difficult track, it is better she is allowed to do so for herself (because she can obtain more desirable lifetime outcomes) and for society (inasmuch as there are collective gains from a more educated community);  (2) changing between tracks is costly for students and for the education system. 
Note that given these two assumptions, even in the absence of negative congestion externalities in the educational process and in the labor market\footnote{For evidence and a discussion of these externalities see \cite{ichino2025}.} it would be sub-optimal to send all students to the high track because it would be costly for some of them not to complete it and have to change to the low track.

Let's divide the population of students into groups --  \emph{principal} strata  \citep{frangakis2002principal} -- that differ by how the education outcome depends on track recommendations.
The first group in which we are interested comprises students for whom the track recommendation determines unequivocally the completed track: if the student is recommended the low track in the first year, she will graduate in the same track at the end of secondary school, while if recommended the high track, this will be her final graduation outcome. These are students for whom the teacher's recommendation is a self-fulfilling prophecy, thus becoming crucially important. In light of this consideration, we call these subjects ``Helpable'' (H) to signal that they are the ones who can be {\it helped} by a high first-year track recommendation instead of a low one.\footnote{They correspond to the group of cases that, in the context of \cite{Imai2023}, are labeled as ``Preventable'' crimes. Similarly, \cite{Oosterveen2022} label them as ``Trapped-in-Track'' (TT).} Adapting the methodology proposed by \cite{Imai2023} to this context, we can establish whether the information provided by test scores helps teachers improve their recommendations by measuring the fraction of students in the  H group that are recommended the high track by teachers and see if it increases when teachers can decide to upgrade based on test score information. In the Principal Stratification framework, this is the Average Principal Causal Effect ($APCE$) of ``changing the source" of a recommendation (i.e., giving test score information to the source, in our context) within the principal stratum of H students. If this $APCE$ is positive, test scores help teachers provide better track advice by upgrading to more challenging tracks students who can effectively complete them.

We are also interested in two other groups of students: those who, independently of the track initially recommended to them, always finish secondary education in the low track, and those who, instead, always graduate from the high track independently of the initial recommendation. Adapting to our setting the notation of \cite{Oosterveen2022}, we call these students \textit{Always Low} (AL) and  \textit{Always High} (AH), respectively. The common characteristic of these two groups is that the final student's outcome is not affected by the teacher's recommendation; however, this outcome can be achieved in different ways, which are more or less costly, depending on the number of track changes they require. In these cases, the goal of a recommendation should be to minimize the number of track changes that these two types of students will experience if they start on a track that differs from the one they will ultimately complete. The best recommendation is then the one that sends all AL students to the low track and all AH students to the high track. Therefore, allowing teachers to upgrade based on test scores produces more desirable outcomes, the larger the $APCE$ is for AH students and the lower it is for their AL peers.\footnote{It is conceivable the existence of a fourth group of students who always finish secondary education in the {\it opposite} track with respect to the one that was initially recommended to them. This is discussed below. }

We estimate an $APCE$ for H students which is not lower than 6\%, suggesting that when teachers can upgrade on the basis of test scores, the quality of their advice improves by at least 6\% as measured by the fraction of these students that are recommended a more difficult track and can complete it successfully, while in the counterfactual case, they would have remained in a lower level track. For AH students, for whom an upgrade is also desirable to reduce the cost of track changes, the $APCE$ is estimated to have approximately the same lower bound. 

In the case of AL students, for whom an upgrade is not desirable, the corresponding estimates are also positive with an $APCE$ of at least 7\%. This suggests that a significant fraction of students who are unable to complete the more challenging curriculum are upgraded, and we cannot rule out that this fraction is even greater than that estimated for the H and AH strata. A possible reason for this finding is that primary school teachers do not bear any consequence related to track changes in secondary schools and are, therefore, less sensitive to the goal of reducing them in the case of AL students.

Given the opposite desirability of the findings concerning the effect of test scores on the quality of teacher advice for H and AH students on one hand and AL students on the other, it is natural to seek a method to obtain a comprehensive evaluation. An approach is offered by the classification framework proposed by \cite{benmichael2024does}. This framework enables us to construct a weighted sum of the losses generated by mistakes in recommendations for the various types of students. We can then test whether this sum decreases when teachers can upgrade based on test scores, depending on the relative weight of the short-term losses suffered by students who must change track because of misplacement versus the lifetime losses of students who are not directed to a more challenging track they can eventually complete. We find that the relative weight of short-term losses would have to be unreasonably high to conclude that teachers should not be allowed to upgrade based on test scores.

Finally, building on \cite{Imai2022}, a corollary contribution of this analysis is that it makes it possible to evaluate whether being allowed to upgrade based on test scores helps teachers improve the {\it fairness} of their recommendations with respect to protected attributes such as SES, race, or gender. For example, suppose that gender is a protected attribute with respect to which we want to assess how fair a recommender is. Then, in the stratum of H students, the recommender is fair if the fraction of high recommendations is the same for female and male students. However, the analogous fraction in other strata could be lower or higher as long as it is equal for both genders. Therefore, in the overall population, females may receive a different fraction of high first-year advice from this recommender; however, this would not constitute a violation of fairness, as it would be due to differences in the gender composition of the strata, not to discrimination based on gender within a stratum.
We find that allowing teachers to upgrade recommendations based on test scores improves fairness. Specifically, it raises the chances that immigrants and low-SES students, who are otherwise discriminated against, are recommended a more difficult track that they can complete relative to natives and high-SES peers, respectively.

The paper is organized as follows. In Section \ref{s:inst}, we describe the Dutch institutional setting and the quasi-randomized experiment that makes our analysis possible. 
The statistical framework and its assumptions follow in Section \ref{s:stat}.
Section \ref{s:res_apce}  presents our results, while Section \ref{s:robust} supports their robustness by showing that they also hold when the Exclusion Restriction assumption is substituted by alternative assumptions. In Section \ref{s:fair} we move to the analysis of the fairness of track recommendations,  and Section \ref{s:conc} concludes.

\section{The Dutch school system and the quasi-experiment}
\label{s:inst}

What needs to be known about the Dutch school system for the purpose of this paper is that it features five main secondary school tracks, of which three are vocational (VMBO-BL, VMBO-KL, VMBO-GT ranked in this order by level of difficulty) and two lead to university studies (HAVO, VWO, similarly ranked). In addition, tracks that are contiguous by level of difficulty can be combined in mixed tracks, so that there are in total nine tracks to which a student can be recommended in the first year of secondary school. These nine tracks are listed from the easiest to the hardest in the row headings of Table \ref{t:tracks}.

\begin{table}[h]
    \caption{Tracks of the Dutch school system}
    \label{t:tracks}
\begin{center}
    \vspace{0.1cm} 
    \begin{tabular}{lccccc}
      \hline \hline
        \\[-2ex]
   \hspace {2cm} & \hspace{2cm}& \hspace{2cm} & \hspace{2cm}
   \\ [-2ex]
Track   &  \multicolumn{2}{c}{\makecell[c]{Students initially \\assigned to the track \\denoted by the row}}  &  \multicolumn{1}{c}{\makecell[c]{Test score cut-\\off for upgrade \\ to next level}} &  \multicolumn{2}{c}{\makecell[c]{Fraction above the cutoff\\ of students eligible\\for upgrade to  next level}} \\
         & 1& 2& 3 & 4 & 5 \\
      \\ [-2ex]
     \hline
       \\ [-2ex]
	V: BL	     &	9,585     & 6,929   &	519	&	\hspace{0.5cm} 0.45 &	0.46\\
	V: BL/KL	 &	3,781     & 3,828	&	526	&	\hspace{0.5cm} 0.31 &	0.30\\
	V: KL	     &	15,698	  & 11,258  &	529 &	\hspace{0.5cm} 0.33 &	0.36\\
	V: KL/GT	 &	3,255     & 4,003	&	529	&	\hspace{0.5cm} 0.49 &	0.54\\
	V: GT	     &	29,117	  & 21,154  &	533	&	\hspace{0.5cm} 0.45 &	0.46\\
	V: GT/HAVO	 &	8,540     & 9,620	&	537	&	\hspace{0.5cm} 0.40 &	0.41\\
    A: HAVO	     &	28,226    &	21,465  &	540	&	\hspace{0.5cm} 0.47 &	0.50\\
	A: HAVO/VWO  &	9,932     & 11,136	&	545	&	\hspace{0.5cm} 0.29 &	0.31\\
    A: VWO       &  27,750    & 24,469  &   n.a &  \hspace{0.5cm} n.a. &  n.a.\\
        \\ [-2ex]
        \hline
        \\ [-2ex]
	Total	     &	135,884	  &	113,862 &	    &  \hspace{0.5cm}  0.42&	0.43\\
     \\ [-2ex]
        \hline
        \\ [-2ex]
    Cohort & 2015 & 2016 & 2015 \& 2016 & \hspace{0.5cm} 2015 & 2016 \\
        \\ [-2ex]
      \hline \hline
    \end{tabular}
\end{center}
\vspace{-0.2cm}
\begin{minipage}{1\linewidth  \setstretch{0.75}}
{\scriptsize Notes: 
The row headings are the names of the secondary school tracks to which a student can be assigned in the Dutch system, ranked by level of difficulty. The first six tracks are vocational, while the last three lead to university studies (academic). Columns 1 and 2 report the number of students initially assigned to each track by their primary education teachers, in view of starting secondary school in the academic year 2015-16 or 2016-17. Column 3 reports the CITO test score cutoffs above which a student in the track denoted by the corresponding row may be upgraded to the next level by the primary school teacher; the test score takes on integer values $\in\{\underline{S}, \underline{S}+1, \underline{S}+2, \dots,  \overline{S}\}$, with  $\underline{S}=501$ and  $\overline{S}=550$. The upgrade is not mandatory. Columns 4 and 5 report, for each initial track, the fraction of students who, in the years considered, may be upgraded due to a sufficiently high CITO test score.}   
\end{minipage}
\end{table}

\begin{figure}[tbp]
	\caption{Time-line for track recommendations and outcomes}
 \label{f:timeline}
 \vspace{-0.5 cm}
	\begin{center}
\includegraphics[width = 14.5cm]{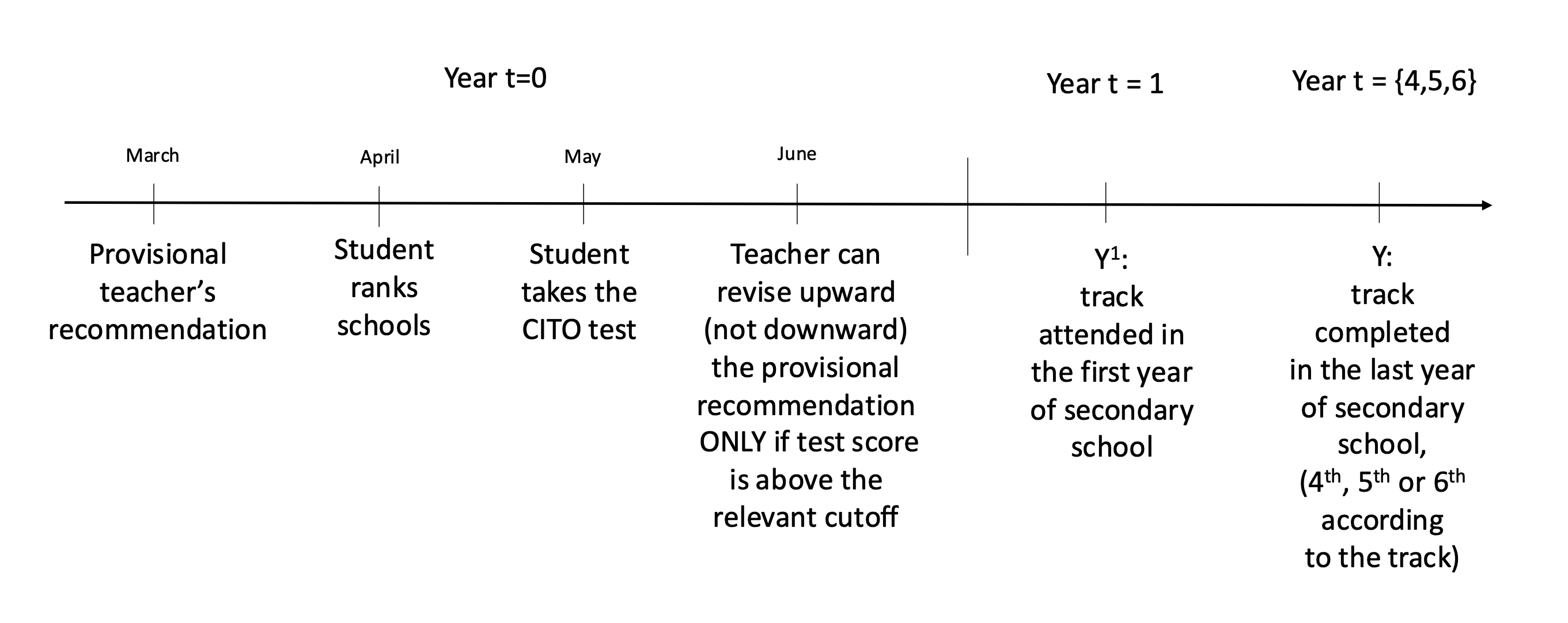}
  \end{center}
 \vspace{-0.6cm}
\end{figure}

Using the database of the Dutch National Cohort Study on Education (NCO), we consider the population of students who begin their secondary school in the academic years  
2015--16 and 2016--17 (hereafter, 2015 and 2016 cohorts).\footnote{This database can be accessed following the instructions provided at this \href{https://www.cbs.nl/nl-nl/onze-diensten/maatwerk-en-microdata/microdata-zelf-onderzoek-doen/microdatabestanden/nco-nationaal-cohortonderzoek-onderwijs}{link}.
For documentation in English, \href{https://www.nationaalcohortonderzoek.nl/sites/nco/files/media-files/20211022_NRO-NCOcodeboek_ENG_def.pdf}{ see here}). The track assignment regulations that generate the quasi-experiment we exploit became effective for students starting secondary school in the 2014-15 academic year, but the data for this cohort do not contain all the necessary information and thus cannot be used.  
}
The timing implied by these regulations is described in Figure \ref{f:timeline}. Let $t=1$ denote the first year of secondary school. 
In March of the previous year, $t=0$, the primary education teachers give their students an initial track recommendation.
The first and second columns of Table \ref{t:tracks} report the number of students in each cohort initially assigned to each track by their primary education teachers as a result of this provisional recommendation.

In April of the same year, $t=0$, the students rank the schools offering the track that was recommended to them (or a less difficult one if they prefer) and are assigned to one of them.\footnote{The more specific method of student allocation differs by municipality. In Amsterdam, for example, a deferred acceptance algorithm is used, with lottery numbers serving as tiebreakers (\citealp{Ketel2023}).} Then, in the following month, they take their end-of-primary education standardized test (CITO), which is graded by a computer with no intervention from primary education teachers.  The scores obtained in this test map into track levels on the basis of nationwide, predetermined test score cutoffs. As in \cite{Oosterveen2022}, we refer to the mapping of a CITO test score to a track level as a ``test-based'' track assignment. If the test-based track assignment is higher than the teacher’s track assignment recorded in March of year $t=0$, the teacher is mandated by the law to consider an upward revision of the provisional recommendation. Such revision is only optional and dependent on the teacher’s judgment.
Column 3 of Table \ref{t:tracks} reports the CITO test score cutoffs above which a student, whose provisional recommendation is the track of the corresponding row, must be re-evaluated by the teacher for a possible upgrade.  Columns 4 and 5 report, for each track, the fraction of students who 
may be upgraded because of a sufficiently high CITO test score.

  Let $R_{i}$ be a dummy variable equal to $0$ if the primary education teacher \textit{finally} recommends the low track to student $i$.
 $R_{i}$ is, instead, equal to $1$ if the teacher recommends an upgrade to a higher track. 
 Note that the 52,219 students whose provisional recommendation is the highest track (VWO in the last row of Table \ref{t:tracks}) cannot be treated with $R_{i}=1$ because there is no higher track to which they can be upgraded. Our analysis focuses on the students for whom the track provisionally recommended in March of year $t=0$ by the teacher is one of those listed in the first 8 rows of Table \ref{t:tracks}, totaling 197,527 students.

 Denote with $K^\tau_i$ the track that student $i$ in cohort $\tau \in \{2015, 2016\}$ 
 is provisionally recommended by the teacher; $K^\tau_i$ can take on values  $\in \{1,8\}$.  Let $S^\tau_i$ be the score obtained by student $i$ in cohort $\tau$ in the CITO test; $S^\tau_i$ can take on integer values $\in\{\underline{S}, \underline{S}+1, \underline{S}+2, \dots,  \overline{S}\}$, with  $\underline{S}=501$ and  $\overline{S}=550$. Let the assignment to treatment be  $Z_i$, defined as follows:
\begin{equation}
Z_i = \begin{cases} 
 0 &  \mbox{if } \hspace{0.3 cm}   
S^\tau_i =  C^\tau_{K^\tau_i} -1 \hspace{0.5 cm}
 \\
 1 &  \mbox{if } \hspace{0.3 cm} 
  S^\tau_i = C^\tau_{K^\tau_i} \hspace{0.5 cm}
\\
 \mbox{else} &  \mbox{if } \hspace{0.3 cm} 
  S^\tau_i \neq \{C^\tau_{K^\tau_i}-1, C^\tau_{K^\tau_i}\} \hspace{0.5 cm}
\end{cases}
\end{equation}
where $C^\tau_{K^\tau_i}$ is the cutoff score that student $i$ in cohort $\tau$ provisionally assigned to track $K^\tau_i$ has to reach in order to be re-evaluated by her primary school teacher for a possible upgrade; ``else''  
denotes students who do not score close to their respective cutoff score. 
If we condition on the students in a particular cohort $\tau$, with a specific value of $K^\tau_i$ and with a value of $Z_i \in \{0,1\}$, we can assume that only randomness, possibly conditional on some covariates to capture the heterogeneity of primary schools as illustrated below, determines whether $S^\tau_i= C^\tau_{K^\tau_i}-1$ or $S^\tau_i=C^\tau_{K^\tau_i}$, that is, whether  
$Z_i=0$ or $Z_i=1$.

This institutional setting generates the quasi-experiment we exploit to answer our research questions. Among students with a realized value of 
$S^\tau_i \in \{C^\tau_{K^\tau_i}-1, C^\tau_{K^\tau_i}\}$, $Z_i$ randomly assigns similar subjects in each of the eight initial tracks to two different ``sources''  of first-year track recommendation: one is the ``teacher alone'', while the other is the ``teacher informed by the CITO test results'', who therefore has the possibility to upgrade students based on such results. This assumption is known in the Regression Discontinuity (RD) literature as ``local randomization'' (LR).\footnote{Local randomization has been first formalized by \cite{LiMatteiMealli2015} and \cite{CattaneoFrandsenTitiunik2015}. The LR approach to the analysis of RD designs has been later discussed in \cite{MatteiMealli2017ObsStudies, BransonMealli2019, CattaneoEtAl_book2020a,CattaneoEtAl_book2020b}, and extended to allow for randomization conditional on covariates in \cite{Forastiere2025}.} One of its advantages is that it allows the researcher to easily deal with discrete running variables without having to rely on the usual continuity-type assumption\footnote{See also \cite{EcklesEtAl2020} and \cite{LiMercatanti2020}}. We formalize the LR assumption for our context in Section \ref{s:stat}, where we also provide evidence to assess its plausibility. 

Next, we denote with $Y_{i}  \in \{0,1\}$ the outcome that we study. We define $Y_{i} = 1$ if student $i$ graduates from a higher track than the one provisionally recommended to him/her in March of $t=0$, before taking the CITO test, and $0$ otherwise. We consider a student as graduating in a given track even if he/she does so in more years than the required minimum.\footnote{Our conclusions do not change qualitatively if we assume instead that a student graduates in a given track only if he/she does so in the regular possible time (results available from the authors). Four years are typically required for a vocational track, while five and six years are needed for the two academic tracks (HAVO and VWO, respectively). Six years are also needed for students who change from VMBO to HAVO, and seven for those who change from HAVO to VWO. 
The release of the NCO data at our disposal follows the 2015 cohort for 7 years, and the 2016 one for only 6 years after enrollment. Graduation is not observed for some students in these cohorts. We thus assume that the students of these cohorts who are still enrolled in secondary school in the last year of observation will graduate in the track in which they are enrolled.}  Our goal is to evaluate which source of track recommendation (``teacher alone'' or ``teacher informed by test scores and allowed to upgrade")
is ``preferable'' according to two criteria: a recommendation is better if it (i) induces a student to enroll in the most difficult track that she/he is able to complete successfully and (ii) reduces to a minimum the chances of costly track changes.

An institutional complication of this quasi-experimental setting is that, on rare occasions, also students who score below the cutoff ($Z_i=0$) may be recommended for an upgrade by their teacher ($R_{i} = 1$).\footnote{As explained by the relevant authorities (\citealp{VanPoNaarVo2025}), 
``[T]here may be situations in which a student does not obtain a higher test recommendation, but a school or teacher still believes that there is more knowledge and skills than expected. In such situations, the school may choose to adjust the provisional school advice. The school's judgment is central here. ... If parents do not agree with the advice, they will consult with the teacher and/or director of the primary school. If they cannot reach an agreement about the level at which the student will enter secondary education, parents can, as a last resort, use the school's complaints procedure.'' \label{foot_def}} 
Moreover, as already mentioned, students with $Z_i=1$  may have $R_{i} = 0$ because their primary school teacher does not want to upgrade them even if an upgrade is possible. Therefore,   as we will show in Section \ref{s:stat}, non-compliance may occur in both assignments to treatment conditions.

\section{Statistical framework}
\label{s:stat}

Consider a sample of $N^\tau_{k}$ units of cohort $\tau$ whose provisional recommendation is track $k$ with $Z_i\in \{0,1\}$, $i \in 1, \cdots, N^\tau_{k}$.
 Let $\mathbf{Z}$ and $\mathbf{R}$ be the $N^\tau_{k}$-dimensional vectors of the possible values of  $Z_i$ and $R_{i}$, with generic element $z_i$ and $r_{i}$ respectively. Denote with $R_{i}(\mathbf{Z}) \in \{0,1\}$ the recommendation that student $i$ receives from the teacher as a function of the assignment vector $\mathbf{Z}$. Similarly, denote with  $Y_{i}(\mathbf{Z}, \mathbf{R}) \in \{0,1\}$ the potential track (low or high) completed by student $i$ at the end of secondary school as a function of the assignment vector $\mathbf{Z}$ and the vector of recommendations $\mathbf{R}$. 
Since different primary schools may have different benchmarks to which they compare their students and decide their provisional track assignments, we must consider the possibility that students with different abilities are assigned to the same provisional track $k$ simply because they have more or less lenient primary teachers.  We denote with $\pi_i$ the benchmark against which student $i$ has been evaluated to decide her/his provisional track assignment.

\subsection{Assumptions}
\label{s:assumptions}

The first assumption we  need is the Stable Unit Treatment Value Assumption (SUTVA, see \citealp{Rubin1980}), which restricts the dependence of the potential outcomes for a student only to the assignment and enrolment of that student, ruling out interference:

\begin{assumption} \label{a:SUTVA}
Stable Unit Treatment Value Assumption: \\
$ R_{i}(\mathbf{Z})=R_{i}(\mathbf{Z'}) 
\hspace{0.5cm} \mbox{if } z_i=z'_i$ \\
\noindent
$ Y_{i}(\mathbf{Z}, \mathbf{R})=Y_{i}(\mathbf{Z'}, \mathbf{R'}) \hspace{0.5cm} \mbox{if  } z_i=z'_i \hspace{0.5cm}\mbox{and 
 } 
 \hspace{0.5cm}
 r_{i}=r'_{i}$.
\end{assumption}
Assumption \ref{a:SUTVA} postulates the existence of two potential versions of $R_{i}(z)$ with $z\in \{0, 1\}$ and four potential versions of $Y_{i}(z, r)$ with $z, r\in \{0, 1\}$. Given SUTVA, the next assumption is:
\begin{assumption} \label{a:rand_sourc}
Local Randomization of the source conditional on primary schools' leniency benchmarks and provisional track assignments:
for every  cohort $\tau$
we have 
$$ \{R_{i}(z), Y_{i}(z,r), X_i, U_i \}  \perp\!\!\!\!\perp Z_i  \hspace{0.1cm}
\vert \hspace{0.1cm} \pi_i , K_i $$
for $z \in \{0, 1\}$ and $r \in \{0, 1\},  $ where $X_i$ and $U_i$ are observable and unobservable covariates, respectively. 
\end{assumption}
This assumption requires some discussion. 
It formally says that for students assigned to the same track and whose teachers have a similar benchmark criterion $\pi_i$ to evaluate beliefs about students' ability, the reasons for them to score just below or at the cutoff are unrelated to either potential outcomes or observed, $X$, or unobserved, $U$, covariates. 
Hence, in their case, $P(Z_i=1|R_{i}(z), Y_{i}(z,r), U_i, X_i, \pi_{i}, K_i)= P(Z_i=1|\pi_{i}, K_i)$.  
If $\pi_i$ is observed, this is a testable implication in that if Assumption \ref{a:rand_sourc} holds, we can expect, within each provisional track assignment and for a given $\pi_i$, the observed characteristics of students to be well-balanced across the score cutoff \citep{MatteiMealli2017ObsStudies,Forastiere2025, {AngristRokkanen2015}}. 
Moreover, if Assumption \ref{a:rand_sourc} holds for all 
tracks, we can view the data as coming from a block-randomized experiment, with blocks defined by the tracks,
and analyze them accordingly \citep{ImbensRubin2015}. 

The problem is that while we observe the initial track assignment and we can condition on scoring just below or at the cutoff, we do not observe the benchmark $\pi_i$ of a student's primary teachers. A solution would be to condition on primary school fixed effects, assuming that teachers in the same school collectively agree on the same benchmark $\pi$. However, we do not have enough students per primary school to conduct such an analysis. What we can do is to proxy $\pi_i$ with the fraction of students in the primary school of $i$ who score above the cutoffs. This fraction measures the average teacher severity in that school because, as more students score above the cutoffs, the stricter their primary school teachers must have been in their provisional track recommendations. We define this variable as
$\tilde{Z}_i = \frac{\sum_{j\in School_{i}}\mathbb{I}(S^\tau_j \geq C^\tau_{K^\tau_j})}{\# School_{i}}$, which takes on the same value for all students of a given school
in a given cohort. $\tilde{Z}_i$ will thus be included in all the specifications of our analysis.

\begin{landscape}

\begin{table}
    \caption{Balancing of covariates}
    \label{t:balance}
    \vspace{-0.1cm}
\begin{center}
\setlength{\tabcolsep}{2pt}
 \begin{tabular}{lccccccccccc}
    \hline \hline
   \\[-2ex]
   & \multicolumn{11}{c}{Panel A: 2015 cohort} \\
    \\[-2ex]
    \hline
      \\[-2ex]
     Track & Female & Immigrant & College & Missing & Household & Urbanity  & Primary & Primary & Student's &  Obs. for & Obs. for \\ 
               &    & origin    & mother  & college & income   & primary & school & school \# & age in &  \(Z=0\) & \(Z=1\) \\
       & & &  & mother &  & school & confession & students & months &  \\  
& 1 & 2 & 3 & 4 & 5 & 6 & 7 & 8 & 9 & 10 &11\\
         \\[-2ex]
    \hline
      \\[-2ex]
 V: BL & 1 & 1 & 1 & 1 & 1 & 1 & 1 & 1 & 1 &   519 &   511 \\ 
   V: BL/KL & 1 & 0.7517 & 1 & 1 & 1 & 1 & 1 & 1 & 1 &   173 &   183 \\ 
   V: KL & 1 & 0.1612 & 1 & 0.2872 & 1 & 0.1856 & 1 & 1 & 1 &  1199 &   860 \\ 
 V: KL/GT & 1 & 1 & 1 & 1 & 1 & 1 & 1 & 1 & 0.4632 &   274 &   219 \\ 
   V: GT & 1 & 1 & 1 & 1 & 1 & 1 & 0.6102 & 1 & 1 &  1942 &  1948 \\ 
   V: GT/HAVO & 1 & 0.8050 & 1 & 1 & 1 & 1 & 1 & 1 & 1 &   614 &   570 \\ 
   A: HAVO  & 1 & 1 & 1 & 1 & 1 & 0.4935 & 1 & 1 & 1 &  2149 &  2172 \\ 
   A: HAVO/VWO & 0.0787 & 1 & 1 & 1 & 1 & 1 & 1 & 1 & 1 &   843 &   769 \\ 
      \\[-2ex]
    \hline
    \\[-2ex]
   & \multicolumn{11}{c}{Panel B: 2016 cohort} \\
    \\[-2ex]
    \hline
      \\[-2ex]
     Track & Female & Immigrant & College & Missing & Household & Urbanity  & Primary & Primary & Student's &  Obs. for & Obs. for \\ 
  &        & origin    & mother  & college & income   & primary & school & school \# & age in &  \(Z=0\) & \(Z=1\) \\
       & & &  & mother &  & school & confession & students & months &  \\   
    & 1 & 2 & 3 & 4 & 5 & 6 & 7 & 8 & 9 & 10 &11\\
            \\[-2ex]
        \hline
      \\[-2ex]
 V: BL & 1 & 0.7578 & 0.7132 & 1 & 1 & 1 & 1 & 1 & 1 &   396 &   458 \\ 
   V: BL/KL  & 1 & 1 & 1 & 1 & 1 & 1 & 1 & 1 & 1 &   195 &   195 \\ 
   V: KL & 0.4012 & 1 & 1 & 1 & 1 & 1 & 1 & 1 & 0.5522 &   656 &   630 \\ 
   V: KL/GT  & 1 & 1 & 0.8141 & 1 & 1 & 1 & 1 & 1 & 1 &   261 &   284 \\ 
   V: GT & 1 & 1 & 0.0193 & 0.9419 & 1 & 1 & 1 & 0.8404 & 1 &  1386 &  1376 \\ 
   V: GT/HAVO  & 1 & 0.0915 & 1 & 1 & 1 & 1 & 1 & 1 & 0.3441 &   923 &   689 \\ 
   A: HAVO & 1 & 1 & 1 & 1 & 1 & 1 & 1 & 1 & 1 &  1643 &  1728 \\ 
   A: HAVO/VWO & 0.0976 & 0.3389 & 1 & 1 & 1 & 0.4697 & 1 & 1 & 0.1691 &   983 &   952 \\ 
      \\[-2ex]
    \hline \hline
\end{tabular}
\end{center}
    \vspace{-0.1cm}
\begin{minipage}{1\linewidth  \setstretch{0.75}}
{\scriptsize Notes: The row headings are the names of the secondary school tracks to which a student can be assigned in the Dutch system, ranked by level of difficulty. Columns 1 to 9 report, for each track and covariate, the Holm-Bonferroni p-value of a test for the equality of means in the two assignment-to-treatment groups defined by $Z_i\in\{0,1\}$. 
The definition of the covariates whose names are not self-explanatory is as follows.
Urbanity of a primary school is a categorical variable indicating the population density at the location of the primary school. Primary school confession is a categorical variable indicating the type of confession in primary schools (Protestant, Catholic, Public, or Other). Age in months refers to the age of the student at the time she/he receive the first recommendation.  
The last two columns report the number of observations in the two assignment groups. Standard errors are always clustered at the primary school level.
}
\end{minipage}
\end{table}
\end{landscape}

Columns 1 to 9 of Table \ref{t:balance} support the validity of Assumption \ref{a:rand_sourc} by showing that nine observed
characteristics of students are balanced in the immediate vicinity of the cutoff. These covariates are: gender, migrant origin, mother's education, missing mother's education, household income, urban location of the primary school, confession of the primary school, number of students in the primary school, and age in months. Under  Assumption \ref{a:rand_sourc}, our estimates should not be (and in fact are not, as we will show below) affected by the inclusion or exclusion of these balanced covariates because they \textit{are not} correlated with $Z_i$, conditional on $\tilde{Z}_i$. However, including them may increase efficiency since they \textit{are} likely to be correlated with $R_i$ and $Y_i$, and for this reason, they will appear in our preferred specification.

 It must be noted that the validity of Assumption \ref{a:rand_sourc} does not imply that the proportion of students scoring above or below the cutoff is around $50\%$, as shown in the last two columns of Table \ref{t:balance}. The CITO test score is the result of the students responding correctly to a series of questions. To simplify this data-generating process, think about the score being generated (approximately) by a binomial distribution and suppose that units in the sub-sample that we consider have the same probability $\chi$ to respond correctly to a question in the test. If so, these units also have the same probability (propensity score) to score at the cutoff ($S^t_i=C^t_{K^\tau_i}$), yet this probability is not equal to the probability of scoring below ($S^\tau_i=C^\tau_{k^\tau_i}-1$) because these probabilities follow a binomial distribution. The difference in these two probabilities will depend on $\chi$ and on the number of questions in the test, but it is by no means an indication of the failure of Assumption \ref{a:rand_sourc}. In the data, we may also observe differences in this probability 
from track to track because the cutoff is different or because the difficulty of the test (i.e., the probability of a correct response to single questions) is different. These findings would not represent violations of Assumption \ref{a:rand_sourc}.\footnote{This binomial example is, of course, over-simplistic; most likely, not all the questions have the same difficulty, and the students may not have the exact same probability of responding correctly to all questions. It is only meant to illustrate why, with a discrete score, the probability of being below or above the cutoff may differ even if the mechanism that generated $Z_i$ is random and unrelated to potential outcomes or covariates.}

Imai et al. (2023) invoke two additional assumptions that are necessary to answer our research question.

\begin{assumption} \label{a:ec}
    Exclusion restriction: \\
    $Y_{i}(z,r) = Y_{i}(z',r) = Y_{i}(r) \hspace{0.3 cm} \mbox{for } z, z',r \in \{0,1\}$.
\end{assumption}    This assumption requires that the potential outcomes $Y_{i}(z,r)$
depend only on the recommendation $r$ and not on the source $z$ of the recommendation so that they can be written as $Y_{i}(r)$.
Intuitively, this means that the track in which the student graduates should only be affected by scoring above or below the cutoff through its effect on the teacher's recommendation. A violation of this restriction would occur, for example, if scoring above the cutoff increased the probability of graduating in the high track regardless of whether teachers decide to upgrade.  In Section \ref{s:robust} we will show that we obtain similar results if we assume a Homogeneity and a Principal Ignorability assumption instead of the Exclusion Restriction (ER)
assumption, and therefore that our conclusions are robust and stable across these alternative identifying assumptions. 

\begin{assumption}
\label{a:strata_monoton}
Monotonicity of $Y_{i}$ with respect to  $R_{i}$: \\
$Y_{i}(1) \geq Y_{i}(0).$ 
\end{assumption}   
    This assumption plausibly excludes the existence of students, whom we call ``Rebels", who would not complete the high track if it were the recommended track, but who would complete it if the low track were recommended. These students would be characterized by the following set of potential outcomes: $(Y(1),Y(0)) = (0,1)$.\footnote{\cite{Oosterveen2022} allow for the existence of a similar stratum, which they label ``Slow Starters". When estimating lower and upper bounds for the corresponding proportion, they cannot reject that this proportion is zero.
    } 
    
Finally, a different type of monotonicity assumption needs to be discussed in our setting. 
\begin{assumption}
\label{a:iv_monoton}
Monotonicity of $R_i$ with respect to  $Z_i$: \\
$R_{i}(1) \geq R_{i}(0).$ 
\end{assumption}   
This assumption is typically made in Instrumental Variable (IV) settings with non-compliance\footnote{ See \cite{Angrist1994} and \cite{Angrist1996}.} and excludes the existence of ``Defiers'', that is, students who would be recommended the low track when scoring at or above the cutoff and the high track when scoring below.  \cite{Imai2023} do not make this assumption. In their context, a judge who decides with the help of an algorithm may release an arrested subject while the same judge, not helped by the algorithm, keeps the same subject in prison, or vice versa. More generally, every set of counterfactual decisions with and without the help of the algorithm is possible in their context. 
To put it differently, they do not have (and do not want to have) any prior on which source of a decision is preferable: the fraction of ``Preventable'' cases kept under arrest may be higher when the judge is helped by the algorithm or when the judge decides alone, and their goal is precisely to assess for which one of the two sources the fraction is higher.

\begin{table}[h]
    \caption{Non-compliance statistics}
    \label{t:non_comp}
\begin{center}
    \vspace{0.1cm} 
        \begin{tabular}{lcccc}
      \hline \hline
   \\ [-2ex]
      TRACK   
       &  $P(R=1|Z=0)$&  $P(R=1|Z=1)$& $P(R=1|Z=0)$&  $P(R=1|Z=1)$\\
 &1 & 2 & 3 & 4 \\
      \\ [-2ex]
      \hline
 \\ [-2ex]
           
V: BL             &       .0039&       .0352&       .0025&       .0677\\
V: BL/KL       &       .0231&       .1148&       .0205&       .1897\\
V: KL           &       .0209&       .1767&       .0122&       .1952\\
V: KL/GT      &       .0109&       .1233&       .0153&       .1655\\
V: GT       &       .0036&        .056&       .0065&       .0574\\
V: GT/HAVO     &       .0163&       .1684&       .0206&       .1858\\
A: HAVO          &       .0051&       .0603&       .0085&       .0758\\
A: HAVO/VWO   &       .0237&       .1821&       .0193&        .188\\ \hline
COHORT      &        2015&        2015&        2016&        2016\\
 \\[-2ex]
\hline \hline
    \end{tabular}
\end{center}
\vspace{-0.2cm}
\begin{minipage}{1\linewidth  \setstretch{0.75}}
{\scriptsize Notes: The row headings are the names of the secondary school tracks to which a student can be assigned in the Dutch system, ranked by level of difficulty. The other columns report the probabilities indicated in the column headings. }   
 \end{minipage}
\end{table}

In our case, the institutional setting is such that an upgrade to a higher track should be possible only if students score at or above the cutoff. Therefore, Defiers should not exist by construction because no student below the cutoff should be eligible for an upgrade. For the same reasons, also Always Takers should not exist. On the other hand, teachers may decide not to upgrade students who score at or above the cutoff, so Never Takers may exist in our setting.\footnote{Table \ref{t:NTandZtilde} in the Online \nameref{ap:NTandZtilde} shows that  $\tilde{Z}_i$ correlates positively with the proportion of Never Takers. This finding supports our choice of $\tilde{Z}_i$ as a proxy for the leniency $\pi_i$ of student $i$ teachers.} As anticipated in Section \ref{s:inst}, however,  there are rare instances of non-compliance for students scoring below the cutoff. This is shown in Table \ref{t:non_comp}: in the first and third columns, the probability that a student is upgraded ($R_{i} = 1$) even if scoring below the cutoff ($Z_i = 0$) is low but positive in all tracks, ranging between 0.3\%  for BL in the 2016 cohort to 2.4\% for HAVO/VWO in the 2015 cohort, and therefore Always Takers do exist in our setting. In the second and fourth columns, the probability that a student is upgraded ($R_{i} =1$) when scoring above the cutoff ($Z_i = 1$) is significantly lower than 1 in all tracks, ranging from 3.5\% for BL in the 2015 cohort to 19.5\% for KL in the 2016 cohort. So, and this is no surprise, Never Takers also exist and are frequent.  In general, bridge tracks show a significantly higher probability of upgrading than the other tracks, which makes intuitive sense as $R_{i}$ is defined to be equal to one if the student is upgraded to the higher component of the mixed track. 

This evidence indicates that non-compliance is certainly present in our setting. However, we do not see reasons to think that in the rare cases in which students scoring below the cutoff are upgraded (see footnote \ref{foot_def} for the possible reasons), these same students would not be upgraded if scoring above, which makes them Always Takers but not Defiers. Hence,   Assumption \ref{a:iv_monoton} of Monotonicity of $R$ with respect to $Z$, excluding the possibility of Defiers, appears plausible. Even if not required to identify causal effects in our setting, we will refer to it for the interpretation of our results.

\begin{table}[ht]
    \caption{First stage}
    \label{t:first_stage}
\begin{center}
    \vspace{0.1cm} 
         \begin{tabular}{lcccc}
      \hline \hline
   \\ [-2ex]
    Track   &  $\delta$ &  $\delta$&  $\delta$ &  $\delta$  \\
     &1 & 2 & 3 & 4 \\
      \\ [-2ex]
      \hline
 & \hspace{1.5cm} & \hspace{1.5cm}    \\ [-2ex]
V: BL          &        .0319&       .0322&         .07&       .0712\\
          &    \footnotesize   (.0093)&  \footnotesize     (.0092) &    \footnotesize   (.0147)&  \footnotesize     (.0146)\\
V: BL/KL       &     .0931&       .0872&       .1722&       .1615\\
          &       \footnotesize (.0265)&       \footnotesize (.0276) &       \footnotesize (.0308)&       \footnotesize (.0304)\\
V: KL          &       .1639&       .1607&        .187&       .1892\\
          &       \footnotesize (.0152)&       \footnotesize (.0149) &       \footnotesize (.0176)&       \footnotesize (.0177)\\
V: KL/GT       &      .1152&         .11&       .1535&       .1564\\
          &       \footnotesize (.0248)&       \footnotesize (.0244)  &       \footnotesize (.0244)&       \footnotesize (.0247)\\
V: GT          &       .0564&       .0561&       .0537&       .0537\\
          &       \footnotesize (.0058)&       \footnotesize (.0058) &       \footnotesize (.0074)&       \footnotesize (.0074)\\
V: GT/HAVO     &     .1579&       .1533&       .1719&       .1675\\
          &       \footnotesize (.0186)&       \footnotesize (.0182)&       \footnotesize (.0162)&       \footnotesize (.0162)\\
A: HAVO        &    .0581&        .058&       .0707&       .0705\\
          &       \footnotesize (.0063)&       \footnotesize (.0063) &       \footnotesize (.0076)&       \footnotesize (.0076)\\
A: HAVO/VWO    &        .168&       .1667&       .1732&       .1764\\
          &       \footnotesize (.0167)&       \footnotesize (.0166)&       \footnotesize (.0158)&       \footnotesize (.0155)\\
 \\[-2ex]
\hline
\footnotesize COVARIATES $X_i$ & \footnotesize NO & \footnotesize YES& \footnotesize NO & \footnotesize YES  \\ 
 \\[-2ex]
\hline
\footnotesize{COHORT} & \footnotesize 2015& \footnotesize 2015& \footnotesize 2016& \footnotesize 2016 \\
      \hline \hline
    \end{tabular}
\end{center}
\vspace{-0.2cm}
\begin{minipage}{1\linewidth  \setstretch{0.75}}
{\scriptsize Notes: The table reports, separately for each track, estimates of parameter $\delta$ in the first stage regression (\ref{e:first_stage}): $R_{i} = \gamma + \delta Z_i + \mu X_i + \xi \tilde{Z}_i + \epsilon_i$. The row headings are the names of the secondary school tracks to which a student can be assigned in the Dutch system, ranked by level of difficulty. The covariates $X$ included in columns 2 and 4  are those described in  Table \ref{t:balance}, respectively, for the two cohorts.
}   
 \end{minipage}
\end{table}

Table \ref{t:first_stage} reports, separately for each track, estimates of the first stage regression
\begin{equation}
\label{e:first_stage}
R_{i} = \gamma + \delta Z_i + \mu X_i + \xi \tilde{Z}_i + \epsilon_i
\end{equation}
where $\tilde{Z}_i$ is the proxy for $\pi_i$ on which we need to condition for the validity of Assumption \ref{a:rand_sourc} and $X_i$ are the nine balanced covariates described in  Table \ref{t:balance}.
As expected, the estimates of $\delta$ confirm the existence of imperfect compliance. Considering the covariates $X_i$, the stability of the first-stage estimates with and without them supports Assumption \ref{a:rand_sourc} that the assignment to treatment is as good as random.

\subsection{Students' strata}
\label{s:strata}

Under Assumption 1 (SUTVA) and 3 (Exclusion restriction), and according to \cite{frangakis2002principal}, students can be classified in four strata based on the joint values of the two binary potential outcomes: $Y_{i}(R_{i}=1), Y_{i}(R_{i}=0)$.
\begin{itemize}
    \item[AH:]  $(Y_{i}(1),Y_{i}(0)) = (1,1)$\\
    These are  {\it Always High} students who always complete the high track independently of the recommendation they receive.
    \item[AL:] $(Y_{i}(1),Y_{i}(0)) = (0,0)$\\
     These are {\it Always Low} students who always complete the low track independently of the recommendation they receive.
     \item[H:] $(Y_{i}(1),Y_{i}(0)) = (1,0)$\\
     These are {\it Helpable} students who complete the high track if this is the track recommended to them by primary school teachers, and who would not do the same otherwise.
        \item[R:] $(Y_{i}(1),Y_{i}(0)) = (0,1)$\\
     These are {\it Rebel} students who complete the high track if recommended for the low track by primary school teachers, and the low track if recommended for the high track.
\end{itemize}
These principal strata, which we denote with  $J\in\{AL, AH, H, R\}$,  will play a crucial role in the definition of the metric we propose in Section \ref{s:metric} to measure the quality of track recommendations.

\begin{landscape}
\begin{table}[ht]
    \caption{The Principal Strata}
    \label{t:strata2}
\begin{center}
    \vspace{0.1cm} 
    \begin{tabular}{ccccccc}
    \hline \hline
     \\ [-2ex]
 Strata  
 & \scriptsize $\underbrace{Y_{i}(0,0)= Y_{i}(1,0)= Y_{i}(0) }$  
& \scriptsize $\underbrace{Y_{i}(0,1)= Y_{i}(1,1)= Y_{i}(1) }$  
 &  Strata
 & \scriptsize $R_{i}(0)$  
 & \scriptsize$R_{i}(1)$ 
 & Notes
\\ 
\scriptsize $R_{i} \rightarrow Y_{i} $  
&  Exclusion restriction
& Exclusion restriction

& \scriptsize $Z_i \rightarrow R_{i} $ 
&   
&  
& 
\\
        \\ [-2ex]
            1 & 2 & 3 & 4 & 5 & 6  &7\\
      \\ [-2ex]
      \hline
       \\ [-2ex]
AL & 0 & 0 & NT  & 0 & 0 
& \\
      \\ [-2ex]
   &   &   & AT & 1 & 1 \\
        \\ [-2ex]
   &   &   & C & 0 & 1\\
            \\ [-2ex]
   &   &   &\color{blue} D  & \color{blue} 1 & \color{blue} 0 
& \color{blue} May $\exists$ w/out Monotonicity  $Z_i \rightarrow R_{i}$ 
   \\
      \\ [-2ex]
            \hline
       \\ [-2ex]
AH & 1 & 1 & NT  & 0 & 0 
& \\
      \\ [-2ex]
   &   &   & AT  & 1 & 1 \\
        \\ [-2ex]
   &   &   & C & 0 & 1\\
            \\ [-2ex]
   &   &   & \color{blue} D  & \color{blue} 1 & \color{blue} 0 
   & \color{blue} May $\exists$ w/out Monotonicity  $Z_i \rightarrow R_{i}$ \\
      \\ [-2ex]
            \hline
       \\ [-2ex]
H & 0 & 1 & NT  & 0 & 0 
& \\
      \\ [-2ex]
   &   &   & AT  & 1 & 1 \\
        \\ [-2ex]
   &   &   & C & 0 & 1\\
            \\ [-2ex]
   &   &   &\color{blue} D  &\color{blue} 1 &\color{blue} 0 
   & \color{blue} May $\exists$ w/out Monotonicity  $Z_i \rightarrow R_{i}$ \\
      \\ [-2ex]
        \hline
       \\ [-2ex]
\color{red} R & 
\color{red} 1 & 
\color{red} 0 &
\color{red} NT  &
\color{red} 0 &
 \color{red} 0 
& 
\color{red} 
$\nexists$ because Monotonicity $R_{i} \rightarrow Y_{i}$ holds
\\
      \\ [-2ex]
   &   &   &  \color{red} AT  &  \color{red} 1 &  \color{red} 1 \\
        \\ [-2ex]
   &   &   &  \color{red} C &  \color{red} 0 &  \color{red} 1\\
            \\ [-2ex]
   &   &   & \color{red}  D  & 
   \color{red}  1 & \color{red}   0
& \\
      \\ [-2ex]       
      \hline \hline
    \end{tabular}
\end{center}
\vspace{-0.2cm}
\begin{minipage}{1\linewidth  \setstretch{0.75}}
{\scriptsize Notes: The four panels refer to the four Principal Strata listed in the first column of the table and defined by the possible values of $R_{i}$ and $Y_{i}$: the Always-Low (AL), the Always-High (AH), the Helpable (H) and the Rebels (R). Columns 2 and 3 describe the values of the potential outcomes $Y_{i}$ when the treatment $R_{i}$ is equal to $0$ or $1$, respectively, and the Exclusion Restriction assumption holds. Each one of these Principal Strata is divided into additional strata defined by the possible values of $Z_i$ and $R_{i}$, as listed in the fourth column: the Never Takers (NT), the Always Takers (AT), the Compliers (C) and the Defiers (D). Columns 5 and 6 describe the values of the potential treatments  $R_{i}$ when the assignment to treatment $Z_i$ is equal to $0$ or $1$,  respectively. The Notes in the first three rows of the last column make clear that Defiers may or may not exist depending on whether the Monotonicity Assumption \ref{a:iv_monoton} of $R_{i}$ with respect to $Z_i$ holds. The last note in the fourth row states that given the maintained Monotonicity Assumption \ref{a:strata_monoton} of  $Y_{i}$ with respect to $R_{i}$, the Rebels do not exist.

}   
 \end{minipage}
\end{table}
\end{landscape}

Under the same assumptions, the more conventional classification in strata that characterizes compliance types is also possible, based on the joint values of the two binary potential treatment values: $R_{i}(Z_i=1),R_{i}(Z_i=0)$:

\begin{itemize}
    \item[AT:]  $(R_{i}(1),R_{i}(0)) = (1,1)$\\
These are {\it Always Takers} who are always recommended the high track, regardless of their score around the cutoff.
    
    \item[NT:] $(R_{i}(1),R_{i}(0)) = (0,0)$\\
 These are  {\it Never Takers}  who are never recommended the high track, independently of where they score around the cutoff.
   
     \item[C:] $(R_{i}(1),R_{i}(0)) = (1,0)$\\
 These are  {\it Compliers}  who are recommended the high track if they score at the cutoff, and the low track if they score below.
 
     \item[D:] $(R_{i}(1),R_{i}(0)) = (0,1)$\\
 These are  {\it Defiers}  who are recommended the low track if they score at the cutoff, and the high track if they score below.
     
\end{itemize}
We denote with $G\in\{AT, NT, C, D\}$ these strata.

Table \ref{t:strata2} describes the relationship between these basic principal strata. Under Assumption \ref{a:strata_monoton} (Monotonicity of $Y_{i}$ with respect to $R_{i}$), the stratum of the  Rebels, $J=R$ in red, does not exist irrespective of the student belonging to any of the strata $G$, while under  Assumption \ref{a:iv_monoton} (Monotonicity of $R_{i}$ with respect to $Z_{i}$), the stratum of Defiers, $G=D$ in blue,  does not exist in any of the strata $J$.

\subsection{The metric to compare recommendations}
\label{s:metric}

 Following \cite{Imai2023}, the metric that we propose to compare recommendations and establish which one is preferable relies on the classification of students in Principal Strata described in the previous section, combined with the following criteria:

\begin{enumerate}
\item If a student is able to complete a more difficult track, it is better that she is allowed to do it for both
    \begin{itemize}   
    \item[1a)]   herself, because she can obtain more desirable lifetime outcomes
   \item[1b)] and for society, if there are collective gains from a more educated community.
    \end{itemize}   
\item Changing tracks is costly for students and for the education system.
\end{enumerate}

The first criterion defines what constitutes a better recommendation for Helpable (H) students. In their case, the goal of teachers should be to recommend the high track to the largest fraction of these students because, as a result of this recommendation, they will complete precisely this track, achieving a better outcome for themselves as well as for society. Therefore, we need to estimate the difference between ``the fraction of H students who are recommended a high first-year track by teachers who can upgrade based on students' test scores" and ``the analogous fraction when teachers do not see test scores". If this difference is positive, test scores help teachers improve the track advising they provide to their students.

The second criterion defines instead what constitutes a better recommendation for the Always Low (AL) and  Always High (AH) students.  In their cases, track advising should aim to minimize the number of track changes that these two types of students experience during secondary school, if they start on a track that differs from the one they will ultimately complete. Therefore, the best recommendation is to send all AL students to the low track and all AH students to the high track.
 
To formalize this intuitive characterization of what constitutes a better recommendation,  consider the following Average Principal Causal Effect:
 \begin{equation}
 \label{e:apce_J}
     APCE_{J} = E(R_{i}(1) - R_{i}(0) | \mbox{ student $i$ is in } J)
 \end{equation}
The $APCE_{J}$ is the difference between the fraction of students in stratum $J$ who are recommended the high track by teachers who can upgrade based on test scores ($Z_i=1$) minus the fraction receiving the same recommendation by teachers who have not seen the test scores ($Z_i=0$). To put it differently, the   $APCE_{J}$ is the Intention To Treat effect of the source of recommendations on their content in stratum $J$.

Note that each  $APCE_{J}$ is also equal to the difference between the fractions of Compliers and Defiers:
\[APCE_{J} 
=P(G_i=\mbox{C}|\mbox{ student $i$ is in } J)-P(G_i=\mbox{D}|\mbox{ student $i$ is in } J).\]
If Monotonicity of $R_{i}$  with respect to $Z_i$  (Assumption \ref{a:iv_monoton}) holds, Defiers do not exist, and the three $APCE_{J}$ are all non-negative by construction, being equal to the proportion of Compliers. In this case we could conclude\footnote{As already mentioned when we introduced Assumption \ref{a:iv_monoton} in the previous section, we emphasize that this assumption allows to interpret $APCE_{J}$ as the proportion of Compliers in stratum $j$, but it is not necessary for identification.}  that test scores help teachers improve their recommendations if the $APCE_H$ and the $APCE_{AH}$ are positive while the $APCE_{AL}$ is equal to zero. Having established that the $APCE_{J}$ are informative about whether test scores help teachers improve the quality of track advising, the next section shows how these population parameters can be identified using data generated by the quasi-experiment under study.

\subsection{Identification of the $APCE_J$}
\label{s:point_est}

\subsubsection*{Estimands for Helpable (H) students}
\label{s:APCE_H_est}

Adapting Theorem 1 of \cite{Imai2023} to this context, it is possible to show that\footnote{For the derivation of this equation and all the others in this section, see the proofs in the Online \nameref{ap:apce_proof}.
}
\begin{equation}
\label{e:apce_h_est}
  APCE_{H} = \frac
  {E(Y_{i}|Z_i=1) - E(Y_{i}|Z_i=0) }
  {Pr(Y_{i}(1)=1) - Pr(Y_{i}(0)=1)}.
\end{equation}
 where the terms in the numerator do not depend on missing potential outcomes and can be estimated with their sample analogs. 
 The denominator is instead the proportion of Helpable students, given that Rebels are excluded because of the Monotonicity Assumption \ref{a:strata_monoton}, and can only be partially identified. In some settings, even if principal strata are latent, their proportions can be point identified under specific assumptions.  This is the case, for example, of the proportion of compliers in IV settings, because $Z$ is random. In our case, because the principal strata are defined by the joint values of $Y(r)$ and $R$ is confounded, the strata proportions are not point-identified unless we impose further assumptions on the distribution of potential outcomes and the recommendation $R$.
 
 To address this problem, \cite{Imai2023} propose to derive non-parametric bounds for the $APCE_{H}$
by bounding the terms $Pr(Y_{i}(R)=1)$ in the denominator of (\ref{e:apce_h_est}). 
Using  Assumption 1 and the Law of Total Probability, it is possible to show that these bounds are:
\begin{eqnarray}
\label{e:bounds_h}
\max_z Pr(Y_{i} = 1,R_{i} =0 | Z_i=z) 
\leq 
& Pr(Y_{i}(0)=1 )&
\leq 
\min_z Pr(Y_{i} = 1 | Z_i=z) 
\\ \nonumber 
\max_z Pr(Y_{i} = 1 | Z_i=z) 
\leq 
& Pr(Y_{i}(1)=1) &
\leq 
1- \max_z Pr(Y_{i} = 0, R_{i}=1 | Z_i=z) 
\end{eqnarray}
These results imply that the lower bound of $APCE_{H}$ is
\begin{equation}
\label{e:lb_apce_h}
\frac{E(Y_{i}|Z_i=1)-E(Y_{i}|Z_i=0)}{1- \max_z Pr(Y_{i} = 0, R_{i}=1 | Z_i=z)  - \max_z Pr(Y_{i} = 1,R_{i} =0 | Z_i=z)}     
\end{equation}
while the upper bound is 
\begin{equation}
\label{e:ub_apce_h}
 \frac{E(Y_{i}|Z_i=1)-E(Y_{i}|Z_i=0)}{\max_z Pr(Y_{i} = 1 | Z_i=z) -\min_z Pr(Y_{i} = 1 | Z_i=z) }.
 \end{equation}

An alternative identification and estimation strategy that we can adopt is instead based on the assumption of unconfoundedness of $R$ with respect to $Y$:
\begin{assumption}
\label{a:unconf}
Unconfoundedness of $R$: \\
$Y_i(r) \perp\!\!\!\!\perp R_i(z) \mid X_i = x, Z_i=z$  \mbox{ and } \\
$0 <Pr(R_i(z) = r \mid X_i = x, Z_i =z) < 1 \quad \forall z \in \{0,1\}, x \in \mathcal{X}, r \in \{0,1\} $
\end{assumption}
This assumption, which allows one to point identify the $APCE_J$, implies that, conditioning on observable covariates $X$, a teacher's decision to upgrade is independent of potential outcomes. In other words, covariates $X$ contain all the information teachers use when making their decision, and the content of the recommendation given this information is random.\footnote{This assumption would be violated
if, for example, teachers also relied on the student’s behavior in class, which we do not observe, when taking their upgrade decisions. As we will see, we achieve similar results under the different identification assumptions we adopt.
}
Identification under this assumption is formally discussed in the Online \nameref{ap:apce_proof}.

\subsubsection*{Estimands for Always High (AH) students}
\label{s:APCE_AH_est}

As for the  AH students, we show in
the Online \nameref{ap:apce_proof}  that
\begin{eqnarray}
\label{e:apce_ah_est}
ACPE_{AH} = \frac{Pr(R_{i}=0,Y_{i}=1 \mid Z_i =0) - Pr(R_{i}=0,Y_{i}=1 \mid Z_i =1)}{Pr(Y_{i}(0) = 1)},  
\end{eqnarray}
 where, once again, the terms in the numerator do not depend on missing potential outcomes and can be estimated using their sample analogs, whereas further assumptions are needed for the denominator. To bound this denominator, note that:
\begin{equation}
       \label{e:bounds_ah}
\begin{aligned}
       \max_{z} Pr(Y_{i} = 1, R_{i} = 0 \mid Z_i = z) & \leq Pr(Y_{i}(0)=1) \leq \min_{z} Pr(Y_{i} = 1 \mid Z_i = z) \\
    \end{aligned}
    \nonumber
\end{equation}

\noindent
Therefore the lower bound of $APCE_{AL}$ is 
\begin{equation}
\label{e:lb_apce_AH}
    \frac{Pr(R_{i}=0,Y_{i}=1 \mid Z_i =0) - Pr(R_{i}=0,Y_{i}=1 \mid Z_i =1)}{\min_{z} Pr(Y_{i} = 1 \mid Z_i = z)}
\end{equation}

\noindent
while the upper bound is 
\begin{equation}
\label{e:ub_apce_AH}
    \frac{Pr(R_{i}=0,Y_{i}=1 \mid Z_i =0) - Pr(R_{i}=0,Y_{i}=1 \mid Z_i =1)}{\max_{z} Pr( R_{i} = 0,Y_{i} = 1 \mid Z_i = z)}
\end{equation}

Also, in the case of AH students, as with their H peers, we will compare the bounds obtained under the above identification strategy with point estimates obtained by assuming the unconfoundedness of $R$ (Assumption \ref{a:unconf}), as shown in the Online Appendix \nameref{ap:apce_proof}.

\subsubsection*{Estimands for Always Low (AL) students}
\label{s:APCE_AL_est}

Finally, considering the AL students, the Online \nameref{ap:apce_proof} shows that
\begin{eqnarray}
\label{e:apce_al_est}
    APCE_{AL}  &=& \frac{Pr(R_{i}=1,Y_{i}=0 \mid Z_i=1) - Pr(R_{i}=1,Y_{i}=0 \mid Z_i=0)}{1 - Pr(Y_{i}(1)=1)}.   
\end{eqnarray}
where, also in this case, the terms in the numerator do not depend on missing potential outcomes and can be estimated with their sample analogs, while the estimation of the denominator requires further assumptions. To bound this denominator, note that:
\begin{equation}
        \label{e:bounds_al}
    \begin{aligned}
        \max_{z}  Pr(Y_{i} = 1 \mid Z_i = z) & \leq Pr(Y_{i}(1) = 1 ) \leq  1 - \max_{z} Pr(Y_{i} = 0, R_{i} = 1 \mid Z_i =z)
    \end{aligned}
    \nonumber
\end{equation}

\noindent
Therefore, the lower bound of $APCE_{AL}$ is 
\begin{equation}
\label{e:lb_apce_AL}
    \frac{Pr(R_{i}=1,Y_{i}=0 \mid Z_i=1) - Pr(R_{i}=1,Y_{i}=0 \mid Z_i=0)}{1 - \max_{z}  Pr(Y_{i}\mid Z_i = z) }
\end{equation}

\noindent
while the upper bound is 
\begin{equation}
\label{e:ub_apce_AL}
    \frac{Pr(R_{i}=1,Y_{i}=0 \mid Z_i=1) - Pr(R_{i}=1,Y_{i}=0 \mid Z_i=0)}{\max_{z} Pr( R_{i} = 1,Y_{i} = 0 \mid Z_i =z)}. 
\end{equation}

Once again, as for the  $APCE_{H}$ and the $APCE_{AH}$, also the $APCE_{AL}$   can be point identified under unconfoundedness of $R$ (Assumption \ref{a:unconf}), as shown in the Online \nameref{ap:apce_proof}.

\section{Test scores and  quality of teachers' advice: evidence}
\label{s:res_apce}

\subsection{Estimates for the three strata}
\label{s:3strata}

\subsubsection*{Estimates for Helpable (H) students}
\label{s:APCE_H}
In the first column of Table \ref{t:APCE_H}, we present estimates of the numerator of the $APCE_H$ estimand in the RHS of equation (\ref{e:apce_h_est}), separately for each track and combined for the three aggregate tracks: Vocational, Academic and All.\footnote{Here and in what follows, each population estimand has been estimated separately for each track and cohort. The estimates reported in the tables for aggregate tracks have been obtained as weighted averages with weights based on the relative sample sizes of each cell. For this reason, the remaining tables report only one set of results, rather than separate results by cohort.
Standard errors (reported in parentheses) are bootstrapped (with 1000 repetitions) separately in each track, and then appropriately aggregated with the same weights. The estimates in the tables reported in the text include the covariates $\tilde{Z}_i$ and $X_i$ (see Table \ref{t:balance}).
The Online \nameref{ap:3strata} reports tables with corresponding estimates obtained without the balanced covariates $X_i$ (but including $\tilde{Z}_i$). The inclusion or exclusion of the balanced covariates does not significantly alter the estimates, as expected. This evidence supports the validity of  Assumption \ref{a:rand_sourc} (Randomization of the source).} 
The point estimates of this numerator are strictly positive in all tracks. Students scoring at the cutoff, who can be upgraded by their primary school teachers based on test scores, are  1.8--to--10\% more likely to complete a higher track than students who score below the cutoff. These point estimates are statistically different from zero except for the BL track. The estimates for the three aggregate tracks in the last rows of the table are also significant and equal to 4.4\%.

\begin{table}
    \caption{Estimates for the $APCE_{H}$, with covariates}
    \label{t:APCE_H}
\vspace{-0.1cm} 
\begin{center}
\begin{tabular}{l*{5}{c}}
\hline\hline
      \\ [-2ex]    
Track &   Numerator& Lower &    Upper &   Lower  & Upper\\
&  $APCE_H$    & bound  &    bound     &   bound &   bound \\
&     & denominator  &  denominator&  $APCE_H$ &  $APCE_H$  \\
&     & $APCE_H$  &  $APCE_H$&  \hspace{1.9cm} &  \hspace{1.9cm}   \\
      & 1& 2& 3 & 4 & 5 \\
      \\ [-2ex]  
\hline
  \\ [-2ex] 
V: BL          &        .0251&       .0251&       .6534&       .0381&           1\\
            &  {\footnotesize (.0167)}&  {\footnotesize (.0154)}&  {\footnotesize (.0126)}&  {\footnotesize (.0263)}&  {\footnotesize -}\\
V: BL/KL       &         .0773&       .0773&       .2877&       .2757&           1\\
            &  {\footnotesize (.0294)}&  {\footnotesize (.0279)}&  {\footnotesize (.0222)}&  {\footnotesize (.1029)}&  {\footnotesize -}\\
V: KL          &       .0996&       .0996&       .6984&       .1426&           1\\
            &  {\footnotesize (.0159)}&  {\footnotesize (.0159)}&  {\footnotesize (.0112)}&  {\footnotesize (.0229)}&  {\footnotesize -}\\
V: KL/GT       &   .0256&       .0256&       .3967&       .0644&           1\\
            &  {\footnotesize (.0238)}&  {\footnotesize (.0217)}&  {\footnotesize (.0232)}&  {\footnotesize (.0564)}&  {\footnotesize -}\\
V: GT          &        .0182&       .0182&       .7607&       .0237&       1\\
            &  {\footnotesize (.0077)}&  {\footnotesize (.0077)}&  {\footnotesize (.0076)}&  {\footnotesize (.0104)}&  {\footnotesize -}\\
V: GT/HAVO     &         .0526&       .0526&       .5089&       .1021&           1\\
            &  {\footnotesize (.0164)}&  {\footnotesize (.0159)}&  {\footnotesize (.0138)}&  {\footnotesize (.0309)}&  {\footnotesize -}\\
A: HAVO        &        .0349&       .0349&       .8082&       .0432&           1\\
            &  {\footnotesize (.0076)}&  {\footnotesize (.0076)}&  {\footnotesize (.0065)}&  {\footnotesize (.0096)}&  {\footnotesize -}\\
A: HAVO/VWO    &       .0642&       .0642&       .5451&        .117&           1\\
            &  {\footnotesize (.0156)}&  {\footnotesize (.0154)}&  {\footnotesize (.0126)}&  {\footnotesize (.0276)}&  {\footnotesize -}\\
\hline
VOCATIONAL  &      .0445&       .0445&       .6486&       .0686&       1\\
            &  {\footnotesize (.0058)}&  {\footnotesize (.0059)}&  {\footnotesize (.005)}&  {\footnotesize (.009)}&  {\footnotesize -}\\
ACADEMIC    &      .0442&       .0442&       .7251&       .0609&           1\\
            &  {\footnotesize (.0072)}&  {\footnotesize (.0071)}&  {\footnotesize (.0059)}&  {\footnotesize (.0099)}&  {\footnotesize -}\\
ALL         &         .0444&        .0444&       .6796&       .0653&       1\\
            &  {\footnotesize (.0045)}&  {\footnotesize (.0045)}&  {\footnotesize (.0038)}&  {\footnotesize (.0066)}&  {\footnotesize -}\\
\hline\hline
\end{tabular}
 
\end{center}
\vspace{-0.2cm}
\begin{minipage}{1\linewidth \setstretch{0.75} } 
{\scriptsize Notes: The row headings are the names of the secondary school tracks to which a student can be assigned in the Dutch system, ranked by level of difficulty.  The last three rows of the table aggregate the Vocational (VMBO*), the Academic (HAVO*), and All tracks, respectively.  With reference to the $APCE_H$ estimand defined in equation (\ref{e:apce_h_est}), the table reports for the outcome $Y$ and separately for each track, estimates of the numerator (in column 1), of the lower and upper bounds of the denominator (in columns 2 and 3), and of the lower bound of the $APCE_H$ (in column 4).
Column 5 reports the upper bound of the $APCE_H$ as equal to 1 because it is the ratio between $E(Y_{i}|Z_i=1)-E(Y_{i}|Z_i=0)$ and  $\max_z Pr(Y_{i} = 1 | Z_i=z) -\min_z Pr(Y_{i} = 1 | Z_i=z)$ which are exactly equal under Assumptions  \ref{a:strata_monoton} and \ref{a:iv_monoton}. See footnote \ref{f:col5} for the intuition.
The definitions of these two last statistics are in equations (\ref{e:lb_apce_h}) and (\ref{e:ub_apce_h}), respectively. The aggregations for the last three rows are performed as follows: point estimates of the numerator and of the lower and upper bounds of the denominator are created by weighing each track in the corresponding aggregate by the relative population in each track. For the lower and upper bounds of the $APCE_H$, the aggregated estimates for the lower and upper bounds of the denominators are used directly. Standard errors are computed using a block bootstrap at the school level with 1000 repetitions; each quantity in every iteration is calculated in the same way. 
The specifications in all columns include the covariates $\tilde{Z}_i$ and $X_i$ (see Table \ref{t:balance}).
The Online \nameref{ap:3strata} reports tables with corresponding estimates obtained without the balanced covariates $X_i$ (but including $\tilde{Z}_i$). The inclusion or exclusion of the balanced covariates does not change the estimates in a relevant way, as expected, supporting the validity of  Assumption \ref{a:rand_sourc} (Randomization of the source).}
 \end{minipage}
\end{table}

Recall that the denominator of the $APCE_H$ estimand in equation (\ref{e:apce_h_est}) is certainly positive and equal to the proportion of Helpable students, given Assumption \ref{a:strata_monoton} of Monotonicity of $Y_{i}$  with respect to $R_{i}$. Therefore, a positive numerator supports the conclusion that the information provided by test scores helps recommenders give better advice in the sense that, thanks to this information, they become better at detecting Helpable students and directing them into the most challenging secondary school track that they can successfully complete.

To confirm this conclusion and assess also quantitatively by how much test scores help teachers achieve this goal, estimates of the $APCE_H$ are needed, but as explained in the previous section, the denominator in the RHS of equation (\ref{e:apce_h_est}), which is the proportion of $H$ students, can only be bounded without additional assumptions. Estimates of the lower and upper bounds of this denominator are reported in columns 2 and 3 of Table \ref{t:APCE_H}. Focusing on the aggregate tracks in the last three rows, the proportion of $H$ ranges between approximately 4\% and 73\%.

As a consequence of the large gap between the estimated bounds of the denominator, also the bounds of the $APCE_H$ reported in columns 4 and 5 of Table \ref{t:APCE_H} are substantially different one from the other, ranging between 6\% and 100\% for the three aggregate tracks.\footnote{Note that the upper bound of the $APCE_H$ derived in equation (\ref{e:apce_h_est}) and reported in column 5 of Table \ref{t:APCE_H} is 1 by construction because it is the ratio between $E(Y_{i}|Z_i=1)-E(Y_{i}|Z_i=0)$ and  $\max_z Pr(Y_{i} = 1 | Z_i=z) -\min_z Pr(Y_{i} = 1 | Z_i=z)$, which are exactly equal under the two Monotonicity Assumptions \ref{a:strata_monoton} and \ref{a:iv_monoton}. The intuition is that the intention-to-treat (${E}[Y_i \mid Z_i=1]-{E}[Y_i\mid Z_i=0]$) at the numerator of this ratio identifies the proportion of students who are compliers \textit{and} are also helpable. Hence, the minimum plausible share of all Helpable students at the denominator of this ratio cannot be smaller than the share of Helpable who are also compliers at the numerator, and the upper bound of the $APCE_H$ must be 1 by definition.
Note also that when the share of Helpable who are also compliers is effectively equal to zero or very small, we may estimate it to be negative because of small sample variability. In tracks where this occurs, we round this proportion to zero. However, even when this proportion is infinitesimally small, it is still the case that the upper bound for the $APCE_H$ is 1 for the reason explained above. 
\label{f:col5}}
However, what matters most from the viewpoint of our research question is the lower bound of the $APCE_H$, which allows us to conclude that the information provided by test scores improves the quality of recommendations by at least 6\% as measured by their capacity to push Helpable students into the high track. 
This is a remarkable finding, particularly if the proportion of Helpable students in the population can be as high as  80\%, as suggested by the estimated upper bound of the $APCE_H$ denominator in column 3 of Table \ref{t:APCE_H}.

\begin{table}
    \caption{Estimates of the $APCE_{AH}$, with covariates}
    \label{t:APCE_AH}
\begin{center}
    \vspace{0.1cm} 
\begin{tabular}{l*{5}{c}}
\hline\hline
      \\ [-2ex]       
Track &   Numerator& Lower &    Upper &   Lower  & Upper\\
&  $APCE_{AH}$    & bound  &    bound     &   bound &   bound \\
&     & denominator  &  denominator&  $APCE_{AH}$ &  $APCE_{AH}$  \\
&     & $APCE_{AH}$  &  $APCE_{AH}$&  \hspace{1.9cm} &  \hspace{1.9cm}   \\
      & 1& 2& 3 & 4 & 5 \\
      \\ [-2ex]  
\hline
  \\ [-2ex] 

V: BL          &      .0004&        .317&       .3189&       .0011&       .0011\\

            &   {\footnotesize (.0118)}&   {\footnotesize (.0111)}&   {\footnotesize (.0111)}&   {\footnotesize (.036)}&   {\footnotesize (.036)}\\
V: BL/KL       &      0227&       .6858&        .705&       .0323&       .0332\\
            &   {\footnotesize (.0243)}&   {\footnotesize (.0209)}&   {\footnotesize (.0214)}&   {\footnotesize (.0336)}&   {\footnotesize (.034)}\\
V: KL          &      .0083&       .2289&       .2392&       .0368&       .0379\\
            &   {\footnotesize (.0101)}&   {\footnotesize (.0086)}&   {\footnotesize (.0091)}&   {\footnotesize (.0401)}&   {\footnotesize (.0416)}\\
V: KL/GT       &            .0745&       .5681&        .581&       .1288&       .1319\\
            &   {\footnotesize ( .0297)}& {\footnotesize (.0206)}&        {\footnotesize (.0198)}&        {\footnotesize (.0499)}&          {\footnotesize (.05)}\\
V: GT          &        .0097&       .1938&       .1938&       .0482&       .0482\\
            &    {\footnotesize (.006)}&   {\footnotesize (.0055)}&   {\footnotesize (.0055)}&   {\footnotesize (.03)}&   {\footnotesize (.03)}\\
V: GT/HAVO     &     .0439&       .4169&       .4262&       .1023&       .1041\\
            &   {\footnotesize (.0165)}&   {\footnotesize (.0113)}&   {\footnotesize (.0115)}&   {\footnotesize (.0375)}&   {\footnotesize (.0379)}\\
A: HAVO        &        0&       .1283&       .1283&           0&           0\\
            &   {\footnotesize (.0006)}&   {\footnotesize (.004)}&   {\footnotesize (.004)}&         {\footnotesize (.0047)}&         {\footnotesize (.0047)}\\
A: HAVO/VWO    &         .0281&       .3667&       .3777&        .074&       .0762\\
            &   {\footnotesize (.0146)}&   {\footnotesize (.0105)}&   {\footnotesize (.0109)}&   {\footnotesize (.037)}&   {\footnotesize (.0379)}\\
  \\ [-2ex] 
  \hline
    \\ [-2ex] 
VOCATIONAL  &       .0188&       .2988&       .3044&       .0618&        .063\\
            &   {\footnotesize (.0049)}&   {\footnotesize (.004)}&         {\footnotesize (.0041)}&   {\footnotesize (.0157)}&   {\footnotesize (.0159)}\\
ACADEMIC    &      .0089&       .2035&        .207&       .0429&       .0436\\
            &   {\footnotesize (.0046)}&   {\footnotesize (.0042)}&   {\footnotesize (.0043)}&        {\footnotesize (.0219)}&       {\footnotesize (.0223)}\\
ALL         &      .0148&       .2601&       .2649&       .0558&       .0568\\
            &   {\footnotesize (.0034)}&   {\footnotesize (.0029)}&        {\footnotesize (.003)}&       {\footnotesize (.0126)}&       {\footnotesize (.0128)}\\
  \\ [-2ex] 
\hline\hline
\end{tabular}
\end{center}
\vspace{-0.2cm}
\begin{minipage}{1\linewidth \setstretch{0.75} } 
{\scriptsize Notes: The row headings are the names of the secondary school tracks to which a student can be assigned in the Dutch system, ranked by level of difficulty.  The last three rows of the table aggregate the Vocational (VMBO*), the Academic (HAVO*), and All tracks, respectively.  With reference to the $APCE_{AH}$ estimand defined in equation (\ref{e:apce_ah_est}), the table reports for the outcome $Y$ and separately for each track, estimates of the numerator (in column 1), of the lower and upper bounds of the denominator (in columns 2 and 3), and of the lower and upper bounds of the $APCE_{AH}$ (in columns 4 and 5). The definitions of these two last statistics are in equations (\ref{e:lb_apce_AH}) and (\ref{e:ub_apce_AH}), respectively. The aggregations for the last three rows are performed as follows: point estimates of the numerator and of the lower and upper bounds of the denominator are created by weighing each track in the corresponding aggregate by the relative population in each track. For the lower and upper bounds of the $APCE_{AH}$, the aggregated estimates for the lower and upper bounds of the denominators are used directly. Standard errors for each quantity are computed using a block bootstrap at the school level with 1000 repetitions; each quantity in every iteration is calculated in the same way.
The specifications in all columns include the covariates $\tilde{Z}_i$ and $X_i$ (see Table \ref{t:balance}).
The Online \nameref{ap:3strata} reports tables with corresponding estimates obtained without the balanced covariates $X_i$ (but including $\tilde{Z}_i$). The inclusion or exclusion of the balanced covariates does not change the estimates in a relevant way, as expected, supporting the validity of  Assumption \ref{a:rand_sourc} (Randomization of the source).}
 \end{minipage}
\end{table}

\subsubsection*{Estimates for Always High (AH) students}
\label{s:APCE_AH}

Moving to the stratum of Always High students, the estimates for them are reported in Table \ref{t:APCE_AH}. In their case, the best source of advice is the one that recommends the high track to the \textit{highest} number of them (possibly to all of them) in order to minimize the cost of track changes they would incur if they started secondary school in the low track. Recall, in fact, that these students always complete the high track no matter where they start.

Focusing for brevity just on Columns 4 and 5 of the table, the gap between the lower and the upper bounds of the $APCE_{AH}$ is remarkably small in almost all tracks, allowing for a very precise estimation of this population parameter.\footnote{In the case of HAVO track, the sample analog of the numerator in the RHS of equation (\ref{e:apce_ah_est}) is close to zero but has a negative sign, possibly due to small sample variability.  We approximate it with 0.} The intuition for why we obtain such a precise result is that we can be confident in the population proportion of AH students (i.e. the denominator) as those with $R = 0$ and $Y = 1$ are for sure AH, and measuring this under the cutoff (i.e. for those with $Z = 0$) gives a precise estimate of the population parameter, showing that the population proportion of AH is approximately 26\%.\footnote{More specifically, because it is extremely rare to have $Z = 0$ and $R = 1$ (which can only occur due to the institutional complications explained in footnote \ref{foot_def}) we know that among those with $Z = 0$, those with $R = 0$ and $Y=1$ are AH for sure, while those with $R = 0$ and $Y = 0$ are for sure not AH, and these two categories together form virtually the whole population. This means that the fraction of AH students can be estimated very precisely, as $Z$ is randomized and the same fraction should exist above the threshold. } In the eight basic tracks, the point estimates of the $APCE$ range from 0 to 13\%. When the tracks are aggregated in the Vocational, Academic, and All tracks (last three rows of the table), the $APCE_{AH}$ is estimated to be about 6.2\%, 4.3\%, and 5.6\%, respectively, and statistically significant. These results indicate that when teachers are allowed to upgrade recommendations based on test scores, they send a significantly higher fraction of AH students to the high track, thereby reducing the costly track changes this group would otherwise incur.

\subsubsection*{Estimates for Always Low (AL) students}
\label{s:APCE_AL}

In the case of Always Low students, instead, our estimates of the $APCE_{AL}$ suggest that test score information does not help teachers give better recommendations. These estimates are

\begin{table}[!]
    \caption{Estimates of the $APCE_{AL}$, with covariates}
    \label{t:APCE_AL}
\begin{center}
    \vspace{0.1cm} 
\begin{tabular}{l*{5}{c}}
\hline\hline
      \\ [-2ex]       
Track &   Numerator& Lower &    Upper &   Lower  & Upper\\
&  $APCE_{AL}$    & bound  &    bound     &   bound &   bound \\
&     & denominator  &  denominator&  $APCE_{AL}$ &  $APCE_{AL}$  \\
&     & $APCE_{AL}$  &  $APCE_{AL}$&  \hspace{1.9cm} &  \hspace{1.9cm}   \\
      & 1& 2& 3 & 4 & 5 \\
      \\ [-2ex]  
\hline
  \\ [-2ex] 
V: BL          &       .0271&        .028&       .6576&       .0409&       .9765\\
          &       \footnotesize{(.0056)}&       \footnotesize{(.0055)}&       \footnotesize{(.0125)}&       \footnotesize{(.0085)}&       \footnotesize{(.0266)}\\
V: BL/KL       &       .0239&       .0266&       .2177&       .1126&       .8981\\
          &       \footnotesize{(.0084)}&        \footnotesize{(.008)}&       \footnotesize{(.0208)}&       \footnotesize{(.0404)}&       .\footnotesize{(1399)}\\
V: KL          &       .0658&       .0727&       .6612&       .0991&       .9026\\
         &       \footnotesize{(.0071)}&       \footnotesize{(.0069)}&       \footnotesize{(.0126)}&       \footnotesize{(.0105)}&         \footnotesize{(.03)}\\
V: KL/GT       &       .0335&       .0352&       .3934&       .0853&       .9569\\
    &       \footnotesize{(.0084)}&       \footnotesize{(.0083)}&       \footnotesize{(.0194)}&       \footnotesize{(.0212)}&       \footnotesize{(.0599)}\\
V: GT          &       .0377&        .041&       .7899&       .0477&       .9193\\
       &       \footnotesize{(.0038)}&       \footnotesize{(.0036)}&       \footnotesize{(.0072)}&       \footnotesize{(.0048)}&       \footnotesize{(.0253)}\\
V: GT/HAVO     &       .0644&       .0742&       .5213&       .1238&       .8682\\
  &       \footnotesize{(.0082)}&       \footnotesize{(.0077)}&       \footnotesize{(.0128)}&       \footnotesize{(.0156)}&       \footnotesize{(.0409)}\\
A: HAVO        &       .0487&       .0539&       .8464&       .0576&       .9035\\
     &       \footnotesize{(.0042)}&        \footnotesize{(.004)}&       \footnotesize{(.0058)}&        \footnotesize{(.005)}&       \footnotesize{(.0267)}\\
A: HAVO/VWO    &       .0788&       .0883&       .5581&        .141&       .8928\\
         &       \footnotesize{(.0077)}&       \footnotesize{(.0074)}&       \footnotesize{(.0126)}&       \footnotesize{(.0133)}&       \footnotesize{(.0278)}\\
  \\ [-2ex] 
  \hline
  \\ [-2ex] 
VOCATIONAL  &       .0458&       .0506&        .652&       .0703&       .9059\\
          &       \footnotesize{(.0027)}&       \footnotesize{(.0025)}&        \footnotesize{(.005)}&        \footnotesize{(.004)}&        \footnotesize{(.016)}\\
ACADEMIC    &       .0582&       .0648&       .7554&       .0771&       .8988\\
         &       \footnotesize{(.0037)}&       \footnotesize{(.0035)}&       \footnotesize{(.0055)}&       .0048&       \footnotesize{(.0192)}\\
ALL         &       .0508&       .0563&        .694&       .0733&       .9026\\
        &       \footnotesize{(.0022)}&       \footnotesize{(.0021)}&       \footnotesize{(.0037)}&       \footnotesize{(.0032)}&       \footnotesize{(.0126)}\\
  \\ [-2ex] 
\hline\hline
\end{tabular}
 
\end{center}
\vspace{-0.2cm}
\begin{minipage}{1\linewidth \setstretch{0.75} } 
{\scriptsize Notes: The row headings are the names of the secondary school tracks to which a student can be assigned in the Dutch system, ranked by level of difficulty.  The last three rows of the table aggregate the Vocational (VMBO*), the Academic (HAVO*), and All tracks, respectively.  With reference to the $APCE_{AL}$ estimand defined in equation (\ref{e:apce_al_est}), the table reports for the outcome $Y$ and separately for each track, estimates of the numerator (in column 1), of the lower and upper bounds of the denominator (in columns 2 and 3), and of the lower and upper bounds of the $APCE_{AL}$ (in columns 4 and 5). The definitions of these two last statistics are in equations (\ref{e:lb_apce_AL}) and (\ref{e:ub_apce_AL}), respectively. The aggregations for the last three rows are performed as follows: point estimates of the numerator and of the lower and upper bounds of the denominator are created by weighing each track in the corresponding aggregate by the relative population in each track. For the lower and upper bounds of the $APCE_{AL}$, the aggregated estimates for the lower and upper bounds of the denominators are used directly. Standard errors for each quantity are computed using a block bootstrap at the school level with 1000 repetitions; each quantity in every iteration is calculated in the same way.
The specifications in all columns include the covariates $\tilde{Z}_i$ and $X_i$ (see Table \ref{t:balance}).
The Online \nameref{ap:3strata} reports tables with corresponding estimates obtained without the balanced covariates $X_i$ (but including $\tilde{Z}_i$). The inclusion or exclusion of the balanced covariates does not change the estimates in a relevant way, as expected, supporting the validity of  Assumption \ref{a:rand_sourc} (Randomization of the source).}
 \end{minipage}
\end{table}

\noindent
reported in Table \ref{t:APCE_AL}.  The best source of advice for these students is the one that recommends the high track to the {\it lowest} number of them (possibly to none), in order to minimize the cost of the track changes they would incur if they started secondary school in the high track. Recall that these students can only complete the low track, no matter where they start. 

Focusing again for brevity just on the last two columns of the table, Column 4 shows that the estimated lower bound of the $APCE_{AL}$ is positive and statistically different from zero in many tracks, ranging from 4\% for BL to 14\% for HAVO/VWO. Considering the aggregate tracks in the last three rows of the table,  the $APCE_{AL}$ is at least as high as 7\% and significantly different from zero in all of them.
 Therefore, it appears that the information provided by test scores leads teachers to upgrade AL students into the high track. The consequence is a deterioration of the quality of recommendations because AL students cannot complete the high track. 
A possible reason for this finding is that primary school teachers do not incur any costs resulting from secondary school track changes and are therefore less sensitive to these consequences of the advice they offer to their students.

\subsection{A comprehensive evaluation}
\label{s:ben-michael}

The evidence in the previous section suggests that, in the case of Helpable and Always High students, test score information helps teachers improve the quality of their recommendations in terms of the two criteria that we have adopted (directing students towards the most difficult track they can successfully complete and reducing track changes). On the contrary, in the case of the Always Low, the positive $APCE_{AL}$ estimate indicates that the same information reduces this quality because it induces AL students to enroll in the high track that they later cannot complete.

The classification framework proposed by \cite{benmichael2024does} provides a comprehensive evaluation of whether it is a good idea to offer teachers the possibility to upgrade based on test score information, depending on how much the policy-maker cares about H and AH students compared to their AL peers. Define with $l_{ry}$ the loss deriving from a recommendation $R=r \in\{0,1\}$ given to a student with potential outcome $Y(1)= y \in\{0,1\}$. The four possible values of this loss are represented in the following Confusion Matrix:

\newpage
\begin{center}
\textbf{Confusion Matrix} \\
\vspace{0.3cm}
\begin{tabular}{c|c|c}
                      & Negative ($R=0$)                                                  & Positive ($R=1$)                                                                       \\ \hline
Negative ($Y(1) = 0$) & \begin{tabular}[c]{@{}c@{}}{True Negative}\\ $l_{00}=0$\end{tabular}  & \multicolumn{1}{c|}{\begin{tabular}[c]{@{}c@{}}{False Positive}\\ $l_{10}>0$\end{tabular}} \\ \hline
Positive ($Y(1) = 1$) & \begin{tabular}[c]{@{}c@{}}{False Negative}\\ $l_{01}>0$\end{tabular} & \multicolumn{1}{c|}{\begin{tabular}[c]{@{}c@{}}{True positive}\\ $l_{11}=0$\end{tabular}} \\ \hline
\end{tabular}
\end{center}
\vspace{0.3cm}
\noindent
The True Negative is a student whom the teacher correctly identifies as Always Low, while the True Positive is correctly identified as Always High or Helpable. In all these cases, the recommendation of the teacher does not cause any loss: $l_{00} = l_{11} = 0$.\footnote{To simplify the analysis, we abstract here from the possibility of negative losses (gains), which may differ between H, AL, and AH students.}  The False Negative is a Helpable or Always High student who is not recognized as such by the teacher and thus receives a low recommendation. In her/his case, the loss $l_{01}$ is positive because the student could complete the high track following the opposite advice (in the case of H) and complete the high track regardless, but with a track change (in the case of AH). Similarly positive is the loss for the last type of student, False Positive $l_{10}$, who is an Always Low not recognized as such by the teacher. Therefore, she/he receives a high recommendation without being later able to complete the high track.

Let's normalize to $l_{01} = 1$ the loss suffered by a False Negative student. It is then reasonable to assume that $l_{10}\leq l_{01}  = 1$.
This is because $l_{10}$ is the ``short-term'' loss generated by the cost of changing track that a False Positive AL student incurs if she/he is recommended a too-difficult track. In light of this, if $l_{01}$ is the loss of a False Negative AH, then it is also a short-term cost of having to change track, which can be assumed to be comparable for AL and AH students, and therefore the above inequality holds weakly: $l_{10} =  l_{01}  = 1$. If, instead, $l_{01}$ is the loss of a False Negative H student, he/she suffers a long-term loss extending well beyond the end of secondary school, because she/he earns, in expectation, the lower lifetime income of a low graduate, while with the opposite advice she/he would earn the higher one of a high graduate. Therefore, in the case of this student, we can assume that  $l_{10} < l_{01} = 1$ strictly.

We can then write the overall expected loss of a recommendation given by a source $Z=z \in \{0,1\}$ as
\begin{equation}
    \label{e:over_loss}
    \mathcal{L}(l_{10},z ) = \Pi_{01}(z) + l_{10}\Pi_{10}(z),
\end{equation}
where $\Pi_{01}(z) = Pr(R(z)=0; Y(1)=1)$ is the probability of a False Negative (H or AH) and $\Pi_{10}(z) = Pr(R(z)=1; Y(1)=0)$ is the probability of a False Positive (AL). Our goal is to compare the overall loss $\mathcal{L}(l_{10}, z )$ when the source is a ``teacher alone'' $(z=0)$ or a ``teacher informed by test scores and allowed to upgrade"'' $(z=1)$, computed at different values of the relative weight $l_{10}$. This weight is at most equal to $1$, when the short-run loss of AL students is considered by the policy-maker as relevant as the long-run loss of H students, so that $l_{10}=1$, and decreases towards $0$ as the former loss becomes less relevant $(l_{10}<1)$.

\begin{figure}[ht]
    \caption{Comprehensive evaluation that test score information improves teachers' recommendations }
    \label{f:ben_michael}
    \centering
\includegraphics[width = 15 cm]{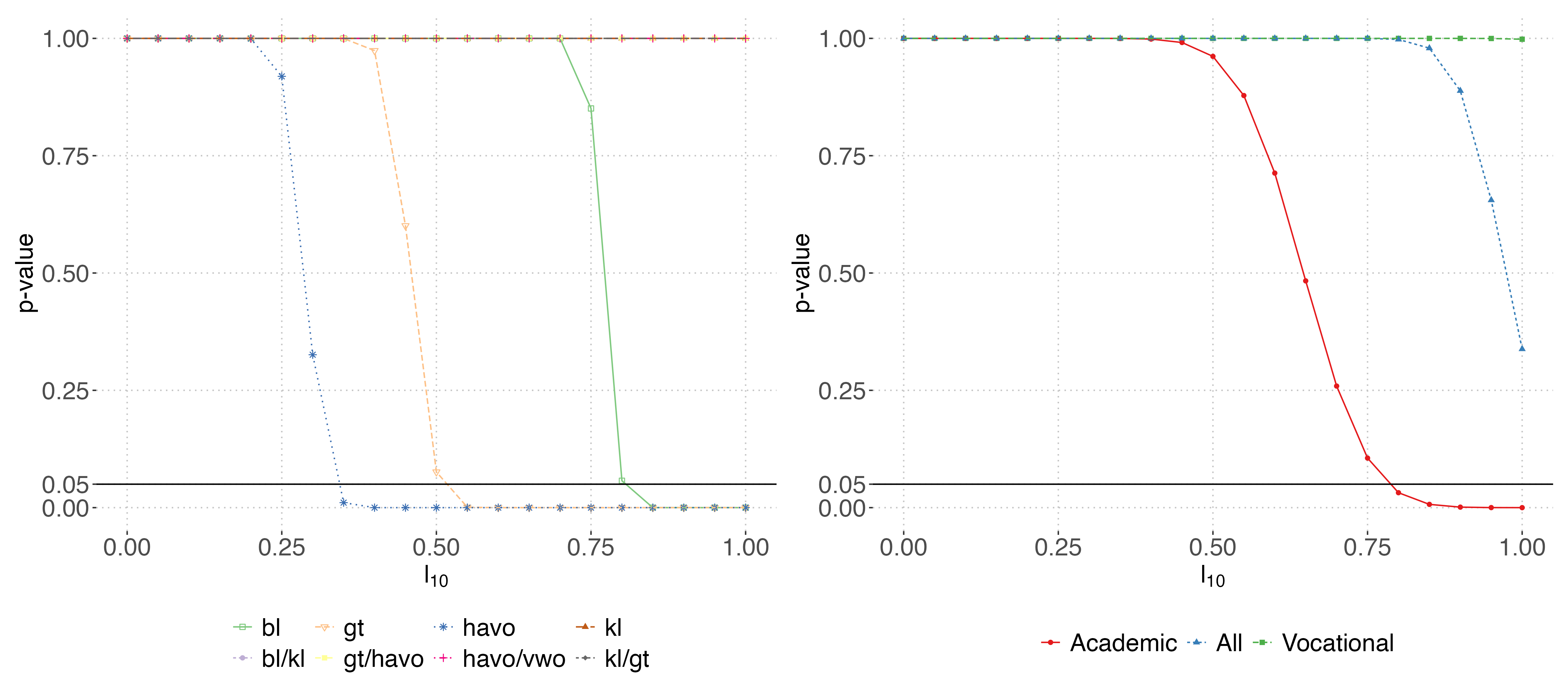}
\begin{minipage}{0.95\linewidth  \setstretch{0.75}}
{\scriptsize Notes: The figure reports p-values of the test that the difference $ \mathcal{L}(l_{01},1) - \mathcal{L}(l_{01},0)$ between the overall losses with and without test score information (equation \ref{e:over_loss_dif}) is negative, at different values of the relative weight $l_{10}$ of AL students. Failure to reject the null hypothesis in this test (high P-value) indicates that we cannot rule out that the loss is larger when teachers do not see the test score information. The solid line indicates the 5\% significance level of the P-values above which the null cannot be rejected with sufficient statistical confidence. }
 \end{minipage}
\end{figure}

In the Online \nameref{ap:ben-michael} we show that 
\begin{equation}
\label{e:over_loss_dif}
    \mathcal{L}(l_{01},1) - \mathcal{L}(l_{01},0) = \Pi_{11}(0) -  \Pi_{11}(1) + l_{10}(\Pi_{10}(1) - \Pi_{10}(0))
\end{equation}
where the RHS is identified and can be estimated thanks to  Assumption \ref{a:rand_sourc} (Randomization of the source). The left panel of  Figure \ref{f:ben_michael} reports for each track the P-values of the tests of the null that this difference is negative for values of the relative weight $l_{01}$ in the set $\{0, 0.05, 0.1, ...,1\}$:
\begin{equation}
\label{test_loss}
    H_0:\mathcal{L}(l_{01},1) - \mathcal{L}(l_{01},0)\leq 0 ;  \hspace{1 cm}
    H_1:\mathcal{L}(l_{01},1) - \mathcal{L}(l_{01},0)> 0  
\end{equation}
Failure to reject the null hypothesis of this test (high P-value)
indicates that we cannot rule out the possibility that allowing teachers to upgrade their recommendations based on test scores is a better (or equally effective) system because we do not have enough evidence to conclude that the provision of test scores leads to larger losses.
Conversely, if we can reject $H_0$ (with a small P-value), we can rule out the possibility that allowing teachers to upgrade based on test scores generates smaller losses and thus improves recommendations.

With the exception of three tracks (V:BL, V:GT and A:HAVO), we cannot reject that the difference is negative even when the short-run loss of AL students is considered as relevant as the long-run loss suffered by H students $(l_{10}= 1)$ and AL students have the maximum relative weight.
 For the remaining V:BL, V:GT and A:HAVO tracks, we reject that the difference is negative only when the short-term loss suffered by AL students has an (implausibly)  high relative weight -- higher than about 0.8, 0.5, and 0.3, respectively, at the 5\% significance level. Similar evidence is reported in the right panel of the figure for the three aggregate tracks. We never reject the null at the same significance level for the All and Vocational tracks, even if the relative weight on the short-term loss of AL students is the highest. In the case of the Academic track, we reject only when this weight is higher than about 0.75. 

In light of this evidence, we conclude that, overall, the possibility of upgrading based on test scores helps primary teachers improve the quality of their secondary school track recommendations. For this not to be the case, the short-term loss of AL students (who have to change track because they are recommended too difficult a track that in the end they cannot complete) would have to be considered implausibly almost as relevant as the life-time loss of H students (who are prevented from completing a high track they would be able to complete).

\section{An alternative to assuming the Exclusion Restriction
}
\label{s:robust}

In this section, we explore the identification and estimation of the Average Principal Causal Effects without relying on the Exclusion Restriction (ER) assumption. This assumption may fail to hold if the causal effect of $Z$ on $Y$ is not mediated solely by $R$.  Consider, for instance, the Never-Takers stratum. Since these students are never upgraded, under the ER their expected value of $Y$ should, in principle, be the same on both sides of the cutoff. 
However, scoring above the cutoff may boost the confidence of these students and encourage their parents to advocate for an upward track change after initial enrollment in secondary school. More generally, it may have a direct positive effect on these students' performance in secondary school, even in the absence of an upgrade at the moment of the initial track enrollment. In these cases, the Exclusion Restriction would not hold.

We study such a possibility by deriving bounds on the APCEs when
a specific form of direct effect of $Z$ on $Y$ for given $R$ is allowed to play a role. Finding that even in the presence of this effect, our results are unchanged would suggest that the ER assumption is not strictly necessary for our conclusions about the preferability of informing teachers about the test score results of their students, or, in other words, that at least this deviation from the Exclusion Restriction assumption is not changing our results (see also \citealp{Angrist1996}).

To this end, we first assume that the direct effect of $Z$ on $Y$ is additive and equal for all students:
\begin{assumption} \label{a:homogeneity} Homogeneity of the direct effect of $Z$ on $Y$ \\
$\mathbb{E}[Y_{i}(1,r)] =\mathbb{E} [Y_{i}(0,r)] + \eta \hspace{0.3 cm} \mbox{for } r \in \{0,1\}$,
\end{assumption}
 Under Assumption \ref{a:homogeneity}, we derive new expressions for the APCEs and their bounds as a function of $\eta$, as described in the Online \nameref{ap:robust}.\footnote{\cite{Mealli2013} study another example of bounds for principal causal effects when Exclusion Restriction assumptions are relaxed.} These expressions generalize the original framework proposed by \cite{Imai2023}. Indeed, it can be shown that as $\eta$ goes to 0, i.e.~the level at which the Exclusion Restriction holds, the new bounds converge to those derived in Section \ref{s:stat} under the ER.
More generally, the bounds of the APCEs that we obtain are functions of the unknown parameter $\eta$, which serves as a \emph{sensitivity} parameter (e.g., \citealp{Imbens2003Sensitivity}), and as such cannot be identified and estimated without invoking further assumptions.
In order to gain some insights into the plausible magnitude of $\eta$, we can invoke the Principal Ignorability assumption described below that allows to identify and estimate $\eta$.
This enables us to establish not only the extent to which the ER assumption is violated but also the stability of our results if the violation indeed occurs and identification is achieved under an alternative assumption.

To understand the nature of this Principal Ignorability assumption, consider the effect of $Z$ on $Y$ for Never-Takers, which under the Homogeneity Assumption \ref{a:homogeneity} is equal to $\eta$:
\begin{equation} \label{e:eta}
    \eta = \mathbb{E}[Y \mid Z=1 , R(1) = 0, R(0) = 0] - \mathbb{E}[Y \mid Z=0 , R(1) = 0, R(0) = 0].
\end{equation}
  The first term on the RHS of equation (\ref{e:eta}) is identified and can be easily estimated because students who score at the cutoff but are not upgraded by the teacher are certainly Never-Takers and are observed.  The second term, however, is not identified because the group of students scoring below the cutoff (and not upgraded) comprises both Compliers and Never Takers, who cannot be distinguished. We solve this problem by invoking the following assumption:

\begin{assumption} \label{a:pi}
    Principal Ignorability: \\
    $Y_i(0) \perp\!\!\!\!\perp R_i(1) \mid R_i(0) = 0, X_i$
\end{assumption}
This assumption implies that, conditional on covariates, the distribution of potential outcomes below the cutoff is the same for Compliers and Never-Takers.\footnote{This assumption is, of course, as debatable as the ER assumption. Moreover, similarly to the unconfoundedness assumption, its plausibility crucially depends on the information contained in the observed covariates (see \citealp{feller2017principal}, \citealp{mattei2023assessing}, and \citealp{ding2017principal}) However, our goal here is simply to assess how stable our results are when invoking this assumption instead of the ER.} Therefore, the second term on the RHS of equation (\ref{e:eta}) is identified and $\eta$ can be estimated.\footnote{Using observations above the cutoff, where Never-Takers are identified, we can estimate the probability of being a Never-Taker conditional on covariates (the estimated principal score $\hat p_{nt}$). Then, $\mathbb{E}[Y_i \mid Z_i=0 , R_i(1) = 0, R_i(0) = 0] $ can be estimated by the average $Y_i$ for units with $Z_i=0$ and $R_i=0$, weighted by the principal score. That is, under Assumption \ref{a:pi}, $\mathbb{E}[Y_i \mid Z_i=0 , R_i(1) = 0, R_i(0) = 0] $ is identified and can be estimated by $ \frac{\sum_{i: Z_{i} = 0, R_{i}=0} \hat p_{i,nt}Y_{i}}{\sum_{i: Z_{i} = 0, R_{i}=0}\hat p_{i,nt}}$. See \cite{mattei2023assessing} for further details.}

Table \ref{t:eta} in the Online \nameref{ap:robust} reports estimates of  $\eta$ obtained with the above procedure. 
Some point estimates of $\eta$ are quantitatively sizeable. For example, $\hat \eta$ for students in track V:BL/KL in the 2016 cohort is 0.1013 (s.e.: 0.0465), which is larger than the estimated numerator of the $APCE_H$ for Helpable students in the same track, i.e.~0.0773. 
In light of this finding, we use the point estimates $\hat \eta$ to estimate the bounds for the APCEs when the Homogeneity Assumption \ref{a:homogeneity} and the Principal Ignorability Assumption \ref{a:pi} hold instead of the ER.

\begin{figure}[tbp]
    \caption{Overview of the $APCE_J$ estimates}
    \label{f:apce_over}
    \centering
\includegraphics[width = 15 cm]{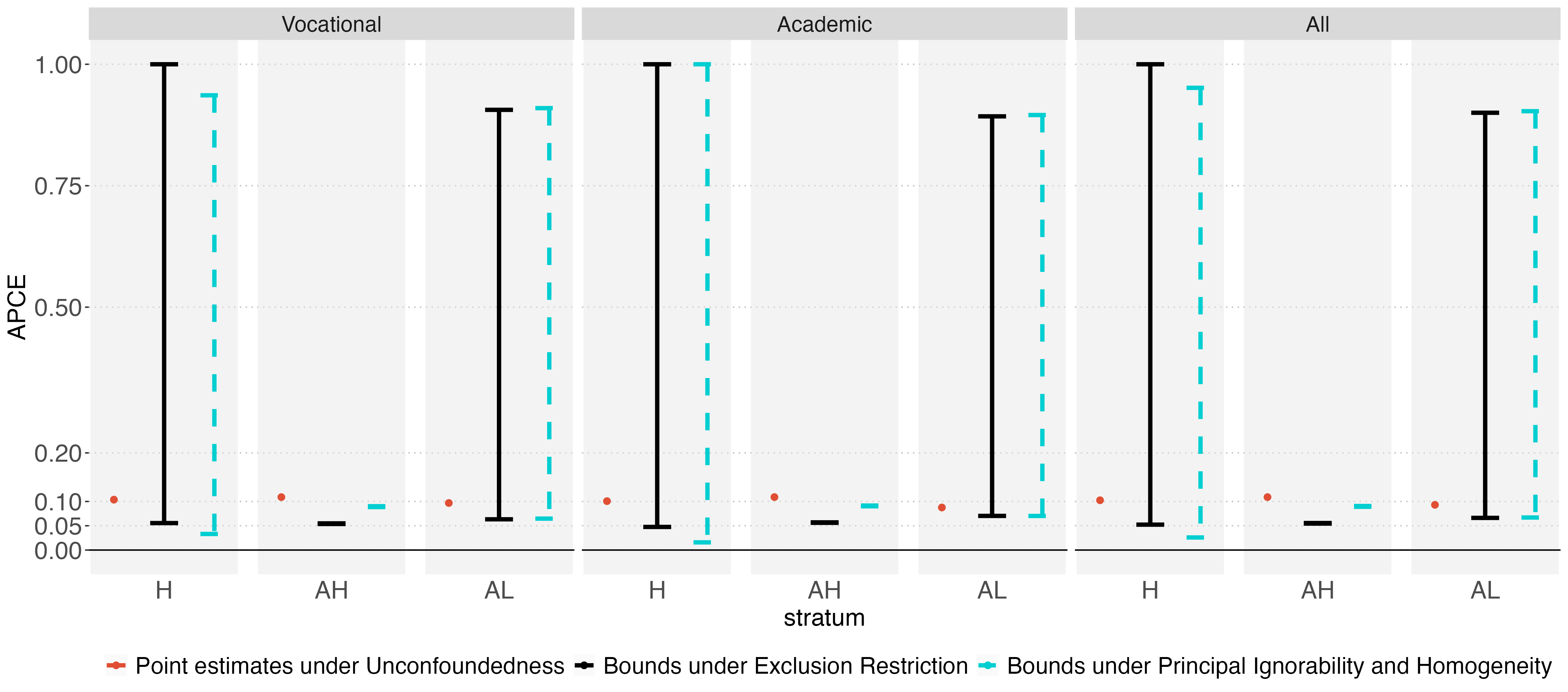}
\vspace{.5cm}
\begin{minipage}{0.9\linewidth  \setstretch{0.75}}
{\scriptsize Notes:  The figure reports, for the three aggregate tracks and for the H, AH and AL strata: the point identified $APCE_J$ estimates obtained under unconfoundedness (Assumption \ref{a:unconf}, red dot), that are derived in the Online \nameref{ap:apce_proof}; the bounds of the $APCE_J$ obtained under the ER (Assumption \ref{a:ec}, solid black), that are derived in Section \ref{s:stat};   and the bounds of the $APCE_J$ obtained under PI and Homogeneity (Assumptions \ref{a:pi} and \ref{a:homogeneity}, dashed light-blue), that are derived in Section \ref{s:robust}.  
}
 \end{minipage}
\end{figure}

Figure \ref{f:apce_over} compares, for the three aggregate tracks and for the H, AH, and AL strata, 
the point identified $APCE_J$ estimates obtained under unconfoundedness (Assumption \ref{a:unconf}, red dot), that are derived in the Online \nameref{ap:apce_proof}; the bounds of the $APCE_J$ obtained under the ER (Assumption \ref{a:ec}, solid black), that are derived in Section \ref{s:stat};   and the bounds of the $APCE_J$ obtained under PI and Homogeneity (Assumptions \ref{a:pi} and \ref{a:homogeneity}, dashed light-blue), that are derived in this section.  
In the case of H and AL students, unconfoundedness delivers estimates that are located near the lower bounds of the partially identified alternative estimates. For the AH students, both with and without unconfoundedness, the $APCE_J$ estimates are precise and close one to the other.

This graphical comparison under the different assumptions indicates that the implications in terms of \textit{lower bounds} of the relevant effects are very similar in all cases. We conclude that our main results are robust with respect to these alternative identification strategies.\footnote{The estimates of all the bounds obtained without the Exclusion Restriction and under the alternative PI and Homogeneity assumptions can be found in Online \nameref{ap:robust}.}

\section{Fairness of teachers' recommendations}
\label{s:fair}

 Within the same statistical framework described in the previous sections, \cite{Imai2022} and \cite{Imai2023} propose the concept of ``Principal Fairness," which can be interestingly used to compare the fairness of two sources of first-year track assignment. 

\begin{definition} Principal Fairness:\\
A source $Z_i = z \in \{0,1\}$ of first-year track assignment $R_{i}(z)$ satisfies Principal Fairness with respect to a protected attribute $B_i$ (e.g., SES, race, gender), if the first-year recommendations deriving from this source are conditionally independent of $B_i$ within each principal stratum $J_i \in \{AH,AL,H\}$
\begin{equation}
\label{e:prin_fair}
Pr(R_{i}(z) | J_i, B_i) = Pr(R_{i}(z) | J_i)
\end{equation}
\end{definition}

According to this definition, a recommender is fair as long as her recommendations are independent of the protected attribute among students in the same stratum. The analogous fraction in other strata can be lower or higher if it is equal for both genders within each stratum. A test for Principal Fairness of a source of secondary school track recommendations can then be designed as follows. Let the two sources of recommendations that we would like to compare be: ``teacher alone" ($z=0$) versus ``teacher informed by test scores and allowed to upgrade" ($z=1$).
Given two values $b$ and $b'$ of a  protected attribute $B_i$, the Principal Fairness of source  $ z$ in  stratum $j$ is given by:
\begin{equation}
\Delta_j(z) =  Pr\{R_i(z)=1 | B_i=b, J_i=j\} - Pr\{R_i(z)=1 | B_i=b', J_i=j\},
\label{e:delta_noabs}
\end{equation}
where $R_i(z) = R \in \{0,1\}$ is the recommendation given by source $z$ to student $i$. Note that  source $z$ is perfectly fair if $ \Delta_j(z)=0$. 
Otherwise,  source $z$ is {\it unfair} because its probability of recommending the high track to students in stratum $j$ changes with the values of  $B_i$. 

It is important to note that, considering the protected attribute $B_i$ without conditioning on the other covariates requires to interpret carefully the reason for a possible lack of fairness. A fairness criterion that does not hold unconditionally may hold conditionally if the covariates correlate with the protected attribute (see \citealp{Imai2022}, for a discussion). For instance, suppose that teachers do not discriminate based on immigrant status but do discriminate based on SES, and immigrants have, in general, lower SES. In this case, if the protected attribute $B_i=b$ in equation (\ref{e:delta_noabs}) is immigrant status, this attribute should actually be considered as a proxy for SES. However, from a descriptive viewpoint, it would still be the case that immigrants are treated differently, even if in each stratum defined by SES they are not. As argued by \cite{Imai2023}, the choice between marginal principal fairness and conditional principal fairness is not statistical. 

The sign of $\Delta_j(z)$ is also important, as it indicates the direction of discrimination. Suppose again that $B_i=b$ denotes immigrant students. Then finding, for example, that  $\Delta_j(1) > 0$ while $\Delta_j(0) <0$, would mean that when recommenders are allowed to upgrade their advice based on test scores, immigrant students are positively discriminated into the high track, while when the recommenders give advice without seeing test scores, immigrants are negatively discriminated. Note that in our setting, where teachers are officially allowed to upgrade only students who score above the cutoff, all the $\Delta_{j}(0)$ should be equal to 0 by definition. Therefore, a $\Delta_{j}(0) = 0$ cannot be interpreted as evidence of fairness in the provisional track recommendation made by teachers without the test scores. On the other hand, a $\Delta_j(0)$ different from 0 would indicate different rates of non-compliance below the cutoff across protected attributes in stratum $j$.

\begin{figure}[ht]
    \caption{Fairness for immigrants}
    \label{f:fair}
    \centering
\includegraphics[width = 15 cm]{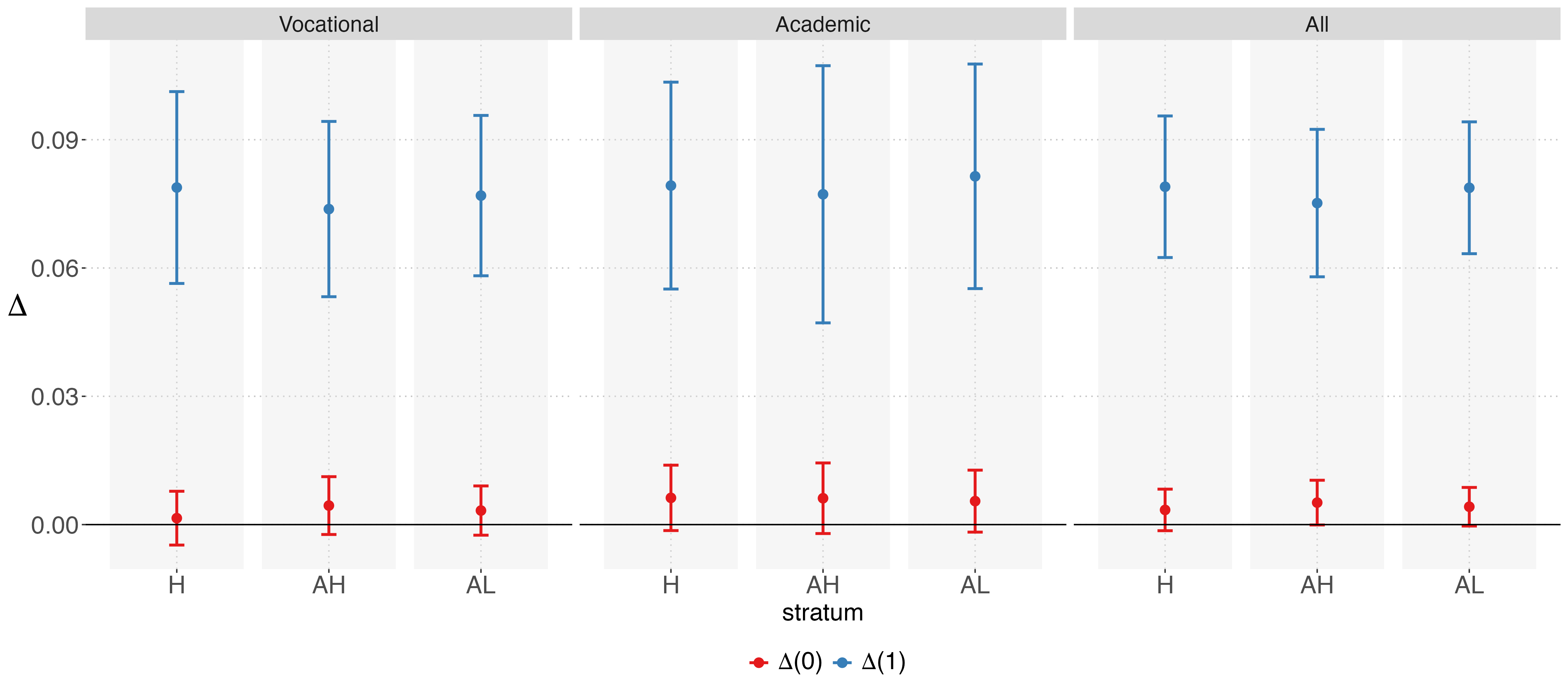}
\vspace{-0.2cm}
\begin{minipage}{0.9\linewidth  \setstretch{0.75}}
\scriptsize  Notes:  the figure plots,  for the three aggregate tracks and for the H, AH and AL strata, the point estimates of the statistic 
$\Delta_{j}(z) = Pr(R(z)=1 \mid \text{Immigrant}) - Pr(R(z) = 1 \mid \text{No Immigrant})$  and the correspondent 95\% confidence intervals.
\end{minipage}
\end{figure}

\begin{figure}[ht]
    \caption{Fairness by SES}
    \label{f:fair_ses}
    \centering
\includegraphics[width = 15 cm]{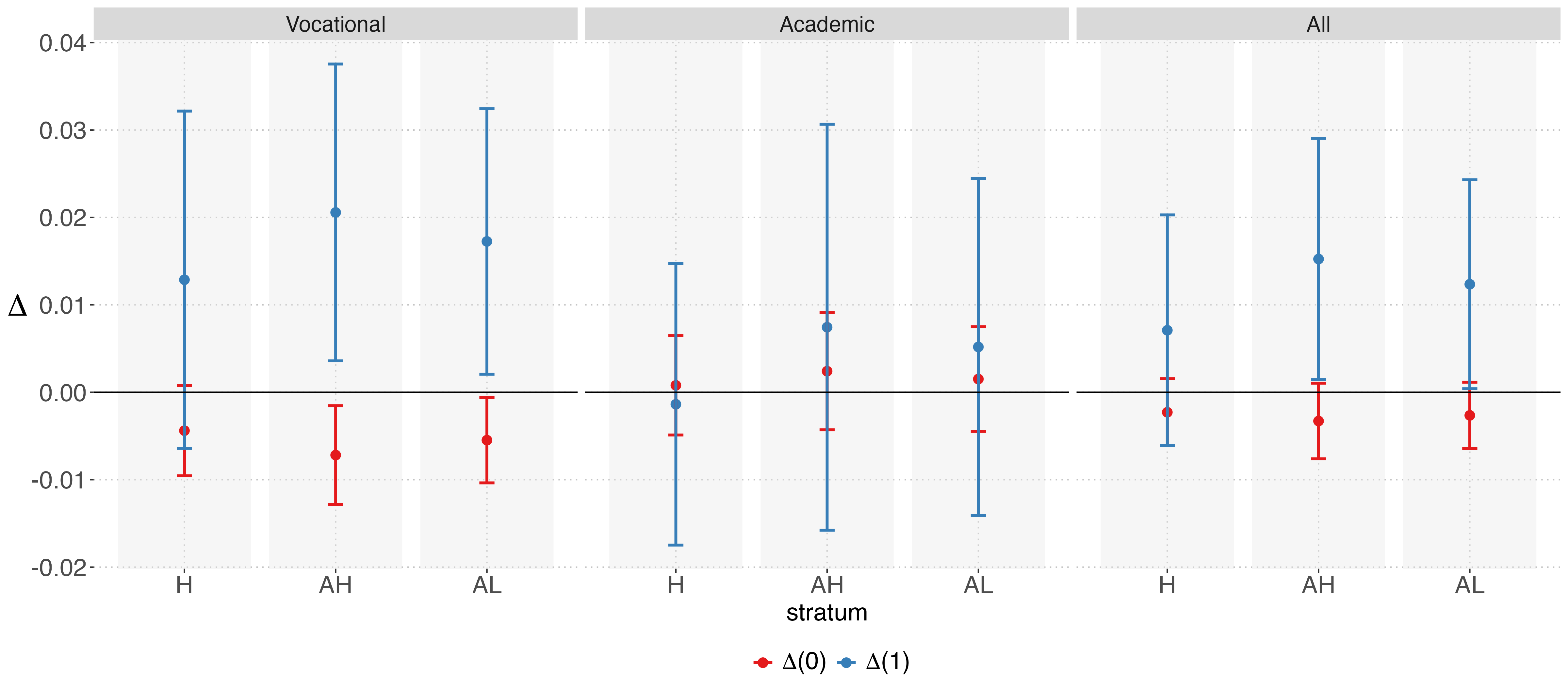}
\vspace{-0.2cm}
\begin{minipage}{0.9\linewidth  \setstretch{0.75}}
\scriptsize  Notes:  the figure plots,  for the three aggregate tracks and for the H, AH and AL strata, the point estimates of the statistic 
$\Delta_{j}(z) = Pr(R(z)=1 \mid \text{Low SES}) - Pr(R(z) = 1 \mid \text{High SES})$  and the correspondent 95\% confidence intervals.
\end{minipage}
\end{figure}

Under assumptions \ref{a:SUTVA} (SUTVA), \ref{a:rand_sourc} (Local Randomization), \ref{a:ec} (Exclusion Restriction), \ref{a:strata_monoton} (Monotonicity of $Y$ with respect to R), and \ref{a:unconf} (Unconfoundedness of R with respect to Y), as shown by \cite{Imai2023}, we can estimate equation \eqref{e:delta_noabs}. Figure \ref{f:fair} plots,  for the three aggregate tracks and for the H, AH, and AL strata, $\Delta_j(1)$ and $\Delta_j(0)$,  considering immigrant status as the protected attribute $B_i=b$.
In all these cases, $\Delta_j(0)$ is not statistically different from zero and is small in size. $\Delta_j(1)$ is instead clearly positive and significantly larger than $\Delta_j(0)$ in all strata. This evidence 
suggests that the information provided by test scores considerably increases the chances that immigrant students are upgraded.   
The interpretation of this finding is that test scores induce a larger revision of the teachers' posterior beliefs regarding the ability of immigrant students compared to those of native students. Note that this outcome is good news for immigrant students in the H and AH strata but not for immigrant students in the AL stratum, for whom this positive discrimination causes track changes.

Figure \ref{f:fair_ses} reports the analogous quantities considering low socioeconomic status (SES) as the protected attribute, where low SES is defined as having a household income below the median. For the academic tracks, none of the $\Delta_{j}(z)$ is significantly different from 0. 
In vocational tracks, however, the $\Delta_{j}(0)$ are negative and the $\Delta_{j}(1)$ are positive and economically significant, albeit smaller in size than in the case of immigrant students (the scale of the vertical axis changes between Figures \ref{f:fair} and \ref{f:fair_ses}). The fact that all the $\Delta_{j}(0)$ are negative suggests a higher rate of non-compliance below the cutoff among high-SES students. A plausible explanation is that parents of high-SES students may be more likely to pressure teachers to upgrade their children when they are provisionally recommended for vocational tracks. On the other hand, the positive $\Delta_{j}(1)$ values indicate that the provision of test scores induces upgrades for low-SES students more often than for high-SES ones. 
We view this finding as a consequence of test scores inducing a larger shift in teachers’ posterior beliefs for low-SES students.

Finally, the analogous evidence for gender is shown in Figure \ref{f:fair_gender} of the Online \nameref{ap:fair}.
We do not find any $\Delta_{j}(z)$ significantly different from 0, meaning that teacher recommendations are fair when the protected attribute is gender.

\section{Conclusions}
\label{s:conc}

Using a quasi-experimental setting offered by Dutch educational institutions, we show that when track recommendations given to students at the end of primary education can be upgraded based on standardized test scores, the quality of advice improves by at least 6\% as measured by the fraction of students that are recommended a more challenging track they can complete successfully, while in the counterfactual case, they would have remained in a lower level track. An improvement of about the same size is also estimated for students who should be recommended to more challenging tracks, as they are always able to complete them independently of where they start. 

We also find, however, that the possibility of upgrading based on test scores results in a significant number of students who are indeed upgraded but are unable to complete a more challenging track. In the case of these students, the goal should be to direct them immediately toward the low track that they can complete successfully, thus minimizing costly track changes. However, an overall evaluation that combines these opposite findings in a weighted objective function indicates that the relative weight of the short-term losses suffered by students who must change track because of misplacement when teachers are given the opportunity to upgrade based on test scores would have to be unreasonably high to conclude that it is better if teachers do not see these scores.

The implications for policy of these results are particularly relevant in countries where tracking is the established way to organize secondary school studies. It is surprising that no experimental evidence exists to guide educational policymakers of these countries in designing the best system of track advising. We fill in this gap in several ways.

First, we define a reasonable metric to evaluate the quality of teachers' recommendations. This metric values two objectives: reducing costly track changes and 
directing towards more challenging tracks students who are able to complete them, but who need to be ``pushed" towards these tracks and be convinced that they can complete them successfully.  Differently from the case of a weather forecast, predicting the school track in which a student would have the best performance and recommending this track may have an effect on the object of the prediction itself, i.e., the choice of the student and her performance in the chosen track, as well as later in life. In other words, track recommendations may be self-fulfilling prophecies, and it is crucial to consider this feature in the optimal design of their implementation. The metric we propose puts this feature at the center stage. 

Second, our results show that allowing teachers to upgrade their recommendations based on standardized test scores helps them to improve the quality of track advising.  
Ideally, one would like to compare recommendations given by teachers with and without the information provided by test scores, which is not what the quasi-experiment we study can offer.
However, our evidence suggests that test scores do help teachers provide better recommendations in general, not just to upgrade previous ones. Therefore, appropriate controlled experiments should be implemented to further study and formally test the more general hypothesis.

Third, a large literature (see footnote \ref{f:bias_teacher}) has demonstrated, in different contexts, that teachers’ recommendations do not typically reflect only the previous academic achievement and the future ability potential of a student, but are also highly correlated with characteristics like SES, gender, race, and behavior in class. The bias affecting recommendations that are determined by the conscious or unconscious use of these variables by teachers is typically deemed unacceptable and has reinforced the opposition to school tracking in general. Our evidence suggests that test score information also has an impact on the fairness of recommendations with respect to protected attributes. We find that the provision of test scores induces more frequent upgrades of recommendations for immigrant and low SES students. We interpret these results as indicating that the provision of test scores information induces a larger shift in teachers' posterior beliefs regarding these students. 
Together with the fact that high-SES students are more likely to be upgraded when originally recommended a vocational track and scoring below the cutoff, 
we conclude that test scores are a valuable tool for increasing the fairness of track recommendations.

 Independent of track-advising, but no less importantly, our results contribute to the recent and growing literature on algorithm-assisted human decisions (see, for example, \citealp{Imai2022}, \citealp{Imai2023}, \citealp{Imai_2024}, and \citealp{Rambacan2024}). It is increasingly common for decision-makers in various fields (justice, medicine, finance, politics, and education, to name a few) to use big data processed by multiple algorithms to make high-stakes decisions in highly uncertain contexts.  However, it is still unclear how to make the best use of these algorithmic and data-driven tools in human decision-making. Our results contribute to this literature by providing novel quasi-experimental evidence and by improving existing methods to assess algorithm-assisted human decisions.

\end{spacing}

\begin{spacing}{0.8}

{
\small
\bibliography{ref_test_teachers}{}
\bibliographystyle{chicago}
}
\end{spacing}

\clearpage

\newpage

\setcounter{page}{1}

\begin{center}
{\bf \LARGE Online Appendix}

\vspace{2cm}
{\large Do Test Scores Help Teachers Give Better Track Advice to Students?

\vspace{0.1cm}
 A Principal Stratification Analysis}

\vspace{1cm}
{\large

\vspace{1cm}

Andrea Ichino, Fabrizia Mealli, Javier Viviens

\vspace{0.2cm}

}

\vspace{1cm}
February 13, 2026
\end{center}

\newpage

\begin{spacing}{1}

\appendix

\renewcommand{\theequation}{A\arabic{equation}}
\renewcommand{\thefootnote}{A\arabic{footnote}}
\renewcommand{\thetable}{A\arabic{table}}
\renewcommand{\thefigure}{A\arabic{figure}}
\setcounter{equation}{0}
\setcounter{footnote}{0}
\setcounter{table}{0}
\setcounter{figure}{0}
\linespread{1}

\clearpage
\newpage
\vspace{-0.4 cm}
\section*{Appendix to Section \ref{s:assumptions}}\label{ap:NTandZtilde}
\begin{table}[ht]
    \caption{Leniency of teachers and frequency of Never Takers}
    \label{t:NTandZtilde}
\begin{center}
    \vspace{0.1cm} 
         \begin{tabular}{lcccc}
      \hline \hline
   \\ [-2ex]
    Track   & $\xi$ & $\xi$ & $\xi$ & $\xi$  \\
      & 1& 2& 3 & 4 \\
      \\ [-2ex]  
\hline
  \\ [-2ex] 
V: BL          &        .014&      -.0187&       .1829&       .1482\\
               &  \footnotesize (.0535)&  \footnotesize (.0527)&  \footnotesize (.0702)&  \footnotesize (.0734)\\
V: BL/KL       &       .1297&       .0148&       .0582&      -.0281\\
               &  \footnotesize (.1073)&  \footnotesize (.1366)&  \footnotesize (.1373)&  \footnotesize (.1531)\\
V: KL          &       .3561&       .2637&       .3123&       .2355\\
               &  \footnotesize (.0846)&  \footnotesize (.0866)&  \footnotesize (.0763)&  \footnotesize (.0776)\\
V: KL/GT       &       .1322&        .172&       .0916&      -.0134\\
               &  \footnotesize (.121)&  \footnotesize (.1195)&  \footnotesize (.1359)&  \footnotesize (.1507)\\
V: GT          &       .1875&       .1448&       .1712&       .1534\\
               &  \footnotesize (.0341)&  \footnotesize (.0338)&  \footnotesize (.0397)&  \footnotesize (.0405)\\
V: GT/HAVO     &       .2028&       .1308&       .3215&       .2633\\
               &  \footnotesize (.096)&  \footnotesize (.0981)&  \footnotesize (.0854)&  \footnotesize (.0898)\\
A: HAVO        &        .142&        .097&       .1365&       .0997\\
               &  \footnotesize (.0348)&  \footnotesize (.0317)&  \footnotesize (.0367)&  \footnotesize (.0387)\\
A: HAVO/VWO    &       .3135&       .2067&       .1839&       .1142\\
               &  \footnotesize (.0694)&  \footnotesize (.0764)&  \footnotesize (.072)&  \footnotesize (.0794)\\
\\ [-2ex]
\hline
\\ [-2ex]
\footnotesize COVARIATES $X_i$ & \footnotesize NO & \footnotesize YES& \footnotesize NO & \footnotesize YES  \\ 
 \\[-2ex]
\hline
\footnotesize{COHORT} & \footnotesize 2015& \footnotesize 2015& \footnotesize 2016& \footnotesize 2016 \\
      \hline \hline
    \end{tabular}
\end{center}
\vspace{-0.2cm}
\begin{minipage}{1\linewidth  \setstretch{0.75}}
{\scriptsize  Notes: the table reports, separately for each track and cohort, estimates of the correlation of the proxy for school leniency $\tilde Z$ with the probability of being a Never-Taker. Specifically, it reports the coefficient $\xi$ from running the following regression using units with $Z = 1$: $1-R_i = \gamma + \xi \tilde Z_i + \mu X_i + \epsilon_i$. The row headings report the names of the secondary school tracks to which a student can be assigned in the Dutch system, ranked by level of difficulty. The covariates $X$ included in the second and fourth columns are those described in  Table \ref{t:balance}.
}   
 \end{minipage}
\end{table}

\newpage
\vspace{-0.4 cm}
\section*{Appendix to Section \ref{s:point_est}}
\label{ap:apce_proof}

\subsection*{Proof of Theorem 1 of  \cite{Imai2023} in our context.}

By Assumption \ref{a:strata_monoton}, we have that:
\begin{equation}
    Pr(Y(0) = 0, Y(1) = 0) = Pr(Y(1) = 0) \label{e:a1}
\end{equation}
and
\begin{equation}
    Pr(Y(0) = 1, Y(1) = 1) = Pr(Y(0) = 1). \label{e:a2}
\end{equation}
Then, we can write:
\begin{equation}
\begin{aligned}
    Pr(R(z) = 1, Y(0) = 0, Y(1) = 0)  & \\ 
  &  = Pr(R(z) = 1, Y(1) =  0) \\
   & = Pr(R(z) = 1, Y(R(z)) = 0 \mid Z = z)  \\
   & = Pr(R =1, Y=0 \mid Z = z ),  \label{e:a3}
    \end{aligned}
\end{equation}
where the second equality comes from Assumption \ref{a:rand_sourc}. Equivalently, we can write
\begin{equation}
    \begin{aligned}
        Pr(R(z) = 1, Y(0) = 1, Y(1) = 1 )  \\
        &= Pr(Y(0)=Y(1) = 1) - Pr(R(z) = 0, Y(0) = Y(1) = 1) \\
        &= Pr(Y(0) = 1) - Pr(R(z) = 0, Y(0) = 1) \\
        &= Pr(Y(0) =1) - Pr(R(z) = 0, Y(R(z))=1 \mid Z=z)  \\
        &= Pr(Y(0) = 1) - Pr(R = 0, Y=1 \mid Z=z)  \label{e:a4}
    \end{aligned}
\end{equation}
and 
\begin{equation}
    \begin{aligned}
        Pr(R(z) = 1, Y(0) = 0, Y(1) = 1) =  \\
       &Pr(R(z) = 1) -  Pr(R(z) = 1, Y(0)=Y(1) = 1)  \\
       &  - Pr(R(z) =1, Y(0)=Y(1)=0) \\
       &-\underbrace{Pr(R(z) = 1, Y(0) =1, Y(1) = 0)}_{=0}
    \end{aligned} \label{e:a5}
\end{equation}
Using equations (\ref{e:a3}) and (\ref{e:a4}):
\begin{equation}
    \begin{aligned}
         Pr(R(z) = 1, Y(0) = 0, Y(1) = 1) \\
         &= Pr(R(z) = 1) - \underbrace{Pr(Y(0) = 1) + Pr(R=0, Y=1 \mid Z = z)}_{\textrm{from equation \eqref{e:a4}}} \\
          &- \underbrace{Pr(R=1, Y=0 \mid Z=z)}_{\textrm{from equation \eqref{e:a3}}}\\
        & =Pr(R=1 \mid Z = z) - Pr(R=1, Y= 0 \mid Z=z) \\
        &+ Pr(R=0,Y=1 \mid Z=z) - Pr(Y(0) = 1) \\
        &= Pr(R=1, Y=1 \mid Z=z) + Pr(R=0, Y=1 \mid Z=z) \\
        & - Pr(Y(0)=1)  \\
        & = Pr(Y=1 \mid Z = z) - Pr(Y(0) = 1) \label{e:a6}
    \end{aligned}
\end{equation} \\
\begin{equation}
\begin{aligned}
    Pr(Y(0)=0, Y(1)=1) &= Pr(Y(1) = 1) - Pr(Y(0) =1, Y(1) = 1 ) = \\
    &= Pr(Y(1) = 1) - Pr(Y(0)=1) \label{e:a7}
    \end{aligned}
\end{equation}

\bigskip
\subsection*{Average Principal Causal Effects (APCEs, equations \ref{e:apce_h_est}, \ref{e:apce_al_est} and \ref{e:apce_ah_est})}
\begin{equation*}
    \begin{aligned}
        APCE_{H} &= E[R(1) - R(0) \mid \textrm{student is in H}]  \\
        &= E[R(1) - R(0) \mid Y(0) = 0, Y(1) = 1] \\
        &= \frac{Pr(R(1)=1, Y(0) = 0, Y(1) = 1) - Pr(R(0) = 1, Y(0) = 0, Y(1) = 1)}{Pr(Y(0) = 0, Y(1) = 1)} 
    \end{aligned}
\end{equation*}
Using equations (\ref{e:a6}) and (\ref{e:a7}):
\begin{equation*}
    \begin{aligned}
        \frac{\overbrace{Pr(R(1)=1, Y(0) = 0, Y(1) = 1)}^{Pr(Y=1 \mid Z = 1) - Pr(Y(0) = 1)} - \overbrace{Pr(R(0) = 1, Y(0) = 0, Y(1) = 1)}^{Pr(Y=1 \mid Z = 0) - Pr(Y(0) = 1)}}{\underbrace{Pr(Y(0) = 0, Y(1) = 1)}_{Pr(Y(1) = 1) - Pr(Y(0)=1)}} 
    \end{aligned}
\end{equation*}
\begin{equation*}
    \begin{aligned}
        APCE_{H} &= \frac{Pr(Y=1 \mid Z=1) - Pr(Y=1 \mid Z=0)}{Pr(Y(1) = 1) - Pr(Y(0)=1)} = \\
        &= \frac{E[Y \mid Z = 1] - E[Y \mid Z=0]}{Pr(Y(1) = 1) - Pr(Y(0)=1)}
    \end{aligned}
\end{equation*}

\begin{equation*}
    \begin{aligned}
        APCE_{AL} &= E[R(1) - R(0) \mid \textrm{student is in AL}] = \\
        &= E[R(1) - R(0) \mid Y(0) = 0, Y(1) = 0] = \\
        &=  \frac{Pr(R(1) = 1, Y(0) =0, Y(1) = 0) - Pr(R(0) = 1, Y(0) =0, Y(1) = 0)}{Pr(Y(0)=0,Y(1)=0)} \\
        & \text{From equation \eqref{e:a3}:} \\
        & = \frac{Pr(R=1,Y=0 \mid Z=1) - Pr(R=1,Y=0 \mid Z=0)}{Pr(Y(0)=0,Y(1)=0)} \\
         & \text{From equation \eqref{e:a1}:} \\
         & = \frac{Pr(R=1,Y=0 \mid Z=1) - Pr(R=1,Y=0 \mid Z=0)}{Pr(Y(1)=0)} \\ 
         & = \frac{Pr(R=1,Y=0 \mid Z=1) - Pr(R=1,Y=0 \mid Z=0)}{1 - Pr(Y(1)=1)} 
    \end{aligned}
\end{equation*}

\begin{equation*}
    \begin{aligned}
        APCE_{AH} &= E[R(1) - R(0) \mid \textrm{student is in AH}] = \\
        &= E[R(1) - R(0) \mid Y(0) = 1, Y(1) = 1] = \\
        &= \frac{Pr(R(1) = 1, Y(0) =1, Y(1) = 1) - Pr(R(0) = 1, Y(0) =1, Y(1) = 1)}{Pr(Y(0)=1,Y(1)=1)} \\
        & \text{From equation \eqref{e:a4}:} \\
        & = \frac{Pr(Y(0)=1)-Pr(R=0,Y=1 \mid Z =1) - Pr(Y(0)=1) + Pr(R=0,Y=1 \mid Z =0)}{Pr(Y(0)=1,Y(1)=1)} \\
         & \text{From equation \eqref{e:a2}:} \\
         &= \frac{Pr(R=0,Y=1 \mid Z =0) - Pr(R=0,Y=1 \mid Z =1)}{Pr(Y(0) = 1)}
    \end{aligned}
\end{equation*}

\subsection*{Derivation of the bounds of the $APCE_J$; equation (\ref{e:bounds_h})}
By assumption \ref{a:rand_sourc}:
\begin{equation}
    Pr(Y(R) = 1 ) = Pr(Y(R) = 1 \mid Z = z) \label{e:a8}
\end{equation}
By the law of total probability:
\begin{equation}
\begin{aligned}
    Pr(Y(R) = 1 \mid Z = z) = \\
    =& Pr(Y = 1 \mid R = R, Z = z) Pr(R = R \mid Z = z ) \\
    & + Pr(Y(R) = 1 \mid R = 1- R, Z = z) Pr(R =1- R \mid Z = z )
    \label{e:a9}
\end{aligned}
\end{equation}
We can rewrite it as:
\begin{equation}
    \begin{aligned}
    Pr(Y(R) = 1 \mid Z = z) = \\
    =& Pr(Y = 1,R = R \mid Z = z)  \\
    & + \underbrace{Pr(Y(R) = 1 \mid R = 1- R, Z = z) Pr(R =1- R \mid Z = z )}_{\geq 0}
    \label{e:a10}
\end{aligned}
\end{equation}
By Assumption \ref{a:strata_monoton}:
\begin{equation}
Pr(Y = 1 \mid R = 0, Z = z ) \leq Pr(Y(1) = 1 \mid R = 0, Z = z ) \leq 1 
\label{e:a11}
\end{equation}
and
\begin{equation}
    0 \leq Pr(Y(0) = 1 \mid R = 1, Z=z) \leq Pr(Y = 1 \mid R = 1, Z = z)
    \label{e:a12}
\end{equation}
From equation (\ref{e:a10})
\begin{equation}
    Pr(Y(0) = 1)  \geq Pr(Y = 1, R = 0 \mid Z= z)
    \label{e:a13}
\end{equation}
and
\begin{equation}
    \begin{aligned}
        Pr(Y(1) = 1 ) = \\
        = & Pr(Y = 1, R = 1 \mid Z=z) \\
        & + \underbrace{Pr(Y(1)=1 \mid R = 0, Z=z)}_{\leq 1}Pr(R=0 \mid Z=z)
        \label{e:a22}.
    \end{aligned}
\end{equation}
So that we can write
\begin{equation}
    \begin{aligned}
         Pr(Y(1) = 1 ) \leq & Pr(Y = 1, R = 1 \mid Z=z) + Pr(R=0 \mid Z=z) \\
         = & 1 - Pr(Y = 0, R = 1 \mid Z=z) 
         \label{e:a23}
    \end{aligned}
\end{equation}
Then, using equations (\ref{e:a11}), (\ref{e:a12}), (\ref{e:a13}) and (\ref{e:a23}), the bounds for $Pr(Y(R) = 1 )$ are:
\begin{equation}
    \begin{aligned}
        \max_{z} Pr(Y = 1, R = 0 \mid Z = z) & \leq Pr(Y(0)=1) \leq \min_{z} Pr(Y = 1 \mid Z = z) \\
        \max_{z}  Pr(Y = 1 \mid Z = z) & \leq Pr(Y(1) = 1 ) \leq  1 - \max_{z} Pr(Y = 0, R = 1 \mid Z =z)
    \end{aligned}
\end{equation}

\bigskip
\subsection*{Upper and Lower Bounds of the $APCE_J$; equations (\ref{e:lb_apce_h}), (\ref{e:ub_apce_h}), (\ref{e:lb_apce_AH}),  (\ref{e:ub_apce_AH}) (\ref{e:lb_apce_AL}), and (\ref{e:ub_apce_AL}) }
Lower bound of $APCE_{H}$:
\begin{equation*}
    \begin{aligned}
      \frac{E[Y \mid Z = 1] - E[Y \mid Z=0]}{1 - \max_{z} Pr(Y = 0, R = 1 \mid Z =z) -  \max_{z} Pr(Y = 1, R = 0 \mid Z = z)}
    \end{aligned}
\end{equation*}
Upper bound of $APCE_{H}$:
\begin{equation*}
     \frac{E[Y \mid Z = 1] - E[Y \mid Z=0]}{ \max_{z}  Pr(Y = 1 \mid Z = z) - \min_{z} Pr(Y = 1 \mid Z = z)}
\end{equation*}
Lower bound of $ APCE_{AH}$:
\begin{equation*}
    \frac{Pr(R=0,Y=1 \mid Z =0) - Pr(R=0,Y=1 \mid Z =1)}{\min_{z} Pr(Y = 1 \mid Z = z)}
\end{equation*}
Upper bound of $ APCE_{AH}$:
\begin{equation*}
    \frac{Pr(R=0,Y=1 \mid Z =0) - Pr(R=0,Y=1 \mid Z =1)}{\max_{z} Pr( R = 0,Y = 1 \mid Z = z)}
\end{equation*}
Lower bound of $ APCE_{AL}  $:
\begin{equation*}
    \frac{Pr(R=1,Y=0 \mid Z=1) - Pr(R=1,Y=0 \mid Z=0)}{1 - \max_{z}  Pr(Y\mid Z = z) }
\end{equation*}
Upper bound of $ APCE_{AL}  $:
\begin{equation*}
    \frac{Pr(R=1,Y=0 \mid Z=1) - Pr(R=1,Y=0 \mid Z=0)}{\max_{z} Pr( R = 1,Y = 0 \mid Z =z)} 
\end{equation*}

\bigskip
\subsection*{Identification and Estimation of $APCE_J$ and Principal Fairness under Unconfoundedness}

Because the goal of fairness analysis is really to understand and compare the size of the $\Delta_j$'s, it is crucial to be able to point-identify and estimate these quantities for fairness comparisons. A natural and useful starting point, that would be more or less plausible depending on the information cointained in the baseline covariates, is assuming Assumption \ref{a:unconf} Unconfoundedness.

Consider the following principal scores, which represents the population proportions for each of the strata in each cell defined by the covariates:
\begin{equation*}
    \begin{aligned}
        e_{AL} = Pr(Y(1) = 0, Y(0) = 0 \mid X =x) \\
        e_{AH} = Pr(Y(1) = 1, Y(0) = 1 \mid X =x)\\
        e_{H} = Pr(Y(1) = 1, Y(0) = 0 \mid X =x).
    \end{aligned}
\end{equation*}
Under assumptions 1, \ref{a:rand_sourc}, \ref{a:ec}, \ref{a:strata_monoton} and \ref{a:unconf}, we can identify these  principal strata proportions as:
\begin{equation*}
    \begin{aligned}
        e_{AL} = &Pr(Y = 0 \mid R = 1, X =x) \\
        e_{AH} = &Pr(Y = 1 \mid R = 0, X =x)\\
        e_{H} = &Pr(Y = 1 \mid R = 1, X = x) -  Pr(Y = 1 \mid R = 0, X =x)
    \end{aligned}
\end{equation*}
\begin{theorem}
    Point identification of APCEs under unconfoundedness: \\
    Under assumptions 1, \ref{a:rand_sourc}, \ref{a:ec}, \ref{a:strata_monoton},  and \ref{a:unconf}, $APCE_{AL}$, $APCE_{AH}$ and $APCE_{H}$ are identified as follows:
    \begin{equation*}
        \begin{aligned}
            APCE_{AL} = &\mathbb{E}[w_{AL}(X)R \mid Z = 1] -\mathbb{E}[w_{AL}(X)R \mid Z = 0]  \\
            APCE_{AH} =&\mathbb{E}[w_{AH}(X)R \mid Z = 1] -\mathbb{E}[w_{AH}(X)R \mid Z = 0]  \\
            APCE_{H} =& \mathbb{E}[w_{H}(X)R \mid Z = 1] -\mathbb{E}[w_{H}(X)R \mid Z = 0] ,
        \end{aligned}
    \end{equation*}
    where
    \begin{equation*}
        \begin{aligned}
            w_{AL} =& \frac{e_{AL}(X)}{\mathbb{E}[e_{AL}(X)]}, \quad
            w_{AH} =& \frac{e_{AH}(X)}{\mathbb{E}[e_{AH}(X)]}, \quad
            w_{H} =& \frac{e_{H}(X)}{\mathbb{E}[e_{H}(X)]}
        \end{aligned}
    \end{equation*}
    Proof: See \cite{Imai2023}.
\end{theorem}
Similarly, consider the following principal scores for a specific value of a protected attribute $B=b$ with $b\in \{0, 1\}$ ($B$ is in general one of the covariates in $X$):
\begin{equation*}
    \begin{aligned}
        e^w_{AL} = Pr(Y(1) = 0, Y(0) = 0 \mid X =x, B=b) \\
        e^w_{AH} = Pr(Y(1) = 1, Y(0) = 1 \mid X =x, B=b)\\
        e^w_{H} = Pr(Y(1) = 1, Y(0) = 0 \mid X =x, B=b).
    \end{aligned}
\end{equation*}

\noindent which can be still identified under Assumption 1, 2, 3, 4 and 7.

\begin{theorem}
    Point identification of fairness under unconfoundedness: \\
    Under assumptions 1, \ref{a:rand_sourc}, \ref{a:ec}, \ref{a:strata_monoton},  and \ref{a:unconf}, $\Delta_{AL}(z)$, $\Delta_{AH}(z)$ and $\Delta_{H}(z)$ are identified as follows:
    \begin{equation*}
        \begin{aligned}
            \Delta_{AL}(z) = &\mathbb{E}[w^1_{AL}(X)R \mid Z = z, B = 1] -\mathbb{E}[w^{0}_{AL}(X)R \mid Z = z, B=0]  \\
            \Delta_{AH}(z) = &\mathbb{E}[w^1_{AL}(X)R \mid Z = z, B = 1] -\mathbb{E}[w^{0}_{AH}(X)R \mid Z = z, B=0]  \\ 
            \Delta_{H}(z) = &\mathbb{E}[w^1_{H}(X)R \mid Z = z, B = 1] -\mathbb{E}[w^{0}_{H}(X)R \mid Z = z, B=0]  \\ ,
        \end{aligned}
    \end{equation*}
    where
    \begin{equation*}
        \begin{aligned}
            w^w_{AL} =& \frac{e^w_{AL}(X)}{\mathbb{E}[e^w_{AL}(X)]}, \quad
            w^w_{AH} =& \frac{e^w_{AH}(X)}{\mathbb{E}[e^w_{AH}(X)]}, \quad
            w^w_{H} =& \frac{e^w_{H}(X)}{\mathbb{E}[e^w_{H}(X)]}
        \end{aligned}
    \end{equation*}
    Proof: See \cite{Imai2023}
\end{theorem}

\clearpage
\newpage
\section*{Appendix to Section \ref{s:3strata}}\label{ap:3strata}

\subsection*{Estimates obtained without covariates}

\begin{table}[ht]
    \caption{Estimates of the $APCE_{H}$, no covariates}
    \label{t:APCE_H_nocov}
\begin{center}
    \vspace{0.1cm} 
\begin{tabular}{l*{5}{c}}
\hline\hline
      \\ [-2ex]       
Track &   Numerator& Lower &    Upper &   Lower  & Upper\\
&  $APCE_H$    & bound  &    bound     &   bound &   bound \\
&     & denominator  &  denominator&  $APCE_H$ &  $APCE_H$  \\
&     & $APCE_H$  &  $APCE_H$&  \hspace{1.9cm} &  \hspace{1.9cm}   \\
      & 1& 2& 3 & 4 & 5 \\
      \\ [-2ex]  
\hline
  \\ [-2ex]  
V: BL          &       .0238&       .0238&        .651&        .036&           1\\
          &       \footnotesize (.0171)&       \footnotesize (.0162)&       \footnotesize (.0137)&        \footnotesize (.027)&       \footnotesize -\\
V: BL/KL       &       .0737&       .0737&       .2872&       .2643&           1\\
          &       \footnotesize (.0298)&       \footnotesize (.0288)&       \footnotesize (.0226)&       \footnotesize (.1019)&        \footnotesize -\\
V: KL          &       .1049&       .1049&       .7016&       .1496&           1\\
          &       \footnotesize (.0159)&       \footnotesize (.0159)&       \footnotesize (.0105)&       \footnotesize (.0229)&           \footnotesize -\\
V: KL/GT       &       .0221&       .0221&       .3946&       .0566&           1\\
          &       \footnotesize (.0229)&       \footnotesize (.0207)&       \footnotesize (.0221)&       \footnotesize (.0553)&       \footnotesize -\\
V: GT          &       .0216&       .0216&       .7624&       .0284&           1\\
          &       \footnotesize (.0089)&       \footnotesize (.0087)&       \footnotesize (.0075)&       \footnotesize (.0118)&       \footnotesize -\\
V: GT/HAVO     &       .0554&       .0554&       .5051&       .1072&           1\\
          &        \footnotesize (.017)&       \footnotesize (.0165)&       \footnotesize (.0146)&       \footnotesize (.0323)&       \footnotesize -\\
A: HAVO        &       .0376&       .0376&       .8068&       .0466&           1\\
          &       \footnotesize (.0079)&       \footnotesize (.0079)&       \footnotesize (.0067)&       \footnotesize (.0101)&           -\\
A: HAVO/VWO    &        .062&        .062&       .5431&       .1124&           1\\
          &        \footnotesize (.016)&       \footnotesize (.0156)&        \footnotesize (.013)&       \footnotesize (.0286)&       \footnotesize -\\
            \\ [-2ex] 
  \hline
  \\ [-2ex]
VOCATIONAL  &       .0469&       .0469&       .6489&       .0723&           1\\
          &       \footnotesize (.0063)&       \footnotesize (.0061)&        \footnotesize (.005)&       \footnotesize (.0097)&       \footnotesize -\\
ACADEMIC    &       .0453&       .0453&       .7236&       .0626&           1\\
          &       \footnotesize (.0076)&       \footnotesize (.0075)&        \footnotesize (.006)&       \footnotesize (.0106)&       \footnotesize -\\
ALL         &       .0463&       .0463&       .6792&       .0681&           1\\
          &        \footnotesize (.005)&       \footnotesize (.0049)&       \footnotesize (.0039)&       \footnotesize (.0073)&       \footnotesize -\\
\hline\hline
\end{tabular}
\end{center}
\vspace{-0.2cm}
\begin{minipage}{1\linewidth  \setstretch{0.75}}
{\scriptsize Notes: The row headings reports the names of the secondary school tracks to which a student can be assigned in the Dutch system, ranked by level of difficulty.  The last three rows of the table aggregate the Vocational (VMBO*), the Academic (HAVO*), and All tracks, respectively.  With reference to the $APCE_H$ estimand defined in equation (\ref{e:apce_h_est}), the table reports for the outcome $Y$ and separately for each track, estimates of the numerator (in column 1), of the lower and upper bounds of the denominator (in columns 2 and 3), and of the lower and upper bounds of the $APCE_H$ (in columns 4 and 5). The definitions of these two last statistics are in equations (\ref{e:lb_apce_h}) and (\ref{e:ub_apce_h}), respectively. The aggregations for the three last rows are performed as follows: point estimates of the numerator and of the lower and upper bounds of the denominator are created by weighing each track in the corresponding aggregate by the relative population in each track. For the lower and upper bounds of the $APCE_H$, the aggregated estimates for the lower and upper bounds of the denominators are used directly. Standard errors for each quantity are computed using a block bootstrap at the school level with 1000 repetitions; each quantity in every iteration is calculated in the same way.}
 \end{minipage}
\end{table}

\begin{table}[ht]
    \caption{Estimates of the $APCE_{AH}$, no covariates}
    \label{t:APCE_AH_nocov}
\begin{center}
    \vspace{0.1cm} 

\begin{tabular}{l*{5}{c}}
\hline\hline
      \\ [-2ex]       
Track &   Numerator& Lower &    Upper &   Lower  & Upper\\
&  $APCE_{AH}$    & bound  &    bound     &   bound &   bound \\
&     & denominator  &  denominator&  $APCE_{AH}$ &  $APCE_{AH}$  \\
&     & $APCE_{AH}$  &  $APCE_{AH}$&  \hspace{1.9cm} &  \hspace{1.9cm}   \\
      & 1& 2& 3 & 4 & 5 \\
      \\ [-2ex]  
\hline
  \\ [-2ex] 
V: BL          &       .0033&       .3188&       .3208&       .0097&       .0098\\
            &  \footnotesize (.0129)&  \footnotesize (.0115)&  \footnotesize (.0116)&  \footnotesize (.0384)&  \footnotesize (.0384)\\
V: BL/KL       &       .0331&       .6843&       .7032&       .0469&       .0482\\
            &  \footnotesize (.0275)&  \footnotesize (.0208)&  \footnotesize (.0211)&  \footnotesize (.0378)&  \footnotesize (.0383)\\
V: KL          &       .0035&       .2256&       .2367&       .0161&       .0166\\
            &  \footnotesize (.0091)&  \footnotesize (.0086)&  \footnotesize (.0092)&  \footnotesize (.0364)&  \footnotesize (.0379)\\
V: KL/GT       &       .0792&       .5698&       .5813&       .1369&       .1399\\
            &  \footnotesize (.0283)&  \footnotesize (.0192)&  \footnotesize (.0186)&  \footnotesize (.0478)&  \footnotesize (.0477)\\
V: GT          &       .0049&       .1923&        .193&       .0249&       .0251\\
            &  \footnotesize (.0054)&  \footnotesize (.0057)&  \footnotesize (.0057)&  \footnotesize (.0272)&  \footnotesize (.0272)\\
V: GT/HAVO     &       .0433&       .4178&       .4269&       .0997&       .1014\\
            &  \footnotesize (.017)&  \footnotesize (.0113)&  \footnotesize (.0115)&  \footnotesize (.0385)&  \footnotesize (.0389)\\
A: HAVO        &           0&       .1283&       .1283&           0&           0\\
            &  \footnotesize (.0004)&  \footnotesize (.004)&  \footnotesize (.004)&  \footnotesize (.0028)&  \footnotesize (.0029)\\
A: HAVO/VWO    &       .0293&       .3684&       .3798&       .0759&       .0783\\
            &  \footnotesize (.0146)&  \footnotesize (.0108)&  \footnotesize (.0113)&  \footnotesize (.0368)&  \footnotesize (.0375)\\
             \\ [-2ex] 
  \hline
  \\ [-2ex]
VOCATIONAL  &       .0169&        .298&       .3038&       .0558&       .0569\\
            &  \footnotesize (.005)&  \footnotesize (.004)&  \footnotesize (.0041)&  \footnotesize (.016)&  \footnotesize (.0162)\\
ACADEMIC    &       .0092&       .2041&       .2077&       .0445&       .0453\\
            &  \footnotesize (.0046)&  \footnotesize (.0043)&  \footnotesize (.0044)&  \footnotesize (.0218)&  \footnotesize (.0221)\\
ALL         &       .0138&       .2599&       .2648&       .0522&       .0532\\
            &  \footnotesize (.0035)&  \footnotesize (.0029)&  \footnotesize (.003)&  \footnotesize (.013)&  \footnotesize (.0132)\\
\hline\hline
\end{tabular}
 
\end{center}
\vspace{-0.2cm}
\begin{minipage}{1\linewidth  \setstretch{0.75}}
{\scriptsize Notes: The row headings report the names of the secondary school tracks to which a student can be assigned in the Dutch system, ranked by level of difficulty.  The last three rows of the table aggregate the Vocational (VMBO*), the Academic (HAVO*), and All tracks, respectively.  With reference to the $APCE_{AH}$ estimand defined in equation (\ref{e:apce_ah_est}), the table reports for the outcome $Y$ and separately for each track, estimates of the numerator (in column 1), of the lower and upper bounds of the denominator (in columns 2 and 3), and of the lower and upper bounds of the $APCE_{AH}$ (in columns 4 and 5). The definitions of these two last statistics are in equations (\ref{e:lb_apce_AH}) and (\ref{e:ub_apce_AH}), respectively. The aggregations for the last three rows are performed as follows: point estimates of the numerator and of the lower and upper bounds of the denominator are created by weighing each track in the corresponding aggregate by the relative population in each track. For the lower and upper bounds of the $APCE_{AH}$, the aggregated estimates for the lower and upper bounds of the denominators are used directly. Standard errors for each quantity are computed using a block bootstrap at the school level with 1000 repetitions; each quantity in every iteration is calculated in the same way.}
 \end{minipage}
\end{table}

\begin{table}[ht]
    \caption{Estimates of the $APCE_{AL}$, no covariates}
    \label{t:APCE_AL_nocov}
\begin{center}
    \vspace{0.1cm} 
\begin{tabular}{l*{5}{c}}
\hline\hline
      \\ [-2ex]       
Track &   Numerator& Lower &    Upper &   Lower  & Upper\\
&  $APCE_{AL}$    & bound  &    bound     &   bound &   bound \\
&     & denominator  &  denominator&  $APCE_{AL}$ &  $APCE_{AL}$  \\
&     & $APCE_{AL}$  &  $APCE_{AL}$&  \hspace{1.9cm} &  \hspace{1.9cm}  \\
      & 1& 2& 3 & 4 & 5 \\
      \\ [-2ex]  
\hline
  \\ [-2ex] 
V: BL          &       .0267&       .0277&       .6579&       .0403&       .9734\\
            &  \footnotesize (.0051)&  \footnotesize (.005)&  \footnotesize (.0137)&  \footnotesize (.0078)&  \footnotesize (.0319)\\
V: BL/KL       &       .0261&       .0285&        .223&       .1212&       .9209\\
            &  \footnotesize (.0087)&  \footnotesize (.0084)&  \footnotesize (.0202)&  \footnotesize (.0417)&  \footnotesize (.1156)\\
V: KL          &       .0663&       .0728&       .6584&       .1005&         .91\\
            &  \footnotesize (.0071)&  \footnotesize (.0068)&  \footnotesize (.0128)&  \footnotesize (.0105)&  \footnotesize (.0276)\\
V: KL/GT       &       .0341&       .0356&       .3966&       .0861&       .9603\\
            &  \footnotesize (.0084)&  \footnotesize (.0082)&  \footnotesize (.0185)&  \footnotesize (.0213)&  \footnotesize (.0536)\\
V: GT          &       .0376&       .0411&       .7896&       .0477&       .9174\\
            &  \footnotesize (.0038)&  \footnotesize (.0037)&  \footnotesize (.0069)&  \footnotesize (.0048)&  \footnotesize (.0261)\\
V: GT/HAVO     &       .0677&       .0771&       .5177&        .131&       .8774\\
            &  \footnotesize (.0086)&  \footnotesize (.0082)&  \footnotesize (.0136)&  \footnotesize (.0161)&  \footnotesize (.0367)\\
A: HAVO        &       .0486&        .054&       .8449&       .0575&       .8994\\
            &  \footnotesize (.0041)&  \footnotesize (.0039)&  \footnotesize (.0061)&  \footnotesize (.0048)&  \footnotesize (.027)\\
A: HAVO/VWO    &       .0792&       .0886&       .5583&       .1416&       .8943\\
            &  \footnotesize (.0077)&  \footnotesize (.0075)&  \footnotesize (.0123)&  \footnotesize (.0135)&  \footnotesize (.0272)\\
            \\ [-2ex] 
  \hline
  \\ [-2ex]
VOCATIONAL  &       .0466&       .0512&       .6512&       .0715&       .9097\\
            &  \footnotesize (.0027)&  \footnotesize (.0026)&  \footnotesize (.0048)&  \footnotesize (.0041)&  \footnotesize (.0149)\\
ACADEMIC    &       .0583&       .0649&       .7544&       .0772&       .8973\\
            &  \footnotesize (.0037)&  \footnotesize (.0036)&  \footnotesize (.0057)&  \footnotesize (.0049)&  \footnotesize (.0188)\\
ALL         &       .0513&       .0568&       .6931&        .074&       .9039\\
            &  \footnotesize (.0022)&  \footnotesize (.0021)&  \footnotesize (.0037)&  \footnotesize (.0031)&  \footnotesize (.0118)\\
\hline\hline

\end{tabular}
 
\end{center}
\vspace{-0.2cm}
\begin{minipage}{1\linewidth  \setstretch{0.75}}
{\scriptsize Notes: The row headings report the names of the secondary school tracks to which a student can be assigned in the Dutch system, ranked by level of difficulty.  The last three rows of the table aggregate the Vocational (VMBO*), the Academic (HAVO*), and All tracks, respectively.  With reference to the $APCE_{AL}$ estimand defined in equation (\ref{e:apce_al_est}), the table reports for the outcome $Y$ and separately for each track, estimates of the numerator (in column 1), of the lower and upper bounds of the denominator (in columns 2 and 3), and of the lower and upper bounds of the $APCE_{AL}$ (in columns 4 and 5). The definitions of these two last statistics are in equations (\ref{e:lb_apce_AL}) and (\ref{e:ub_apce_AL}), respectively. The aggregations for the last three rows are performed as follows: point estimates of the numerator and of the lower and upper bounds of the denominator are created by weighing each track in the corresponding aggregate by the relative population in each track. For the lower and upper bounds of the $APCE_{AL}$, the aggregated estimates for the lower and upper bounds of the denominators are used directly. Standard errors for each quantity are computed using a block bootstrap at the school level with 1000 repetitions; each quantity in every iteration is calculated in the same way.}
 \end{minipage}
\end{table}

\clearpage

\newpage
\section*{Appendix to Section \ref{s:ben-michael}} \label{ap:ben-michael}

\subsection*{Derivation of the Loss Function}

We can estimate
\begin{equation*}
    \Pi_{11}(z) = Pr(R(z)=1; Y(1)=1)
\end{equation*}
and
\begin{equation*}
    \Pi_{10}(z) = Pr(R(z)=1; Y(1)=0),
\end{equation*}
but we cannot identify
\begin{equation*}
    \Pi_{01}(z) = Pr(R(z)=0; Y(1)=1)
\end{equation*}
and
\begin{equation*}
    \Pi_{00}(z) = Pr(R(z)=0; Y(1)=0).
\end{equation*}
From Assumption \ref{a:rand_sourc}, it follows that the distribution of potential outcomes is the same at both sides of the cutoff:
\begin{equation*}
    Pr(Y(1) = 1 \mid Z=1) = Pr(Y(1) = 1 \mid Z=0)
\end{equation*}
We can rewrite it as:
\begin{align*}
    \sum_{r} Pr(R=r , Y(1) = 1 \mid Z=1) &=\sum_{r} Pr(R=r ,Y(1) = 1\mid Z=0) \\
    \Pi_{11}(1) + \Pi_{01}(1) &= \Pi_{11}(0) + \Pi_{01}(0) \\
     \Pi_{01}(1)-\Pi_{01}(0)  &= \Pi_{11}(0)-\Pi_{11}(1) 
\end{align*}
From equation (\ref{e:over_loss}) we had that:
\begin{equation*}
    \mathcal{L}(l_{01},z ) = \Pi_{01}(z) + l_{10}\Pi_{10}(z),
\end{equation*}
Consequently,
\begin{align*}
    \mathcal{L}(l_{01},1)- \mathcal{L}(l_{01},0)& = \Pi_{01}(1) -\Pi_{01}(0) + l_{10}\Pi_{10}(1) - l_{10}\Pi_{10}(0) \\
    & = \Pi_{11}(0)-\Pi_{11}(1)  - l_{10}(\Pi_{10}(1) - \Pi_{10}(0))
\end{align*}

\clearpage

\newpage
\section*{Appendix to Section \ref{s:robust}}\label{ap:robust}

\subsection*{Identification of APCEs Without Assuming Exclusion Restriction}
Without the exclusion restriction, we have to index the potential outcomes using both $Z$ and $R$, $Y(z,r)$. Nevertheless, under Assumptions \ref{a:homogeneity} and \ref{a:pi}, we can normalize $\mathbb{E}[Y(0,r)]$ to $\mathbb{E}[Y(r)]$ and $\mathbb{E}[Y(1,r)] = \mathbb{E}[Y(r)] + \eta$.

From equations (\ref{e:a3}) and (\ref{e:a4}), and under the exclusion restriction, we know that:
\begin{align*}
     Pr(R(z) = 1, Y(0) = 0, Y(1) = 0)  = Pr(R =1, Y=0 \mid Z = z ) \\
      Pr(R(z) = 1, Y(0) = 1, Y(1) = 1 ) = Pr(Y(0) = 1) - Pr(R = 0, Y=1 \mid Z=z)
\end{align*}
Replacing the exclusion restriction with assumptions \ref{a:homogeneity} and  \ref{a:pi}, this becomes:
\begin{equation}
     Pr(R(0) = 1, Y(0) = 0, Y(1) = 0)  = Pr(R =1, Y=0 \mid Z = 0 ) \label{e:a17} 
\end{equation}
\begin{align}
      Pr(R(1) = 1, Y(0) = 0, Y(1) = 0)  = Pr(R =1, Y=0 \mid Z = 1 ) + \eta Pr(R=1\mid Z=1) \label{e:a18}
\end{align}
\begin{equation}
     Pr(R(0) = 1, Y(0) = 1, Y(1) = 1 ) = Pr(Y(0) = 1) - Pr(R = 0, Y=1 \mid Z=0)\label{e:a19} 
\end{equation}
\begin{align}
      Pr(R(1) = 1, Y(0) = 1, Y(1) = 1 ) = Pr(Y(0) = 1) - Pr(R = 0, Y=1 \mid Z=1) + \eta Pr(R=0\mid Z=1) \label{e:a20}
\end{align}
Similarly, from the proof in  \nameref{ap:apce_proof}, we known that
\begin{align}
     Pr(R(z) = 1, Y(0) = 0, Y(1) = 1) &\notag \\
     & = Pr(R(z) = 1) -   Pr(R(z) = 1, Y(0) = 0, Y(1) = 0)  \\
     & -  Pr(R(z) = 1, Y(0) = 1, Y(1) = 1 )\label{e:a21}.
\end{align}
Therefore, using the results from (\ref{e:a17}-\ref{e:a20}), it follows that
\begin{align}
    Pr(R(0 ) = 1, Y(0) = 0, Y(1) = 1) &=\notag\\ 
    & = Pr(R = 1 \mid Z = 0) -    Pr(R =1, Y=0 \mid Z = 0 )\notag  \\
     & -  Pr(Y(0) = 1) + Pr(R = 0, Y=1 \mid Z=0)\notag \\
     &  = 
    Pr(Y=1 \mid Z = 0) - Pr(Y(0) = 1) 
\end{align}
and
\begin{align}
    Pr(R(1) = 1, Y(0) = 0, Y(1) = 1) &=\notag\\ 
    & = Pr(R = 1 \mid Z = 1) -    Pr(R =1, Y=0 \mid Z = 1 ) \\
    & - \eta Pr(R=1\mid Z=1)\notag  \\
     & -  Pr(Y(0) = 1) + Pr(R = 0, Y=1 \mid Z=1) - \eta Pr(R=0\mid Z=1)\notag \\
     &  = 
    Pr(Y=1 \mid Z = 0) - Pr(Y(0) = 1) - \eta
\end{align}
From equations (\ref{e:a21}) and (\ref{e:a22}), we have that:
\begin{align*}
    APCE_{H} &= E[R(1) - R(0) \mid \textrm{student is in H}]  \\
    & = \frac{Pr(R(1)=1, Y(0) = 0, Y(1) = 1) - Pr(R(0) = 1, Y(0) = 0, Y(1) = 1)}{Pr(Y(0) = 0, Y(1) = 1)}  \\
    &= \frac{E[Y \mid Z = 1] - E[Y \mid Z=0] - \eta}{Pr(Y(1) = 1) - Pr(Y(0)=1)}.
\end{align*}
Similarly, from equations (\ref{e:a17}) - (\ref{e:a20}),
\begin{align*}
    APCE_{AL} &= E[R(1) - R(0) \mid \textrm{student is in AL}] = \\
     &=  \frac{Pr(R(1) = 1, Y(0) =0, Y(1) = 0) - Pr(R(0) = 1, Y(0) =0, Y(1) = 0)}{Pr(Y(0)=0,Y(1)=0)} \\ 
      & = \frac{Pr(R=1,Y=0 \mid Z=1) + \eta Pr(R=1 \mid Z= 1) - Pr(R=1,Y=0 \mid Z=0)}{1 - Pr(Y(1)=1)} 
\end{align*}

\begin{align*}
     APCE_{AH} &= E[R(1) - R(0) \mid \textrm{student is in AH}] = \\
        &= \frac{Pr(R(1) = 1, Y(0) =1, Y(1) = 1) - Pr(R(0) = 1, Y(0) =1, Y(1) = 1)}{Pr(Y(0)=1,Y(1)=1)} \\
        &= \frac{Pr(R=0,Y=1 \mid Z =0) - Pr(R=0,Y=1 \mid Z =1) + \eta Pr(R=0\mid Z=1)}{Pr(Y(0) = 1)}
\end{align*}
Finally, we derive the bounds on $Pr(Y(0) = 1)$ and $Pr(Y(1) = 1)$ without the exclusion restriction. Following the same intuition, and using the results from the section with the exclusion restriction, we have that:
\begin{align*}
            \max_{}\{ Pr(Y = 1, R = 0 \mid Z = 0),Pr(Y = 1, R = 0 \mid Z = 1)-\eta Pr(R=0\mid Z=1)\} \\
            \leq Pr(Y(0)=1) \leq \min \{Pr(Y = 1 \mid Z = 0),Pr(Y = 1 \mid Z = 1) - \eta\}
\end{align*}
\begin{align*}
        \max \{ Pr(Y = 1 \mid Z = 0),Pr(Y = 1 \mid Z = 1) - \eta \} \leq Pr(Y(1) = 1 ) \\
        \leq  1 - \max \{Pr(Y = 0, R = 1 \mid Z =0),Pr(Y = 0, R = 1 \mid Z =1) +\eta Pr(R=1\mid Z=1)\}
\end{align*}
All in all, the estimators of the bounds of the APCEs are given by the sample analogs of the following expressions:
\begin{landscape}

Lower bound of $APCE_{H}$:

\begin{equation}\label{noER_APCE_H_LB}\scalebox{0.67}{$
\begin{aligned}
      \frac{Pr[Y \mid Z = 1] - Pr[Y \mid Z=0] -\eta}{1 - \max\bigg\{Pr(Y = 0, R = 1 \mid Z =1) {+ \eta Pr(R=1 \mid Z=1)}, Pr(Y = 0, R = 1 \mid Z =0)\bigg\} -  \max \bigg\{Pr(Y = 1, R = 0 \mid Z = 1) - {\eta Pr(R=0 \mid Z=1)},Pr(Y = 1, R = 0 \mid Z = 0)\bigg\}}
    \end{aligned}
    $}
\end{equation}

Upper bound of $APCE_{H}$:
\begin{equation}\label{noER_APCE_H_UB}
     \frac{Pr[Y \mid Z = 1] - Pr[Y \mid Z=0]{-\eta}}{ \max \bigg\{ Pr(Y = 1 \mid Z = 1) {-\eta } ,Pr(Y = 1 \mid Z = 0)\bigg \} - \min \bigg\{ Pr(Y = 1 \mid Z = 1) {-\eta } ,Pr(Y = 1 \mid Z = 0)\bigg \} }
\end{equation}
Lower bound of $ APCE_{AH}$:
\begin{equation}\label{noER_APCE_AH_LB}
    \frac{Pr(R=0,Y=1 \mid Z =0) - Pr(R=0,Y=1 \mid Z =1) {+\eta Pr(R=0 \mid Z=1)}}{\min \bigg\{ Pr(Y = 1 \mid Z = 1) {-\eta } ,Pr(Y = 1 \mid Z = 0)\bigg \} }
\end{equation}
Upper bound of $ APCE_{AH}$:
\begin{equation}\label{noER_APCE_AH_UB}
    \frac{Pr(R=0,Y=1 \mid Z =0) - Pr(R=0,Y=1 \mid Z =1){+\eta Pr(R=0 \mid Z=1)}}{\max \bigg\{Pr( R = 0,Y = 1 \mid Z = 1) {-\eta Pr(R=0 \mid Z=1)},Pr( R = 0,Y = 1 \mid Z = 0)\bigg\}}
\end{equation}
Lower bound of $ APCE_{AL}  $:
\begin{equation}\label{noER_APCE_AL_LB}
    \frac{Pr(R=1,Y=0 \mid Z=1) - Pr(R=1,Y=0 \mid Z=0){+\eta Pr(R=1 \mid Z=1)}}{1 - \max \bigg\{ Pr(Y = 1 \mid Z = 1) {-\eta } ,Pr(Y = 1 \mid Z = 0)\bigg \}  }
\end{equation}
Upper bound of $ APCE_{AL}  $:
\begin{equation}\label{noER_APCE_AL_UB}
    \frac{Pr(R=1,Y=0 \mid Z=1) - Pr(R=1,Y=0 \mid Z=0){+\eta Pr(R=1 \mid Z=1)}}{\max \bigg\{ Pr( R = 1,Y = 0 \mid Z =1) {+ \eta Pr(R=1 \mid Z=1)} , Pr( R = 1,Y = 0 \mid Z =0)\bigg \}} 
\end{equation}
\end{landscape}

\subsection*{Estimates of Scoring Above Cutoff for Never-Takers ($\eta$)}

Table \ref{t:eta} shows the estimated values of $\eta$, as outlined in Section \ref{s:robust}. 

\begin{table}[H]
    \caption{Effect of scoring above the cutoff for never-takers.}
    \label{t:eta}
\begin{center}
    \vspace{0.1cm} 
         \begin{tabular}{lcc}
      \hline \hline
   \\ [-2ex]
    Track   &    $\eta$  &    $\eta$  \\
      & 1& 2 \\
      \\ [-2ex]  
\hline
  \\ [-2ex] 
V: BL          &            -.0097	& .0258\\
          &     \footnotesize     (.0299) & \footnotesize (.0332) \\
V: BL/KL       &       .0336&       .1013\\
            &  \footnotesize (.0521)&  \footnotesize (.0465)\\
V: KL          &       .0492&        .036\\
            &  \footnotesize (.0216)&  \footnotesize (.025)\\
V: KL/GT       &       .0269&      -.0338\\
            &  \footnotesize (.044)&  \footnotesize (.0444)\\
V: GT          &       .0195&      -.0057\\
            &  \footnotesize (.0128)&  \footnotesize (.0154)\\
V: GT/HAVO     &      -.0302&       .0639\\
            &  \footnotesize (.0305)&  \footnotesize (.0265)\\
A: HAVO        &       .0242&       .0278\\
            &  \footnotesize (.0106)&  \footnotesize (.0122)\\
A: HAVO/VWO    &       .0064&       .0551\\
            &  \footnotesize (.0268)&  \footnotesize (.0226)\\
                        \\ [-2ex] 
  \hline
  \\ [-2ex]
VOCATIONAL  &       .0278&       .0338\\
            &  \footnotesize (.0092)&  \footnotesize (.009)\\
ACADEMIC    &       .0196&      -.0215\\
            &  \footnotesize (.0106)&  \footnotesize (.0099)\\
ALL         &       .0245&       .0108\\
            &  \footnotesize (.007)&  \footnotesize (.0068)\\

 \\[-2ex]
\hline
\footnotesize COHORT &  \footnotesize 2015 & \footnotesize 2016 \\
\footnotesize COVARIATES &  \footnotesize YES&  \footnotesize YES \\
      \hline \hline
    \end{tabular}
\end{center}
\vspace{-0.2cm}
\begin{minipage}{1\linewidth  \setstretch{0.75}}
{\scriptsize Notes: The row headings report the names of the secondary school tracks to which a student can be assigned in the Dutch system, ranked by level of difficulty.  The last three rows of the table aggregate the Vocational (VMBO*), the Academic (HAVO*), and All tracks, respectively. Columns 1 and 2 show the estimates of $\eta$ for each track and cohort, as explained in section \ref{s:robust}.
}   
 \end{minipage}
\end{table}

\subsection*{Estimates of APCEs Without Assuming Exclusion Restriction}
\begin{table}[H] 
    \caption{Estimates of the $APCE_{H}$, without exclusion restriction}
\begin{center}
    \vspace{0.1cm} 

\begin{tabular}{l*{5}{c}}
\hline\hline
      \\ [-2ex]       
Track &   Numerator& Lower &    Upper &   Lower  & Upper\\
&  $APCE_{H}$    & bound  &    bound     &   bound &   bound \\
&     & denominator  &  denominator&  $APCE_{H}$ &  $APCE_{H}$  \\
&     & $APCE_{H}$  &  $APCE_{H}$&  \hspace{1.9cm} &  \hspace{1.9cm}   \\
      & 1& 2& 3 & 4 & 5 \\
      \\ [-2ex]  
\hline
  \\ [-2ex] 
V: BL          &       .0187&       .0187&       .6513&       .0289&           1\\
               &  \footnotesize (.0094)&  \footnotesize (.0076)&  \footnotesize (.0164)&  \footnotesize (.0147)&  \footnotesize (.4627)\\
V: BL/KL       &       .0083&       .0083&       .2764&       .0318&           1\\
               &  \footnotesize (.019)&  \footnotesize (.0128)&  \footnotesize (.0239)&  \footnotesize (.069)&  \footnotesize (.6564)\\
V: KL          &       .0555&       .0555&       .6907&       .0803&           1\\
               &  \footnotesize (.0088)&  \footnotesize (.0088)&  \footnotesize (.0112)&  \footnotesize (.0125)&  \footnotesize (0)\\
V: KL/GT       &       .0306&       .0344&       .3982&       .0731&       .8895\\
               &  \footnotesize (.0163)&  \footnotesize (.0135)&  \footnotesize (.021)&  \footnotesize (.0414)&  \footnotesize (.4774)\\
V: GT          &       .0064&       .0072&       .7677&       .0082&       .8888\\
               &  \footnotesize (.0064)&  \footnotesize (.0061)&  \footnotesize (.009)&  \footnotesize (.0124)&  \footnotesize (.4177)\\
V: GT/HAVO     &       .0285&       .0285&       .5041&       .0566&           1\\
               &  \footnotesize (.0098)&  \footnotesize (.0091)&  \footnotesize (.0135)&  \footnotesize (.0191)&  \footnotesize (.337)\\
A: HAVO        &       .0091&       .0091&       .8266&       .0111&           1\\
               &  \footnotesize (.0029)&  \footnotesize (.0028)&  \footnotesize (.006)&  \footnotesize (.0034)&  \footnotesize (.1736)\\
A: HAVO/VWO    &       .0313&       .0313&       .5387&       .0583&           1\\
               &  \footnotesize (.0087)&  \footnotesize (.0086)&  \footnotesize (.0128)&  \footnotesize (.0157)&  \footnotesize (.1429)\\
\\ [-2ex]
\hline
\\ [-2ex]
VOCATIONAL     &       .0232&       .0237&       .6484&       .0357&       .9772\\
               &  \footnotesize (.004)&  \footnotesize (.0037)&  \footnotesize (.0054)&  \footnotesize (.0064)&  \footnotesize (.0761)\\
ACADEMIC       &       .0161&       .0161&       .7357&       .0219&           1\\
               &  \footnotesize (.0034)&  \footnotesize (.0033)&  \footnotesize (.0059)&  \footnotesize (.0046)&  \footnotesize (.0193)\\
ALL            &       .0203&       .0206&       .6838&       .0297&       .9844\\
               &  \footnotesize (.0027)&  \footnotesize (.0025)&  \footnotesize (.004)&  \footnotesize (.004)&  \footnotesize (.0538)\\

            \hline 
COVARIATES         &       YES&       YES&       YES&      YES&       YES\\            
\hline\hline
\end{tabular}
 
\end{center}
\vspace{-0.2cm}
\begin{minipage}{1\linewidth  \setstretch{0.75}}
{\scriptsize Notes: The row headings report the names of the secondary school tracks to which a student can be assigned in the Dutch system, ranked by level of difficulty.  The last three rows of the table aggregate the Vocational (VMBO*), the Academic (HAVO*), and All tracks, respectively. With reference to the $APCE_H$ estimand defined in equation (\ref{e:apce_h_est}), the table reports for the outcome $Y$ and separately for each track, estimates of the numerator (in column 1), of the lower and upper bounds of the denominator (in columns 2 and 3), and of the lower and upper bounds of the $APCE_H$ (in columns 4 and 5), without assuming the exclusion restriction, as given by equations equations (\ref{noER_APCE_H_LB}) and (\ref{noER_APCE_H_UB}), respectively. The aggregations for the three last rows are performed as follows: point estimates of the numerator and of the lower and upper bounds of the denominator are created by weighing each track in the corresponding aggregate by the relative population in each track. For the lower and upper bounds of the $APCE_H$, the aggregated estimates for the lower and upper bounds of the denominators are used directly. Standard errors for each quantity are bootstrapped with 1000 repetitions; each quantity in every iteration is calculated in the same way.}
 \end{minipage}
\end{table}

\begin{table}[H]
    \caption{Estimates of the $APCE_{AH}$, without exclusion restriction}
\begin{center}
    \vspace{0.1cm} 

\begin{tabular}{l*{5}{c}}
\hline\hline
      \\ [-2ex]       
Track &   Numerator& Lower &    Upper &   Lower  & Upper\\
&  $APCE_{AH}$    & bound  &    bound     &   bound &   bound \\
&     & denominator  &  denominator&  $APCE_{AH}$ &  $APCE_{AH}$  \\
&     & $APCE_{AH}$  &  $APCE_{AH}$&  \hspace{1.9cm} &  \hspace{1.9cm}  \\
      & 1& 2& 3 & 4 & 5 \\
      \\ [-2ex]  
\hline
  \\ [-2ex] 
V: BL          &       .0028&       .3187&       .3187&       .0125&       .0125\\
               &  \footnotesize (.0088)&  \footnotesize (.0159)&  \footnotesize (.0158)&  \footnotesize (.0283)&  \footnotesize (.0283)\\
V: BL/KL       &       .0805&       .6858&        .705&       .1138&       .1167\\
               &  \footnotesize (.0213)&  \footnotesize (.0248)&  \footnotesize (.0247)&  \footnotesize (.0304)&  \footnotesize (.0303)\\
V: KL          &       .0433&       .2282&       .2392&       .1827&        .191\\
               &  \footnotesize (.0055)&  \footnotesize (.0101)&  \footnotesize (.0101)&  \footnotesize (.0216)&  \footnotesize (.0224)\\
V: KL/GT       &       .0709&       .5681&       .5791&       .1224&       .1248\\
               &  \footnotesize (.0169)&  \footnotesize (.022)&  \footnotesize (.0225)&  \footnotesize (.0292)&  \footnotesize (.0293)\\
V: GT          &       .0106&       .1908&       .1916&        .054&       .0542\\
               &  \footnotesize (.0063)&  \footnotesize (.0077)&  \footnotesize (.0077)&  \footnotesize (.024)&  \footnotesize (.024)\\
V: GT/HAVO     &       .0632&       .4169&       .4262&       .1486&       .1522\\
               &  \footnotesize (.0081)&  \footnotesize (.0131)&  \footnotesize (.0132)&  \footnotesize (.0183)&  \footnotesize (.0185)\\
A: HAVO        &       .0037&       .1176&       .1186&       .0315&       .0318\\
               &  \footnotesize (.0022)&  \footnotesize (.0052)&  \footnotesize (.0052)&  \footnotesize (.018)&  \footnotesize (.0181)\\
A: HAVO/VWO    &       .0546&       .3667&       .3777&       .1455&       .1498\\
               &  \footnotesize (.0076)&  \footnotesize (.0116)&  \footnotesize (.0117)&  \footnotesize (.0192)&  \footnotesize (.0196)\\
\\ [-2ex]
\hline
\\ [-2ex]
VOCATIONAL     &       .0323&       .2977&       .3034&       .1064&       .1084\\
               &  \footnotesize (.0036)&  \footnotesize (.005)&  \footnotesize (.005)&  \footnotesize (.0117)&  \footnotesize (.0119)\\
ACADEMIC       &       .0198&       .1962&       .2004&       .0988&       .1009\\
               &  \footnotesize (.0028)&  \footnotesize (.0052)&  \footnotesize (.0052)&  \footnotesize (.0136)&  \footnotesize (.0138)\\
ALL            &       .0272&       .2565&       .2616&        .104&       .1061\\
               &  \footnotesize (.0024)&  \footnotesize (.0036)&  \footnotesize (.0037)&  \footnotesize (.0091)&  \footnotesize (.0092)\\
             \hline 
COVARIATES         &       YES&       YES&       YES&      YES&       YES\\   
\hline\hline
\end{tabular}
 
\end{center}
\vspace{-0.2cm}
\begin{minipage}{1\linewidth  \setstretch{0.75}}
{\scriptsize Notes: The row headings report the names of the secondary school tracks to which a student can be assigned in the Dutch system, ranked by level of difficulty.  The last three rows of the table aggregate the Vocational (VMBO*), the Academic (HAVO*), and All tracks, respectively. With reference to the $APCE_H$ estimand defined in equation (\ref{e:apce_h_est}), the table reports for the outcome $Y$ and separately for each track, estimates of the numerator (in column 1), of the lower and upper bounds of the denominator (in columns 2 and 3), and of the lower and upper bounds of the $APCE_H$ (in columns 4 and 5), without assuming the exclusion restriction, as given by equations equations (\ref{noER_APCE_AL_LB}) and (\ref{noER_APCE_AL_UB}), respectively. The aggregations for the three last rows are performed as follows: point estimates of the numerator and of the lower and upper bounds of the denominator are created by weighing each track in the corresponding aggregate by the relative population in each track. For the lower and upper bounds of the $APCE_H$, the aggregated estimates for the lower and upper bounds of the denominators are used directly. Standard errors for each quantity are bootstrapped with 1000 repetitions; each quantity in every iteration is calculated in the same way.}
 \end{minipage}
\end{table}

\begin{table}[H]
    \caption{Estimates of the $APCE_{AL}$, without exclusion restriction}
\begin{center}
    \vspace{0.1cm} 

\begin{tabular}{l*{5}{c}}
\hline\hline
      \\ [-2ex]       
Track &   Numerator& Lower &    Upper &   Lower  & Upper\\
&  $APCE_{AL}$    & bound  &    bound     &   bound &   bound \\
&     & denominator  &  denominator&  $APCE_{AL}$ &  $APCE_{AL}$  \\
&     & $APCE_{AL}$  &  $APCE_{AL}$&  \hspace{1.9cm} &  \hspace{1.9cm}  \\
      & 1& 2& 3 & 4 & 5 \\
      \\ [-2ex]  
\hline
  \\ [-2ex] 
V: BL          &       .0277&       .0286&        .664&       .0411&       .9776\\
               &  \footnotesize (.0053)&  \footnotesize (.0053)&  \footnotesize (.0172)&  \footnotesize (.0078)&  \footnotesize (.0266)\\
V: BL/KL       &       .0351&       .0378&       .2867&       .1237&       .9394\\
               &  \footnotesize (.0103)&  \footnotesize (.0099)&  \footnotesize (.0253)&  \footnotesize (.0338)&  \footnotesize (.0884)\\
V: KL          &       .0741&       .0811&       .7053&       .1049&       .9132\\
               &  \footnotesize (.0078)&  \footnotesize (.0076)&  \footnotesize (.0131)&  \footnotesize (.0105)&  \footnotesize (.0266)\\
V: KL/GT       &       .0321&       .0337&       .3865&       .0825&       .9603\\
               &  \footnotesize (.0098)&  \footnotesize (.0095)&  \footnotesize (.0244)&  \footnotesize (.0236)&  \footnotesize (.0584)\\
V: GT          &       .0382&       .0416&       .8012&       .0477&       .9197\\
               &  \footnotesize (.0039)&  \footnotesize (.0038)&  \footnotesize (.0095)&  \footnotesize (.0048)&  \footnotesize (.0256)\\
V: GT/HAVO     &       .0691&       .0789&       .5453&       .1264&       .8734\\
               &  \footnotesize (.0091)&  \footnotesize (.0086)&  \footnotesize (.0157)&  \footnotesize (.0152)&  \footnotesize (.0395)\\
A: HAVO        &       .0506&       .0558&       .8722&       .0579&       .9066\\
               &  \footnotesize (.0042)&  \footnotesize (.004)&  \footnotesize (.0059)&  \footnotesize (.0047)&  \footnotesize (.0249)\\
A: HAVO/VWO    &       .0851&       .0946&        .591&       .1442&       .8999\\
               &  \footnotesize (.0087)&  \footnotesize (.0085)&  \footnotesize (.0144)&  \footnotesize (.0134)&  \footnotesize (.0258)\\
\\ [-2ex]
\hline
\\ [-2ex]
VOCATIONAL     &        .049&       .0538&       .6731&       .0728&       .9115\\
               &  \footnotesize (.0029)&  \footnotesize (.0027)&  \footnotesize (.0062)&  \footnotesize (.0041)&  \footnotesize (.0148)\\
ACADEMIC       &       .0615&        .068&       .7835&       .0785&       .9036\\
               &  \footnotesize (.004)&  \footnotesize (.0039)&  \footnotesize (.0063)&  \footnotesize (.0049)&  \footnotesize (.0176)\\
ALL            &       .0541&       .0596&       .7179&       .0753&       .9079\\
               &  \footnotesize (.0024)&  \footnotesize (.0023)&  \footnotesize (.0044)&  \footnotesize (.0032)&  \footnotesize (.0113)\\
             \hline 
COVARIATES         &       YES&       YES&       YES&      YES&       YES\\   
\hline\hline
\end{tabular}
 
\end{center}
\vspace{-0.2cm}
\begin{minipage}{1\linewidth  \setstretch{0.75}}
{\scriptsize Notes: The row headings report the names of the secondary school tracks to which a student can be assigned in the Dutch system, ranked by level of difficulty.  The last three rows of the table aggregate the Vocational (VMBO*), the Academic (HAVO*), and All tracks, respectively. With reference to the $APCE_H$ estimand defined in equation (\ref{e:apce_h_est}), the table reports for the outcome $Y$ and separately for each track, estimates of the numerator (in column 1), of the lower and upper bounds of the denominator (in columns 2 and 3), and of the lower and upper bounds of the $APCE_H$ (in columns 4 and 5), without assuming the exclusion restriction, as given by equations equations (\ref{noER_APCE_AH_LB}) and (\ref{noER_APCE_AH_UB}), respectively. The aggregations for the three last rows are performed as follows: point estimates of the numerator and of the lower and upper bounds of the denominator are created by weighing each track in the corresponding aggregate by the relative population in each track. For the lower and upper bounds of the $APCE_H$, the aggregated estimates for the lower and upper bounds of the denominators are used directly. Standard errors for each quantity are bootstrapped with 1000 repetitions; each quantity in every iteration is calculated in the same way.}
 \end{minipage}
\end{table}

\newpage
\subsection*{Estimates of the $APCE_J$ based on unconfoundedness}

Table \ref{t:APCE_unconfoundedness} shows the point estimates of $APCE_J$ from assuming unconfoundedness. As can be seen, the results show that the estimates for the three strata are similar in all tracks.

\begin{table}[ht]
    \caption{Estimates of the $APCE_{J}$ based on unconfoundedness assumption}
    \label{t:APCE_unconfoundedness}
\begin{center}
    \vspace{0.1cm} 
\begin{tabular}{l*{5}{c}}
\hline\hline
      \\ [-2ex]       
Track &   $APCE_{AL}$& $APCE_{H}$ & $APCE_{AH}$ \\
      & 1& 2& 3  \\
      \\ [-2ex]  
\hline
  \\ [-2ex] 
V: BL          &       0.0444&       0.0481&       0.0508\\
               & {\footnotesize (0.0071)}& {\footnotesize (0.0076)}& {\footnotesize (0.0089)}\\
V: BL/KL       &       0.1360&       0.1357&       0.1311\\
               & {\footnotesize (0.0306)}& {\footnotesize (0.0236)}& {\footnotesize (0.0207)}\\
V: KL          &       0.1560&       0.1682&       0.1807\\
               & {\footnotesize (0.0107)}& {\footnotesize (0.0108)}& {\footnotesize (0.0129)}\\
V: KL/GT       &       0.1247&       0.1288&       0.1379\\
               & {\footnotesize (0.0206)}& {\footnotesize (0.0175)}& {\footnotesize (0.0182)}\\
V: GT          &       0.0497&       0.0556&       0.0584\\
               & {\footnotesize (0.0040)}& {\footnotesize (0.0047)}& {\footnotesize (0.0057)}\\
V: GT/HAVO     &       0.1530&       0.1607&       0.1659\\
               & {\footnotesize (0.0122)}& {\footnotesize (0.0117)}& {\footnotesize (0.0126)}\\
A: HAVO        &       0.0574&       0.0700&       0.0776\\
               & {\footnotesize (0.0041)}& {\footnotesize (0.0051)}& {\footnotesize (0.0063)}\\
A: HAVO/VWO    &       0.1531&       0.1674&       0.1770\\
               & {\footnotesize (0.0099)}& {\footnotesize (0.0103)}& {\footnotesize (0.0116)}\\
              \\ [-2ex] 
  \hline
  \\ [-2ex]
VOCATIONAL  &       0.0969&       0.1037&       0.1090\\
            & {\footnotesize (0.0040)}& {\footnotesize (0.0039)}& {\footnotesize (0.0045)}\\
ACADEMIC    &       0.0876&       0.1007&       0.1090\\
            & {\footnotesize (0.0042)}& {\footnotesize (0.0048)}& {\footnotesize (0.0057)}\\
ALL         &       0.0931&       0.1025&       0.1090\\
            & {\footnotesize (0.0029)}& {\footnotesize (0.0030)}& {\footnotesize (0.0035)}\\
\hline\hline
\end{tabular}

\end{center}
\vspace{-0.2cm}
\begin{minipage}{1\linewidth  \setstretch{0.75}}
{\scriptsize Notes: The row headings report the names of the secondary school tracks to which a student can be assigned in the Dutch system, ranked by level of difficulty.  The last three rows of the table aggregate the Vocational (VMBO*), the Academic (HAVO*), and All tracks, respectively. The columns show point estimates obtained by assuming unconfoundedness. Standard errors for each quantity are bootstrapped with 1000 repetitions; each quantity in every iteration is calculated in the same way.}
 \end{minipage}
\end{table}

\clearpage

\newpage
\section*{Appendix to Section \ref{s:fair}}\label{ap:fair}
\subsection*{Estimates of fairness by gender}

\begin{figure}[h]
    \caption{Fairness by gender}
    \label{f:fair_gender}
    \centering
\includegraphics[width = 15 cm]{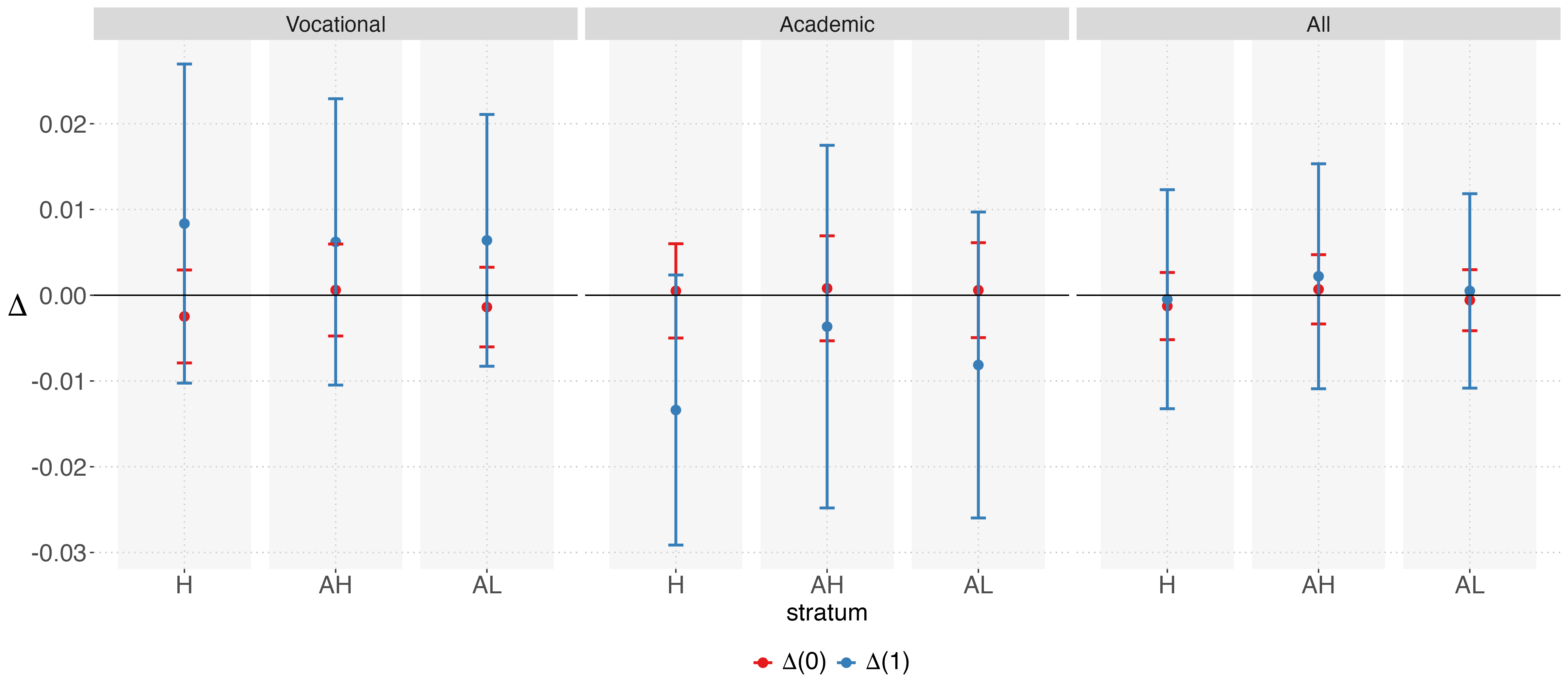}
\vspace{-0.2cm}
\begin{minipage}{0.9\linewidth  \setstretch{0.75}}
\scriptsize  Notes:  the figure plots,  for the three aggregate tracks and for the H, AH and AL strata, the point estimates of the statistic 
$\Delta_{j}(z) = Pr(R(z)=1 \mid \text{Girl}) - Pr(R(z) = 1 \mid \text{Boy})$  and the correspondent 95\% confidence intervals.
\end{minipage}
\end{figure}

\end{spacing}
\end{document}